\documentclass[pre,aps,preprint,amsmath,amssymb]{revtex4}
\usepackage{graphicx}
\usepackage{epsfig}
\include{graphics}
\usepackage{epstopdf} 
\begin{document}

\title{Highly controllable stabilization and switching of multiple colliding soliton sequences 
with generic Ginzburg-Landau gain-loss}

\author{Avner Peleg$^{1}$ and Toan T. Huynh$^{2}$}

\affiliation{$^{1}$ Department of Mathematics, Azrieli College of Engineering, 
Jerusalem 9371207, Israel}
\affiliation{$^{2}$ Department of Mathematics, University of Medicine and Pharmacy 
at Ho Chi Minh City, Ho Chi Minh City, Vietnam}


\begin{abstract} 
We investigate propagation of $J$ soliton sequences in a nonlinear optical waveguide 
array with generic weak Ginzburg-Landau (GL) gain-loss and nearest-neighbor (NN) interaction. 
The propagation is described by a system of $J$ perturbed coupled nonlinear Schr\"odinger (NLS) 
equations. The NN interaction property leads to the elimination of 
collisional three-pulse interaction effects, which prevented the observation of 
stable multisequence soliton propagation with $J>2$ sequences in the presence of 
generic GL gain-loss in all previous studies. We show that the dynamics of soliton amplitudes 
can be described by a generalized $J$-dimensional Lotka-Volterra (LV) model. 
Stability and bifurcation analysis for the equilibrium points of the LV model, 
which is augmented by an application of the Lyapunov function method, 
is used to develop setups that lead to robust and scalable transmission stabilization 
and switching for a general $J$ value. The predictions of the LV model are confirmed 
by extensive numerical simulations with the perturbed coupled-NLS model with $J=3$, $4$, 
and $5$ soliton sequences. Furthermore, soliton stability and the agreement between the 
LV model's predictions and the simulations are independent of $J$. Therefore, our 
study provides the first demonstration of robust control of multiple colliding 
sequences of NLS solitons in the presence of generic weak GL gain-loss with an 
arbitrary number of sequences. Due to the robustness and scalability 
of the results, they can have important applications in stabilization and switching 
of broadband soliton-based optical waveguide transmission. 
\end{abstract}

\maketitle

\section{Introduction}
\label{intro}

The cubic nonlinear Schr\"odinger (NLS) equation, which describes propagation of waves   
in the presence of second-order dispersion and cubic (Kerr) nonlinearity, 
is one of the most extensively used nonlinear wave models in science and engineering. 
It describes a variety of nonlinear wave phenomena in plasmas \cite{Malomed89,Asano69,Horton96}, 
water wave dynamics \cite{Zakharov84,Newell85}, Bose-Einstein condensates \cite{Dalfovo99,BEC2008}, 
and propagation of pulses of light in nonlinear optical waveguides \cite{Agrawal2019,Hasegawa95,Iannone98}.    
The fundamental NLS solitons are the most notable solutions of the cubic NLS equation 
due to their stability and shape preserving properties. Because of these properties, 
fundamental NLS solitons are being considered for applications in many nonlinear optical 
waveguide systems, including optical waveguide communication lines, optical switches, 
pulsed waveguide lasers, and pulse compression \cite{Agrawal2019,Iannone98,Mollenauer2006,Agrawal2020}.

The application of fundamental NLS solitons in nonlinear optical waveguide 
communication systems is considered by many as one of the most important applications for solitons 
of a nonlinear wave model \cite{Agrawal2019,Iannone98,Mollenauer2006,Hasegawa2022}. 
The rates of transmission of information in these optical communication systems 
can be substantially increased by {\it multisequence transmission}, i.e., by sending 
many pulse sequences through the same optical waveguide \cite{Agrawal2019,Iannone98,Mollenauer2006,Multisequence}. Thus, in multisequence transmission, 
the pulses in each sequence propagate with the same central frequency and group velocity, 
but the central frequency and group velocity are different for pulses from different 
sequences \cite{Agrawal2019,Iannone98,Mollenauer2006}. Since pulses from different 
sequences propagate with different group velocities, intersequence pulse collisions 
are very frequent, and can therefore cause significant amplitude shifts, pulse distortion 
due to radiation emission, transmission destabilization, and transmission errors.  
 For this reason, significant research efforts have been devoted to the study of  
intersequence pulse collisions in general \cite{Agrawal2020,Tkach97,Essiambre2010}, 
and to the investigation of intersequence collisions of NLS solitons in particular \cite{Agrawal2019,Hasegawa95,Iannone98,Mollenauer2006}.

In several earlier works \cite{NP2010,PNC2010,PC2012,CPJ2013,NPT2015,CPN2016,PNT2016,
PNH2017A,PC2018A}, we developed general methods for stabilizing multisequence propagation 
of NLS solitons against the harmful effects of intersequence pulse collisions. 
The methods combined stabilization against collision-induced amplitude shifts 
with stabilization against radiation emission effects. Stabilization against 
collision-induced amplitude shifts was realized by showing that the dynamics 
of soliton amplitudes in $J$-sequence transmission systems can be described by 
generalized $J$-dimensional Lotka-Volterra (LV) models. The specific form of 
the LV model is determined by the dissipative perturbation terms in the cubic NLS model, 
which describe the dissipative processes in the optical waveguide. 
Stability and bifurcation analysis for the equilibrium points of the LV models 
was used to develop waveguide setups that lead to robust transmission stabilization 
\cite{CPJ2013,NPT2015,CPN2016,PNT2016,PNH2017A,PC2018A} and to robust transmission 
switching \cite{CPJ2013,NPT2015,PNH2017A}. Stabilization against radiation emission 
was accomplished by three main methods. In the first method, we employed 
perturbation-induced shifting of the soliton's frequency (e.g., due to delayed 
Raman response) along with frequency-dependent linear gain-loss \cite{CPN2016,PNT2016}. 
In the second method, we used nonlinear waveguides with a weak Ginzburg-Landau (GL) gain-loss profile, 
consisting of linear loss, cubic gain, and quintic loss \cite{PC2012,CPJ2013,NPT2015}. 
In the third method, the transmission was stabilized by combining perturbation-induced 
shifting of the soliton's frequency with weak GL gain-loss \cite{PNH2017A}.  
The application of these stabilization methods enabled the observation of stable 
multisequence soliton transmission over distances of 1000 dispersion lengths 
or more \cite{CPJ2013,NPT2015,CPN2016,PNT2016,PNH2017A,PC2018A} and the realization 
of efficient transmission switching of multiple soliton sequences 
\cite{CPJ2013,NPT2015,PNH2017A}.

Despite the impressive progress in transmission stabilization that was achieved 
in Refs. \cite{NP2010,PNC2010,PC2012,CPJ2013,NPT2015,CPN2016,PNT2016,PNH2017A,PC2018A}, 
these works suffer from some very important shortcomings. First, transmission quality 
and stability in all these works decreased significantly with the increase in the 
number of soliton sequences. Second, stabilization in waveguides with weak GL gain-loss 
was either limited to two-sequence transmission \cite{PC2012,CPJ2013,NPT2015}, 
or to transmission in the presence of nongeneric (narrowband) GL gain-loss \cite{PNH2017A}, 
where the cubic gain and the quintic loss did not affect the collision-induced 
amplitude changes at all. This limitation is a consequence of the complex nature 
of three-pulse interaction in three-soliton collisions 
in the presence of quintic loss \cite{PC2012,PNG2014}. Indeed, the 
complex nature of collisional three-pulse interaction creates a serious obstacle 
for constructing LV models for amplitude dynamics in multisequence transmission 
systems with generic (broadband) GL gain-loss and more than two soliton sequences. 
In the absence of an appropriate LV model, it is completely unclear how to stabilize 
the dynamics of soliton amplitudes against collision-induced amplitude shifts. 
For this reason, transmission stabilization and switching in waveguides with 
a GL gain-loss profile have been so far limited to two-sequence systems 
\cite{PC2012,CPJ2013,NPT2015}, or to systems with nongeneric GL gain-loss \cite{PNH2017A}.

In the current paper, we overcome the aforementioned key shortcomings of all previous 
works on transmission stabilization and switching with multiple sequences of NLS 
solitons. For this purpose, we investigate propagation of $J$ colliding soliton sequences 
in a nonlinear optical waveguide array with weak generic (broadband) GL gain-loss and 
nearest-neighbor (NN) interaction. The propagation is described by a system of $J$ 
weakly perturbed coupled-NLS equations. The NN interaction property leads to the complete 
elimination of collisional three-pulse interaction effects, and in this manner, enables 
the first investigation of robust transmission stabilization and switching with an 
arbitrary number of soliton sequences in the presence of generic weak GL gain-loss.

We derive the reduced ordinary differential equation (ODE) model for the dynamics 
of soliton amplitudes in $J$-sequence transmission systems, and show that it has 
the form of a generalized $J$-dimensional LV model with NN interaction. 
We then carry out linear stability analysis and bifurcation analysis for the 
equilibrium points of the LV model and determine the regions in parameter space, 
which are suitable for transmission stabilization and transmission switching.   
Additionally, we use an auxiliary uncoupled nonlinear ODE model and the Lyapunov 
function method for the full LV model to determine the regions in phase space, 
where transmission switching can be realized. The predictions of the LV model are confirmed 
by extensive numerical simulations with the weakly perturbed coupled-NLS model with 3, 4, 
and 5 soliton sequences. Furthermore, soliton stability and the agreement between the LV 
model's predictions and the coupled-NLS simulations are independent of the number of sequences $J$, 
which is a drastic improvement compared with all previous studies of multisequence 
soliton transmission. Based on these results we conclude that robust transmission 
stabilization and transmission switching with an arbitrary number of soliton sequences 
can be achieved in nonlinear waveguide arrays with generic weak GL gain-loss and NN interaction. 
Moreover, the results clearly show that the design of the waveguide setups can be founded on 
stability and bifurcation analysis for the equilibrium points of the LV model.

Our results are also important in the context of research on systems described 
by the complex GL equation, which is another central model in nonlinear 
science \cite{Hohenberg92,Kramer2002}. The complex GL equation describes, 
for example, instabilities, convection, and pattern formation in fluids 
\cite{Kramer2002,Newell93,Stewartson71,Malomed90}, mode-locked lasers 
\cite{Moores93,Akhmediev96,Kutz2006,Wise2008}, and pattern formation 
in diffusion-reaction systems \cite{Kuramoto75,Meron2002}. 
In this context, our previous work in Ref. \cite{PNH2017A} provided the 
first observation of stable long-distance multisequence propagation 
with more than two soliton sequences in a system described by the 
complex GL equation. However, the results in Ref. \cite{PNH2017A} 
were quite restricted, since a nongeneric narrowband GL gain-loss 
profile was considered, and since the cubic gain and quintic loss 
had no effect on the collision-induced changes in soliton amplitudes 
in this work. In the current work, we significantly extend the results 
of Ref. \cite{PNH2017A} by providing the first demonstration of stable  
long-distance propagation of an arbitrary number of soliton sequences 
in a complex GL system with {\it generic} (broadband) gain-loss.     
Furthermore, in contrast to the situation in Ref. \cite{PNH2017A}, 
in the current paper, the cubic gain and quintic loss affect both 
the amplitude changes due to single-soliton propagation and the 
amplitude changes due to intersequence soliton collisions.

The other sections of the paper are organized in the following manner. 
In Section \ref{NLS_model}, we present the perturbed coupled-NLS propagation 
model and discuss its significance. In Section \ref{LV_models}, we obtain 
the corresponding $J$-dimensional LV model for dynamics of soliton amplitudes. 
In Section \ref{stability}, we carry out stability and bifurcation  analysis 
for the equilibrium points of the LV model, and use the results to find the 
regions in parameter space and in phase space, where robust transmission 
stabilization and transmission switching can be realized. 
In Section \ref{simu}, we present the results of numerical simulations 
with the perturbed coupled-NLS model for transmission stabilization and 
switching with 3, 4, and 5 soliton sequences. We also present a careful 
comparison of the simulations results with the predictions of the LV model. 
Our conclusions are presented in Section \ref{conclusions}. 
In Appendix \ref{appendA}, we describe the calculation of the pulse-pattern 
quality integrals.

\section{Perturbed coupled-NLS and Lotka-Volterra models}
\label{models}

\subsection{The perturbed coupled-NLS model for multisequence propagation}
\label{NLS_model}

\begin{figure}[ptb]
\begin{center}
\epsfxsize=10cm  \epsffile{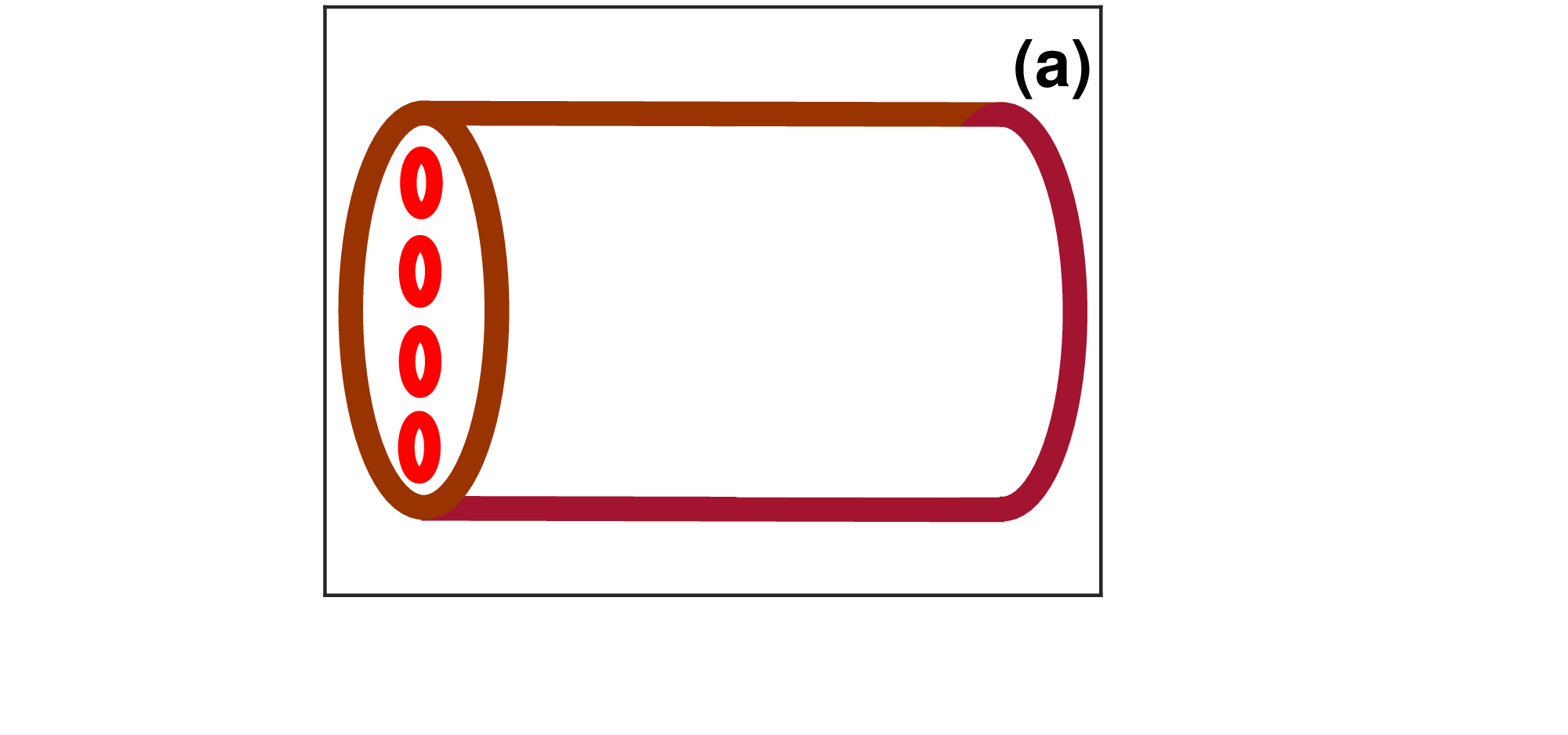}\\
\epsfxsize=10cm  \epsffile{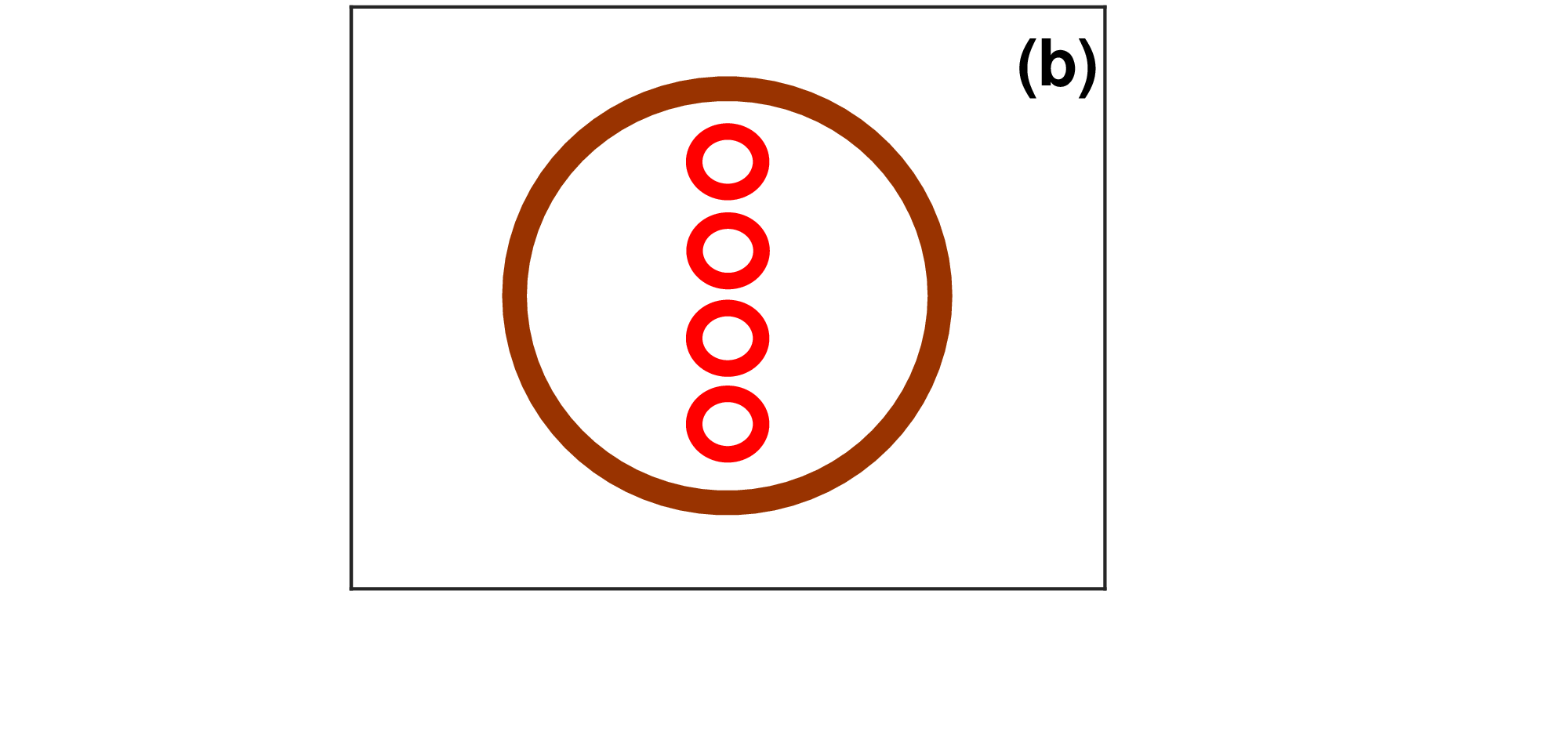}
\end{center}
\caption{(Color online) 
A sketch of a waveguide array with 4 waveguides. 
(a) A side view of the array. (b) The cross section.}
\label{fig1}
\end{figure}

We consider the propagation of $J$ sequences of optical pulses 
in a nonlinear optical waveguide array consisting of $J$ close waveguides. 
A sketch of the nonlinear waveguide array is shown in Fig. 1. 
Each pulse sequence propagates inside its own waveguide in the 
presence of second-order dispersion, broadband cubic (Kerr) nonlinearity, 
and a broadband (generic) weak GL gain-loss profile consisting of weak linear gain-loss, 
cubic gain, and quintic loss. The linear gain-loss is the difference 
between linear amplifier gain and linear waveguide loss, where amplifier gain 
can be realized, for example, by distributed Raman amplification 
\cite{Islam2004,Agrawal2005}. Due to the broadband (generic) nature of 
the cubic nonlinearity and the cubic and quintic gain 
and loss, the pulses in each sequence interact with pulses from other 
sequences during intersequence collisions. However, we assume that the 
magnitude of the electric field of the pulses from a given sequence 
decays sufficiently fast with increasing distance from the pulse sequence's 
waveguide, such that only the interaction between pulses from NN waveguides 
is significant, while all other intersequence interactions are negligible. 
We denote the dimensionless envelope of the electric field for the pulse 
sequence in the $j$th waveguide by $\psi_{j}$, and the dimensionless distance 
and time by $z$ and $t$. The propagation is then described by the following 
system of weakly perturbed coupled-NLS equations:       
\begin{eqnarray}&&
\!\!\!\!\!\!\!\!\!
i\partial_z\psi_{j} + \partial^2_t\psi_{j}
+ 2|\psi_{j}|^2\psi_{j} + 
4\sigma \left(|\psi_{j-1}|^2 + |\psi_{j+1}|^2\right)\psi_{j}=
ig_j\psi_{j}/2 + i\epsilon_3|\psi_{j}|^2\psi_{j}
\nonumber \\ &&
\!\!\!\!\!\!\!\!\!
-i\epsilon_5|\psi_{j}|^4\psi_{j} 
+2i\sigma\epsilon_3 \left(|\psi_{j-1}|^2 + |\psi_{j+1}|^2\right)\psi_{j}
-3i\sigma \epsilon_5 \left(|\psi_{j-1}|^4 + |\psi_{j+1}|^4 \right)\psi_{j}
\nonumber \\ &&
\!\!\!\!\!\!\!\!\!
-6i\sigma \epsilon_5 
\left(|\psi_{j-1}|^2 + |\psi_{j+1}|^2\right) |\psi_{j}|^2 \psi_{j},
\label{gl1}
\end{eqnarray}
for $2 \le j \le J-1$,
\begin{eqnarray}&&
\!\!\!\!\!\!\!\!\!\!\!\!\!\!\!
i\partial_z\psi_{1} + \partial^2_t\psi_{1}
+ 2|\psi_{1}|^2\psi_{1} + 
4\sigma |\psi_{2}|^2\psi_{1}=
ig_j\psi_{1}/2 + i\epsilon_3|\psi_{1}|^2\psi_{1}
-i\epsilon_5|\psi_{1}|^4\psi_{1} 
\nonumber \\ &&
\!\!\!\!\!\!\!\!\!\!\!\!\!\!\!
+2i\sigma\epsilon_3 |\psi_{2}|^2\psi_{1}
-3i\sigma\epsilon_5 |\psi_{2}|^4\psi_{1}
-6i\sigma \epsilon_5 |\psi_{2}|^2 |\psi_{1}|^2 \psi_{1},
\label{gl2}
\end{eqnarray}
for $j=1$, and 
\begin{eqnarray}&&
\!\!\!\!\!\!\!\!\!\!\!\!\!\!\!
i\partial_z\psi_{J} + \partial^2_t\psi_{J}
+ 2|\psi_{J}|^2\psi_{J} + 
4\sigma |\psi_{J-1}|^2\psi_{J}=
ig_j\psi_{J}/2 + i\epsilon_3|\psi_{J}|^2\psi_{J}
-i\epsilon_5|\psi_{J}|^4\psi_{J} 
\nonumber \\ &&
\!\!\!\!\!\!\!\!\!\!\!\!\!\!\!
+2i\sigma\epsilon_3 |\psi_{J-1}|^2\psi_{J}
-3i\sigma\epsilon_5 |\psi_{J-1}|^4\psi_{J}
-6i\sigma \epsilon_5 |\psi_{J-1}|^2 |\psi_{J}|^2 \psi_{J},
\label{gl3}
\end{eqnarray}
for $j=J$. The linear gain-loss, cubic gain, and quintic loss coefficients
in Eqs. (\ref{gl1})-(\ref{gl3}), $g_{j}$, $\epsilon_{3}$, and $\epsilon_{5}$, 
satisfy $|g_{j}| \ll 1$,  $0 < \epsilon_{3} \ll 1$, and $0 < \epsilon_{5} \ll 1$.     
The coefficient $\sigma$ characterizes the reduction in the strength of 
intersequence interaction compared with intrasequence interaction. 
It is associated with the reduction in the magnitude of the electric field 
of the $j$th sequence with increasing distance from the $j$th waveguide.       
The second terms on the left hand sides of Eqs. (\ref{gl1})-(\ref{gl3}) are 
due to second-order dispersion. The third and fourth terms on the left hand 
sides of these equations describe intrasequence and intersequence interaction 
due to cubic nonlinearity. The first terms on the right hand sides of 
Eqs. (\ref{gl1})-(\ref{gl3}) are due to linear gain-loss, while the second and 
third terms represent intrasequence interaction due to cubic gain and 
quintic loss, respectively. Additionally, the fourth terms on the right hand sides of these 
equations describe intersequence interaction due to cubic gain, while the fifth 
and sixth terms represent intersequence interaction due to quintic loss.       
Note that since the cubic nonlinearity, the cubic gain, and the quintic loss 
are generic, i.e. broadband, we take into account both intrasequence and 
intersequence interaction for all three processes.

We point out that somewhat similar perturbed coupled-NLS models with a weak 
GL gain-loss profile were considered by us in several earlier works 
\cite{PC2012,CPJ2013,NPT2015,PNH2017A}. However, the perturbed coupled-NLS 
model considered in the current paper is the first that takes into account 
a generic (broadband) GL gain-loss profile for a general number of soliton 
sequences $J$. The limitations of the perturbed coupled-NLS models of 
Refs. \cite{PC2012,CPJ2013,NPT2015,PNH2017A} are associated with the 
complex nature of three-pulse interaction in generic three-soliton 
collisions in the presence of quintic loss (see Refs. \cite{PC2012,PNG2014}). 
Due to the complex nature of the collisional three-pulse interaction effects, 
it is very difficult to construct LV models for amplitude dynamics in multisequence 
soliton transmission in the presence of a generic GL gain-loss profile 
for $J>2$ sequences. In the absence of a $J$-dimensional LV model, 
it is unclear how to stabilize the transmission against the collision-induced 
amplitude shifts. In the current paper, we circumvent these difficulties 
by considering multisequence propagation in waveguide arrays with NN interaction. 
The NN interaction property leads to the complete elimination of the 
three-pulse interaction effects, and in this manner, enables the construction 
of $J$-dimensional LV models for amplitude dynamics for a general $J$ value. 
This opens the way for developing waveguide setups for transmission stabilization 
and transmission switching with a general $J$ value.

The dimensionless physical quantities are related to the dimensional quantities  
by the standard scaling relations for NLS solitons \cite{Agrawal2019}. The same 
scaling rules were used in our previous works on multisequence propagation of 
NLS solitons \cite{NPT2015,PNH2017A,PC2020}. In particular, the dimensionless 
distance $z$ in Eqs. (\ref{gl1})-(\ref{gl3}) is $z=X/(2L_{D})$, where $X$ is the 
dimensional distance, $L_{D}=\tau_{0}^{2}/|\tilde\beta_{2}|$ is the dispersion length,    
$\tau_{0}$ is the soliton width, and $\tilde\beta_{2}$ is the second-order
dispersion coefficient. The dimensionless time is $t=\tau/\tau_{0}$, 
where $\tau$ is time. $\psi_{j}=(\gamma_{3} \tau_{0}^{2}/|\tilde\beta_{2}|)^{1/2}E_{j}$, 
where $E_{j}$ is the electric field of the $j$th pulse sequence and $\gamma_{3}$ 
is the cubic nonlinearity coefficient. The coefficients $g_{j}$, $\epsilon_{3}$, 
and $\epsilon_{5}$ are related to the dimensional linear gain-loss, cubic gain, 
and quintic loss coefficients $\rho_{1j}$, $\rho_{3}$, and $\rho_{5}$ by:  
$g_{j}=2\rho_{1j}\tau_{0}^{2}/|\tilde\beta_{2}|$, $\epsilon_{3}=2\rho_{3}/\gamma_{3}$, 
and $\epsilon_{5}=2\rho_{5}|\tilde\beta_{2}|/(\gamma_{3}^{2}\tau_{0}^{2})$.

In the absence of gain and loss and intersequence interaction, the propagation 
of the $j$th pulse sequence is described by the unperturbed cubic NLS equation
\begin{eqnarray} &&
i\partial_z\psi_{j}+\partial_{t}^2\psi_{j}+2|\psi_{j}|^2\psi_{j}=0 .
\label{gl4}
\end{eqnarray}        
The fundamental soliton solution of Eq. (\ref{gl4}) with group velocity 
$2\beta_{j}$ is $\psi_{sj}(t,z)=\eta_{j}\exp(i\chi_{j})\mbox{sech}(x_{j})$,
where $x_{j}=\eta_{j}\left(t-y_{j}-2\beta_{j} z\right)$, 
$\chi_{j}=\alpha_{j}+\beta_{j}(t-y_{j})+
\left(\eta_{j}^2-\beta_{j}^{2}\right)z$, 
and $\eta_{j}$, $y_{j}$, and $\alpha_{j}$ 
are the soliton amplitude, position, and phase, respectively. 
Due to the large group velocity differences between the soliton sequences, 
the pulses undergo a large number of fast intersequence collisions. 
The energy exchange in the collisions due to cubic gain and quintic loss 
can lead to significant amplitude shifts and to emission of radiation. 
Radiation is also emitted due to the effects of cubic nonlinearity on the 
collisions and due to the effects of cubic gain and quintic loss on 
single-soliton propagation. All these unwanted effects can cause 
destabilization of the soliton sequences and severe transmission degradation. 
However, it might be possible to counteract these destabilizing effects 
by linear gain-loss with properly chosen $g_{j}$ coefficients. 
In the current paper we demonstrate both theoretically and by 
numerical simulations that such stabilization of multisequence soliton 
propagation with a general number of sequences $J$ can indeed be realized 
in a robust manner. Furthermore, we show that changes in the value of the 
ratio $\epsilon_{3}/\epsilon_{5}$ can be used to induce transmission switching 
of $M$ out of the $J$ soliton sequences for general values of $J$ and $M$.

\subsection{The generalized Lotka-Volterra models for amplitude dynamics}
\label{LV_models}

Highly useful insight about pulse dynamics in a system with $J$ soliton 
sequences can be obtained by deriving generalized $J$-dimensional LV models for the dynamics 
of soliton amplitudes \cite{NP2010,PNC2010,PC2012,CPJ2013,NPT2015,PNT2016,PNH2017A,PC2018A}. 
We first derive the LV model for amplitude dynamics in typical multisequence  
nonlinear waveguide transmission links, and comment on some straightforward 
extensions to this derivation further below.

In typical $J$-sequence soliton-based transmission systems, the frequency 
spacing between two adjacent sequences $\Delta\beta$ is a large constant, i.e., 
$\Delta\beta=|\beta_{j+1}(z)-\beta_{j}(z)|\gg 1$ for $1 \le j \le J-1$ 
\cite{MM98,Nakazawa2000,PNH2017B}. To derive the LV model for dynamics 
of soliton amplitudes in these systems, we employ the following assumptions, which 
were also used in Refs. \cite{NP2010,PNC2010,PC2012,CPJ2013,NPT2015,PNT2016,PNH2017A,PC2018A}.    
(1) The temporal separation $T$ between neighboring solitons 
in each sequence (the time-slot width) is a constant satisfying $T \gg 1$ \cite{ConditionT}.  
Additionally, the amplitudes are equal 
for all solitons from the same sequence, but are not necessarily equal for 
solitons from different sequences. This setup corresponds, for example, 
to phase-shift-keyed soliton transmission. (2) The sequences are either 
(a) subject to periodic temporal boundary conditions or (b) infinitely long. 
Setup (a) corresponds to waveguide-loop experiments and setup 
(b) approximates long-distance transmission. (3) Since $T\gg 1$, intrasequence 
interaction is exponentially weak and is neglected. 
(4) High-order effects due to radiation emission are also neglected.

Under assumptions (1)-(4), the solitons sequences remain periodic throughout 
the propagation. Therefore, the amplitudes of all pulses in a given sequence 
follow the same dynamics. We derive the LV model by taking into account amplitude shifts due 
to the effects of cubic gain and quintic loss on collisions between solitons 
from NN waveguides. We also take into account amplitude shifts due to the 
effects of linear gain-loss, cubic gain, and quintic loss on single-soliton 
propagation. The nonlinear interaction terms in the LV model are obtained by using 
the expressions for the amplitude shifts in a single fast two-soliton collision 
in the presence of weak cubic gain and quintic loss \cite{PNC2010,PC2012}, 
collision-rate calculations similar to the ones in Refs. \cite{NP2010,PNC2010,PC2012},  
and the NN interaction property. The linear and nonlinear non-interaction terms 
in the LV model are obtained by employing the adiabatic perturbation theory for 
the NLS soliton \cite{Hasegawa95,Iannone98,PC2020,Kaup91}. These calculations 
yield the following system of nonlinear equations for the dynamics of 
soliton amplitudes: 
\begin{eqnarray} &&
\frac{d \eta_{j}}{dz}=
\eta_{j}
\left\{g_{j}+\frac{4}{3}\epsilon_{3}\eta_{j}^{2}
-\frac{16}{15}\epsilon_{5}\eta_{j}^{4}
+\frac{8\sigma}{T}\epsilon_{3}
\left(\eta_{j-1} + \eta_{j+1}\right) 
\right.
\nonumber\\&&
\left.
-\frac{8\sigma}{T}\epsilon_{5}
\left[ 2\eta^2_{j}(\eta_{j-1} + \eta_{j+1})
+ \eta^{3}_{j-1} + \eta_{j+1}^3\right]
\right\}  
\label{gl8}
\end{eqnarray}
for sequences $2 \le j \le J-1$, 
\begin{eqnarray} &&
\frac{d \eta_{1}}{dz}=
\eta_{1}\left[g_{1}+\frac{4}{3}\epsilon_{3}\eta_{1}^{2}
-\frac{16}{15}\epsilon_{5}\eta_{1}^{4}
+\frac{8\sigma}{T}\epsilon_{3}\eta_{2} 
-\frac{8\sigma}{T}\epsilon_{5}\eta_{2}\left(2\eta_{1}^{2}+\eta_{2}^{2}\right)
\right]  
\label{gl9}
\end{eqnarray}
for sequence $j=1$, and  
\begin{eqnarray} &&
\!\!\!\!\!\!\!\!\!\!\!\!\!\!\!
\frac{d \eta_{J}}{dz}=
\eta_{J}\left[g_{J}+\frac{4}{3}\epsilon_{3}\eta_{J}^{2}
-\frac{16}{15}\epsilon_{5}\eta_{J}^{4}
+\frac{8\sigma}{T}\epsilon_{3}\eta_{J-1} 
-\frac{8\sigma}{T}\epsilon_{5}\eta_{J-1}\left(2\eta_{J}^{2}+\eta_{J-1}^{2}\right)
\right]  
\label{gl10}
\end{eqnarray}
for sequence $j=J$.

In multisequence optical waveguide systems it is typically desired to realize 
stable steady-state transmission with constant equal amplitudes for 
all sequences \cite{Agrawal2019,PNC2010}. We therefore look for an 
equilibrium point of the system (\ref{gl8})-(\ref{gl10}) in the form
$\eta^{(eq)}_{j}=\eta>0$ for $1 \le j \le J$. We obtain: 
\begin{eqnarray} &&
g_{j}=4\epsilon_{5}\eta
\left(-\frac{\kappa}{3}\eta + \frac{4}{15}\eta^3
- \frac{4\sigma\kappa}{T} 
+ \frac{12\sigma}{T}\eta^2 \right)
\label{gl11}
\end{eqnarray}   
for $2 \le j \le J-1$, and 
\begin{eqnarray} &&
g_{j}=4\epsilon_{5}\eta
\left(- \frac{\kappa}{3}\eta
+ \frac{4}{15}\eta^3
- \frac{2\sigma\kappa}{T} 
+ \frac{6\sigma}{T}\eta^2 \right)
\label{gl12}
\end{eqnarray}   
for $j=1$ and $j=J$, where $\kappa=\epsilon_{3}/\epsilon_{5}$, and 
$\epsilon_{5}\ne 0$. Substituting relations (\ref{gl11})-(\ref{gl12}) 
into Eqs. (\ref{gl8})-(\ref{gl10}), we arrive at the following generalized 
LV model for amplitude dynamics: 
\begin{eqnarray} &&
\frac{d \eta_{j}}{dz}=\epsilon_{5}\eta_{j} 
\left\{ \frac{4\kappa}{3}(\eta_{j}^2 - \eta^{2})
-\frac{16}{15}(\eta_{j}^4 - \eta^{4})
+\frac{8\sigma\kappa}{T}\left(\eta_{j-1} + \eta_{j+1} - 2\eta \right)
\right.
\nonumber\\&&
\left. 
-\frac{8\sigma}{T} \left[2\eta_{j}^{2}(\eta_{j-1} + \eta_{j+1}) 
+(\eta^{3}_{j-1} + \eta_{j+1}^{3}) - 6\eta^{3}\right]
 \right\}
\label{gl13}
\end{eqnarray}
for $2 \le j \le J-1$,    
\begin{eqnarray} &&
\frac{d \eta_{1}}{dz}=
\epsilon_{5}\eta_{1}
\left\{\frac{4\kappa}{3}(\eta_{1}^{2}-\eta^{2}) 
-\frac{16}{15}(\eta_{1}^{4}-\eta^{4})
+\frac{8\sigma\kappa}{T}(\eta_{2}-\eta)
\right.
\nonumber\\&&
\left. 
-\frac{8\sigma}{T}\left[\eta_{2}\left(2\eta_{1}^{2}+\eta_{2}^{2}\right)
-3\eta^{3}\right]\right\}, 
\label{gl14}
\end{eqnarray}     
and 
\begin{eqnarray} &&
\frac{d \eta_{J}}{dz}=
\epsilon_{5}\eta_{J}
\left\{\frac{4\kappa}{3}(\eta_{J}^{2}-\eta^{2}) 
-\frac{16}{15}(\eta_{J}^{4}-\eta^{4})
+\frac{8\sigma\kappa}{T}(\eta_{J-1}-\eta)
\right.
\nonumber\\&&
\left. 
-\frac{8\sigma}{T}\left[\eta_{J-1}\left(2\eta_{J}^{2}+\eta_{J-1}^{2}\right)
-3\eta^{3}\right]\right\}.
\label{gl15}
\end{eqnarray}     
Note that Eqs. (\ref{gl13})-(\ref{gl15}) are the first generalized $J$-dimensional 
LV model for amplitude dynamics in the presence of a generic (broadband) GL gain-loss 
profile with a general $J$ value. The derivation of the model is made possible 
by the NN interaction property of the waveguide array. Indeed, the NN interaction 
property leads to the complete elimination of the complex three-pulse interaction 
effects in intersequence soliton collisions. As a result, only two-pulse interaction 
effects should be taken into account in the model, and the derivation of the 
$J$-dimensional LV model with a general $J$ value is enabled.

We point out that some of the aforementioned assumptions that were used in 
the derivation of the LV model (\ref{gl13})-(\ref{gl15}) can be relaxed without 
substantial changes in the form of the model. In particular, the form of the 
LV model is unchanged when the frequency spacing between adjacent sequences 
varies with the sequence index $j$. Furthermore, when the time slot width 
depends on $j$, the third and fourth terms inside the curly brackets on 
the right hand side of Eq. (\ref{gl13}) change in a simple way to 
$8\sigma\kappa\left[(\eta_{j-1}-\eta)/T_{j-1} + (\eta_{j+1}-\eta)/T_{j+1}\right]$
and $-8\sigma \left[(2\eta_{j}^{2}\eta_{j-1}+\eta^{3}_{j-1}-3\eta^{3})/T_{j-1} 
+(2\eta_{j}^{2}\eta_{j+1}+\eta^{3}_{j+1}-3\eta^{3})/T_{j+1}\right]$,  
respectively. Similar simple changes occur in the nonlinear interaction terms 
on the right hand sides of Eqs. (\ref{gl14}) and (\ref{gl15}).

\section{Stability and bifurcation analysis for the generalized Lotka-Volterra models}
\label{stability}

\subsection{Introduction: transmission switching and its applications}
\label{stability_1} 

The waveguide setups for transmission stabilization and transmission switching are determined 
by stability and bifurcation analysis for the equilibrium points of the generalized 
LV model of Eqs. (\ref{gl13})-(\ref{gl15}). More specifically, in transmission 
stabilization, we require that the equilibrium point $(\eta, \dots, \eta)$ 
is asymptotically stable, such that the amplitude values tend to $\eta$ with 
increasing $z$. Additionally, we require that the equilibrium point at the origin 
is asymptotically stable, such that radiative instability due to growth 
of small amplitude waves is suppressed \cite{CPJ2013,NPT2015,PNH2017A}.

By transmission switching we refer to the turning on or off of the propagation 
of $M$ out of $J$ soliton sequences \cite{CPJ2013,NPT2015,PNH2017A}. 
The switching is based on bifurcations of the equilibrium point $(\eta, \dots, \eta)$, 
which can be realized by changes in the value/s of one or more physical 
parameters \cite{CPJ2013,NPT2015,PNH2017A}. 
In particular, in the current paper, the switching is achieved by 
changes in the value of the parameter $\kappa$. To explain switching in 
a more precise manner,  we denote by $\eta_{th}$ the value of the decision 
level that distinguishes between on and off transmission states of a given 
soliton sequence. Thus, the $j$th sequence is in an on state if $\eta_{j}>\eta_{th}$,  
and in an off state if $\eta_{j}<\eta_{th}$. We then say that off-on switching 
of $M$ out of $J$ sequences occurs when the value of one of the physical parameters 
(e.g. $\kappa$) changes at the switching distance $z_{s}$, such that $(\eta, \dots, \eta)$ 
turns from unstable to asymptotically stable \cite{PNH2017A}. As a result, before the switching, 
soliton amplitudes tend to values smaller than $\eta_{th}$ in $M$ sequences 
and to values larger than $\eta_{th}$ in $J-M$ sequences, while after the switching, 
soliton amplitudes in all $J$ sequences tend to $\eta$, where $\eta>\eta_{th}$.           
We say that on-off switching of $M$ sequences occurs when the value of 
a physical parameter (e.g. $\kappa$) changes at $z=z_{s}$, such that 
$(\eta, \dots, \eta)$ turns from asymptotically stable to unstable, 
while another equilibrium point with $M$ components smaller than $\eta_{th}$ is 
asymptotically stable \cite{PNH2017A}. Therefore, before the switching, soliton amplitudes 
in all $J$ sequences tend to $\eta$, where $\eta>\eta_{th}$, while after the switching, 
soliton amplitudes tend to values smaller than $\eta_{th}$ in $M$ sequences and to values 
larger than $\eta_{th}$ in $J-M$ sequences. Similar to transmission stabilization, we also 
require that the equilibrium point at the origin is asymptotically stable, such that 
radiative instability due to growth of small amplitude waves is suppressed.

The switching method that we study in the current paper (and also in Refs. 
\cite{CPJ2013,NPT2015,PNH2017A}) is different from the switching methods 
that are traditionally considered in linear and nonlinear optics 
(see Refs. \cite{Agrawal2019,Agrawal2020} for a description of the latter methods). 
In particular, in our switching method, the switching is carried out on all 
pulses within the waveguide loop, and therefore it can be implemented with an 
arbitrary number of pulses. In contrast, in traditional methods, the switching 
is applied on a single pulse or on a few pulses \cite{Agrawal2019,Agrawal2020}.
As a result, our switching approach has a great advantage on the traditional approach, 
since it can be significantly faster (see Ref. \cite{PNH2017A} for details).

Note that in our switching method, the switching affects all the pulses within the 
same sequence in the same manner. We can therefore refer to our method as sequence 
switching. Our sequence switching approach can be employed in any application, in which 
the same information processing operation such as amplification, filtering, routing, 
etc. should be performed on all the pulses in the same sequence \cite{PNH2017A}.  
To explain this, we denote by $p_{j}$ the transmission state of the $j$th sequence 
for the purpose of information processing. That is, $p_{j}=0$ if the $j$th sequence 
is off and $p_{j}=1$ if the $j$th sequence is on. The $J$-component vector 
$(p_{1}, ..., p_{j}, ..., p_{J})$, where $1 \le j \le J$, represents the 
transmission state of the full $J$-sequence system. We can use this vector 
to encode information about the processing that should be performed  
on different sequences in the next information processing station 
in the transmission line \cite{PNH2017A}. After this processing has been performed, 
the transmission state of the system can be switched to a new state, 
$(q_{1}, ..., q_{j}, ..., q_{J})$, which represents the type of information processing 
that should be performed in the next processing station.

\subsection{Stability analysis for the equilibrium points $(0,0,\dots,0)$ 
and $(\eta,\eta,\dots,\eta)$}
\label{stability_2}

The Jacobian matrix for the linearization of the $J$-dimensional LV model 
(\ref{gl13})-(\ref{gl15}) about $(0,0,\dots,0)$ is diagonal with eigenvalues 
$\lambda_{j}=g_{j}$ for $1 \le j \le J$, where the $g_{j}$ are given by 
Eqs. (\ref{gl11}) and (\ref{gl12}). Linear stability is guaranteed when 
$\lambda_{j}<0$ for $1 \le j \le J$. We therefore find that the 
equilibrium point at the origin is stable when 
\begin{eqnarray}&& 
\kappa > \kappa_{th} = \frac{\eta^2 (4\eta T + 180 \sigma)}{5(\eta T + 12 \sigma)},
\label{gl21}
\end{eqnarray}
regardless of the value of $J$. Note that $\kappa_{th}$ is the bifurcation 
value at which $(0,0,\dots,0)$ turns from unstable to asymptotically stable.

The Jacobian matrix for the linearization of the LV system (\ref{gl13})-(\ref{gl15}) 
around $(\eta,\eta,\dots,\eta)$ is 
\begin{equation} 
{\cal J}(\eta,\eta,\dots,\eta)=
\epsilon_{5}
\left( 
{\begin{array}{*{20}{c}}
a&b&0&0&\dots&0&0&0\\
b&a-c_1&b&0&\dots&0&0&0\\
0&b&a-c_1&b&\dots&0&0&0\\
\vdots&\vdots& & & & & &\vdots\\ 
0&0&0&0&\dots&b&a-c_1&b\\
0&0&0&0&\dots&0&b&a
\end{array}} 
\right),
\label{gl22}
\end{equation}
where 
\begin{equation} 
a=8\eta^2\left(\frac{\kappa}{3} - \frac{8\eta^{2}}{15}-\frac{4\sigma\eta}{T}\right),
\;\;\;\; 
b=\frac{8\sigma\eta}{T}(\kappa - 5\eta^{2}),
\;\;\;\;
c_{1}=\frac{32\sigma\eta^{3}}{T},  
\label{gl23}
\end{equation} 
and the dots in Eq. (\ref{gl22}) stand for zeros. Since linear stability of  
$(\eta,\eta,\dots,\eta)$ is not affected by $\epsilon_{5}$, it is useful to 
define the auxiliary matrix $\tilde{{\cal J}}(\eta,\eta,\dots,\eta)$ by 
$\tilde{{\cal J}}(\eta,\eta,\dots,\eta)={\cal J}(\eta,\eta,\dots,\eta)/\epsilon_{5}$.

The equation for the eigenvalues of $\tilde{{\cal J}}(\eta,\eta,\dots,\eta)$ 
has a different form for even and odd $J$ values. For even $J$ values, $J=2K$, 
the equation is 
\begin{equation} 
|{\cal A}_{K}|^{2} - b^{2}|{\cal A}_{K-1}|^{2} = 0,  
\label{gl24}
\end{equation}       
where $K=2,3,4,\dots \,$, ${\cal A}_{K}$ is the $K\times K$ matrix  
\begin{equation} 
{\cal A}_{K}=
\left( 
{\begin{array}{*{20}{c}}
a-\lambda&b&0&\dots&0&0&0\\
b&a-\lambda-c_1&b&\dots&0&0&0\\
\vdots& & & & & &\vdots\\ 
0&0&0&\dots&b&a-\lambda-c_1&b\\
0&0&0&\dots&0&b&a-\lambda-c_1
\end{array}} 
\right),
\label{gl25}
\end{equation}
${\cal A}_{1}=(a-\lambda)$, ${\cal A}_{0}\equiv 1$, and $|{\cal A}_{K}|$ is the 
determinant of ${\cal A}_{K}$. For odd $J$ values, $J=2K+1$, the equation for 
the eigenvalues of $\tilde{{\cal J}}(\eta,\eta,\dots,\eta)$ takes the form 
\begin{equation} 
|{\cal A}_{K}| 
\left[(a-\lambda)|{\cal B}_{K}|  
-b^{2}\left(|{\cal A}_{K-1}| + |{\cal B}_{K-1}|\right)\right]= 0,  
\label{gl26}
\end{equation}  
where $K=1,2,3,\dots \,$. In Eq. (\ref{gl26}),  
${\cal B}_{K}$ is the $K\times K$ matrix  
\begin{equation} 
\!\!\!\!
{\cal B}_{K}=
\left( 
{\begin{array}{*{20}{c}}
a-\lambda-c_1&b&0&\dots&0&0&0\\
b&a-\lambda-c_1&b&\dots&0&0&0\\
\vdots& & & & & &\vdots\\ 
0&0&0&\dots&b&a-\lambda-c_1&b\\
0&0&0&\dots&0&b&a-\lambda-c_1
\end{array}} 
\right),
\label{gl27}
\end{equation}
where ${\cal B}_{1}=(a-\lambda-c_1)$, and ${\cal B}_{0}\equiv 1$.

Since the explicit form of the characteristic equation for 
$\tilde{{\cal J}}(\eta,\eta,\dots,\eta)$ is known for a general $J$ value, 
we can find all the eigenvalues either numerically or analytically for 
any value of $J$ and for any given set of physical parameter values. 
Furthermore, by repeating the eigenvalues calculation for different 
values of $\kappa$ while all other parameter values are fixed, we can 
determine the interval of $\kappa$ values on which $(\eta,\eta,\dots,\eta)$ 
is linearly stable for any $J$ value, and the bifurcation value $\kappa_{c}$, 
at which $(\eta,\eta,\dots,\eta)$ turns from asymptotically stable to unstable. 
In what follows, we discuss in some detail the expressions for the eigenvalues 
and the conditions for linear stability of $(\eta,\eta,\dots,\eta)$ 
for $J=3$, $J=4$, and $J=5$.

{\it Stability condition for $J=3$.}
The characteristic equation is 
\begin{equation}
(a-\lambda) \left[(a-\lambda)(a-\lambda-c_1) - 2b^{2}\right] = 0.   
\label{gl28}
\end{equation}  
Therefore, the eigenvalues are 
\begin{eqnarray} 
\,
\lambda_{1}=a,  \;\; 
\lambda_{2}=a - \frac{c_{1}}{2} 
- \frac{c_{1}}{2}\left(1+8b^{2}/c_{1}^{2}\right)^{1/2},  
\;\;
\nonumber \\ 
\!\!\!\!\!
\lambda_{3}=a - \frac{c_{1}}{2} 
+\frac{c_{1}}{2}\left(1+8b^{2}/c_{1}^{2}\right)^{1/2}.  
\label{gl29}
\end{eqnarray} 
Since $\lambda_{2} < \lambda_{1} < \lambda_{3}$, the condition for 
linear stability is $\lambda_{3} < 0$. This condition can be expressed as 
\begin{equation}  
\kappa < \frac{8}{5}\eta^{2} + \frac{6\sigma\eta}{T}
\left\{3-\left[1+\frac{(\kappa-5\eta^{2})}{2\eta^{4}} \right]^{1/2} \right\}.
\label{gl30}
\end{equation}

{\it Stability condition for $J=4$.} 
The  characteristic equation is 
\begin{equation}
\left[(a-\lambda)(a-\lambda-c_1) - b^{2}\right]^{2} 
-b^{2}(a-\lambda)^{2} = 0.   
\label{gl31}
\end{equation}  
It follows that the eigenvalues are 
\begin{eqnarray} 
\lambda_{1}=a - \frac{1}{2}(c_{1}+b)
-\frac{1}{2}|c_{1}+b|\left[1+\frac{4b^{2}}{(c_{1}+b)^{2}}\right]^{1/2},  
\nonumber \\ 
\lambda_{2}=a - \frac{1}{2}(c_{1}+b)
+\frac{1}{2}|c_{1}+b|\left[1+\frac{4b^{2}}{(c_{1}+b)^{2}}\right]^{1/2},  
\nonumber \\ 
\lambda_{3}=a - \frac{1}{2}(c_{1}-b)
-\frac{1}{2}|c_{1}-b|\left[1+\frac{4b^{2}}{(c_{1}-b)^{2}}\right]^{1/2}, 
\nonumber \\ 
\lambda_{4}=a - \frac{1}{2}(c_{1}-b)
+\frac{1}{2}|c_{1}-b|\left[1+\frac{4b^{2}}{(c_{1}-b)^{2}}\right]^{1/2} .  
\label{gl32}
\end{eqnarray} 
It is clear that $\lambda_{2} > \lambda_{1}$ and $\lambda_{4} > \lambda_{3}$ 
for almost all values of the physical parameters. Additionally, it is 
straightforward to show that in the interval of $\kappa$ values that is  
most relevant for optical waveguide transmission, $\eta^{2} < \kappa < 5\eta^{2}$, 
$\lambda_{2} > \lambda_{4}$. Therefore, the condition for linear stability 
of $(\eta,\eta,\eta,\eta)$ for $\eta^{2} < \kappa < 5\eta^{2}$ is $\lambda_{2}<0$.

{\it Stability condition for $J=5$.} 
The  characteristic equation is  
\begin{eqnarray} &&  
\!\!\!\!\!\!\!\!\!\!\!\!\!\!\!\!\!\!\!\!
\left[(a-\lambda)(a-\lambda-c_1) - b^{2}\right] 
\nonumber \\ &&
\!\!\!\!\!\!\!\!\!\!\!\!\!\!\!\!\!\!\!\!
\times
\left\{(a-\lambda-c_1)\left[(a-\lambda)(a-\lambda-c_1) - b^{2}\right]
-2b^{2}(a-\lambda)\right\}  = 0.  
\label{gl33}
\end{eqnarray}  
Therefore, the first two eigenvalues are 
\begin{eqnarray} 
\!\!\!
\lambda_{1}= a - \frac{c_{1}}{2} 
- \frac{c_{1}}{2}\left(1+4b^{2}/c_{1}^{2}\right)^{1/2},  
\;\;\:
\lambda_{2}= a - \frac{c_{1}}{2} 
+ \frac{c_{1}}{2}\left(1+4b^{2}/c_{1}^{2}\right)^{1/2}.
\label{gl34}
\end{eqnarray} 
The other three eigenvalues are roots of the cubic equation 
\begin{equation}
(a-\lambda-c_1)^{2}(a-\lambda) -b^{2}(a-\lambda-c_1) -2b^{2}(a-\lambda) = 0.   
\label{gl35}
\end{equation}  
Using Cardan's formula \cite{Herstein75}, we find 
\begin{eqnarray} &&
\!\!\!\!\!\!\!\!\!\!\!\!\!\!\!\!
\lambda_{3}=a- \frac{2}{3}c_{1} - p\cos(\varphi/3),  \;\;\;\;
\lambda_{4}=a- \frac{2}{3}c_{1} + \frac{1}{2}p\cos(\varphi/3) + \frac{3^{1/2}}{2}p\sin(\varphi/3),  
\nonumber \\ &&
\!\!\!\!\!\!\!\!\!\!\!\!\!\!\!\!
\lambda_{5}=a- \frac{2}{3}c_{1} + \frac{1}{2}p\cos(\varphi/3) - \frac{3^{1/2}}{2}p\sin(\varphi/3),  
\label{gl36}
\end{eqnarray} 
where  
\begin{eqnarray} &&
\!\!\!\!\!\!\!\!\!\!\!\!\!\!\!\!\!\!\!\!\!\!
p = \frac{2}{3}(c_{1}^{2}+9b^{2})^{1/2},   \;\;\;\;
\varphi = \arctan\left[\frac{3^{3/2}b(108b^{4} + 9b^{2}c_1^{2} + 8c_1^{4})^{1/2}}
{c_1(27b^{2} - 2c_1^{2})}\right]. 
\label{gl37}
\end{eqnarray}  
In Section \ref{simu}, we use Eqs. (\ref{gl34}) and (\ref{gl36}) to find the condition  
for linear stability of $(\eta,\eta,\eta,\eta,\eta)$ for the parameter values used in the 
numerical simulations with Eqs. (\ref{gl1})-(\ref{gl3}) for $J=5$.

\subsection{Properties of the uncoupled ODE model and their relevance for transmission 
stabilization and switching}
\label{stability_3}

It is useful to consider the uncoupled nonlinear ODE model that corresponds to the 
full weakly coupled LV model (\ref{gl13})-(\ref{gl15}). This uncoupled ODE 
model takes the form 
\begin{eqnarray} &&
\frac{d \eta_{j}}{dz}=4\epsilon_{5}\eta_{j} 
\left[\frac{\kappa}{3}(\eta_{j}^2 - \eta^{2})
-\frac{4}{15}(\eta_{j}^4 - \eta^{4}) \right] 
\label{gl41}
\end{eqnarray}
for $1 \le j \le J$. We note that the coupling constant $\sigma/T$ 
in the full LV model is another small parameter, 
in addition to $\epsilon_{3}$ and $\epsilon_{5}$. As a result, a study 
of the stability properties of the equilibrium points of the uncoupled 
ODE model (\ref{gl41}) and their bifurcations can provide an approximate picture 
of the stability properties and the bifurcations of the equilibrium points 
of the full LV model. In particular, the stability and bifurcation analysis 
for the uncoupled ODE model can be used as the leading-order approximation 
to the stability and bifurcation analysis for the full coupled LV model. 
It follows that the simple analysis of the uncoupled ODE model can be employed 
as a general {\it approximate guide} for designing optical waveguide setups 
for transmission stabilization and switching.

Another important reason for considering the uncoupled ODE model (\ref{gl41}) 
is related to the stability properties of its equilibrium points. 
More specifically, stability analysis for the equilibrium points of the 
uncoupled ODE model shows that the stability is stronger than mere linear 
stability. Due to the smallness of the coupling constant $\sigma/T$, 
this property is expected to be valid in the full LV model 
(\ref{gl13})-(\ref{gl15}) as well. Furthermore, it is possible to construct 
Lyapunov functions \cite{Lyapunov92,Smale74,Meiss2007} for the equilibrium points of the 
uncoupled ODE model. These Lyapunov functions are also useful for the full 
coupled LV model, as they can be used to provide estimates for the trapping regions 
of the stable equilibrium points of the latter model (see Section \ref{stability_5}). 
This information can then provide important insight into the design 
of waveguide setups for robust transmission stabilization and switching.

\begin{figure}[ptb]
\begin{center}
\epsfxsize=10cm  \epsffile{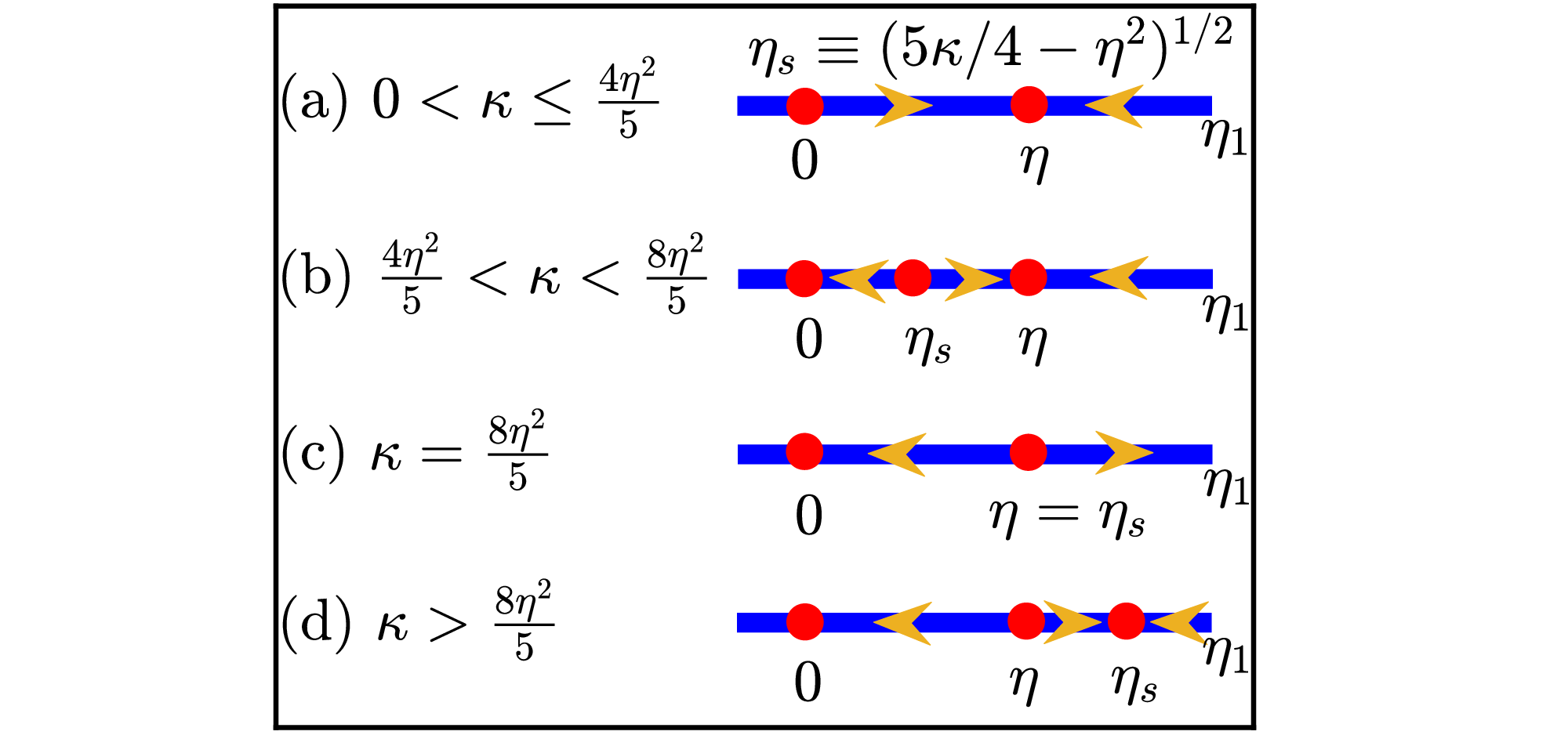}
\end{center}
\caption{(Color online) 
The dynamic flow on the $\eta_{1}$ axis for the 1-dimensional uncoupled 
ODE model. Note: $\eta_{s} \equiv (5\kappa/4-\eta^{2})^{1/2}$.}
\label{fig2}
\end{figure}

We start by considering the 1-dimensional uncoupled ODE model 
$d \eta_{1}/dz=4\epsilon_{5}\eta_{1} \left[\kappa(\eta_{1}^2 - \eta^{2})/3
-4(\eta_{1}^4 - \eta^{4})/15 \right]$. The equation has three equilibrium 
points with nonnegative $\eta_{1}$ values at $\eta_{1}^{(eq1)}=0$, 
$\eta_{1}^{(eq2)}=\eta$, and $\eta_{1}^{(eq3)}=\eta_{s} \equiv (5\kappa/4-\eta^{2})^{1/2}$. 
The first two equilibrium points exist for any $\kappa > 0$, while the third 
equilibrium point exists for $\kappa > 4\eta^{2}/5$. The point $\eta_{1}^{(eq1)}=0$ 
is unstable for $0 < \kappa \le 4\eta^{2}/5$ and stable for $\kappa > 4\eta^{2}/5$. 
The point $\eta_{1}^{(eq2)}=\eta$ is stable for $0 < \kappa < 8\eta^{2}/5$ and 
unstable for $\kappa \ge 8\eta^{2}/5$. The point $\eta_{1}^{(eq3)}=\eta_{s}$ is 
unstable for $4\eta^{2}/5 < \kappa \le 8\eta^{2}/5$ and stable for $\kappa > 8\eta^{2}/5$.   
Additionally, $\eta > \eta_{s}$ for $4\eta^{2}/5 < \kappa < 8\eta^{2}/5$, 
$\eta = \eta_{s}$ for $\kappa = 8\eta^{2}/5$, and $\eta < \eta_{s}$ for 
$\kappa > 8\eta^{2}/5$. The dynamic flow on the $\eta_{1}$ axis 
is summarized in Fig. \ref{fig2}. It follows that two bifurcations occur, 
one at $\kappa = 4\eta^{2}/5$ and another at $\kappa = 8\eta^{2}/5$. 
We also point out that stability of the 
equilibrium points can be established by considering changes in the sign 
of the function $h_{1}(\eta_{1})=\left[\kappa(\eta_{1}^2 - \eta^{2})/3
-4(\eta_{1}^4 - \eta^{4})/15 \right]$. Consequently, stability of the 
equilibrium points is stronger than mere linear stability.

Based on the discussions in the preceding paragraph and in 
Section \ref{stability_1}, we can relate the stability properties 
and the bifurcations in the 1-dimensional uncoupled ODE model to the 
approximate guiding principles for designing waveguide setups for 
transmission stabilization and switching. First, in the interval 
$4\eta^{2}/5 < \kappa < 8\eta^{2}/5$ both equilibrium points   
$\eta_{1}^{(eq1)}=0$ and $\eta_{1}^{(eq2)}=\eta$ are stable. 
Therefore, one should consider this interval as the leading-order 
approximation to the $\kappa$-interval, on which transmission 
stabilization of soliton-sequence 1 can be realized. 
Second, one can use the bifurcation of the uncoupled ODE model 
at $\kappa=8\eta^{2}/5$ for transmission switching. More specifically, 
the value $\kappa=8\eta^{2}/5$ can be used as the leading-order approximation 
for the exact bifurcation value $\kappa_{c}$ (in the full LV model), 
which governs transmission switching.    
That is, when the value of $\kappa$ is decreased from above $\kappa_{c} \simeq 8\eta^{2}/5$ 
to below $\kappa_{c} \simeq 8\eta^{2}/5$, $\eta_{1}^{(eq2)}=\eta$ becomes stable,  
while $\eta_{1}^{(eq3)}=\eta_{s}$ becomes unstable and $\eta_{1}^{(eq1)}=0$ 
remains stable. Therefore, off-on switching of soliton-sequence 1 can be realized by 
this change in the value of $\kappa$. On the other hand, when the value of              
$\kappa$ is increased from below $\kappa_{c} \simeq 8\eta^{2}/5$ 
to above $\kappa_{c} \simeq 8\eta^{2}/5$, $\eta_{1}^{(eq2)}=\eta$ becomes unstable,  
while $\eta_{1}^{(eq3)}=\eta_{s}$ becomes stable and $\eta_{1}^{(eq1)}=0$ 
remains stable. Therefore, in this process, on-off switching of soliton-sequence 1 
can be realized.

Let us discuss the properties of the equilibrium points of the $J$-dimensional  
uncoupled ODE model and their relevance for the design of waveguide setups for 
transmission stabilization and switching. We first note that there are $3^{J}$ 
possible equilibrium points for the $J$-dimensional uncoupled ODE model, 
including the points $(\eta,\eta,\dots,\eta)$ and $(0,0,\dots,0)$. The point 
$(\eta,\eta,\dots,\eta)$ is stable for $0 < \kappa < 8\eta^{2}/5$ and 
unstable for $\kappa \ge 8\eta^{2}/5$. The point $(0,0,\dots,0)$ is stable 
for $\kappa > 4\eta^{2}/5$ and unstable for $0 < \kappa \le 4\eta^{2}/5$. 
It follows that in the leading-order approximation for the full $J$-dimensional LV model, 
transmission stabilization and off-on switching can be realized in the 
$\kappa$-interval $4\eta^{2}/5 < \kappa < 8\eta^{2}/5$. The trapping regions 
in phase space can also be estimated with the help of the $J$-dimensional uncoupled ODE model. 
In particular, in the relevant $\kappa$-interval $4\eta^{2}/5 < \kappa < 8\eta^{2}/5$, 
the trapping region for $(\eta,\eta,\dots,\eta)$ is 
$\eta_{j} > \eta_{s}=(5\kappa/4-\eta^{2})^{1/2}$ for $1 \le j \le J$. 
Therefore, in the leading-order approximation for the full coupled LV model, 
the region in phase space, where transmission stabilization and off-on switching 
can be realized is evaluated as $\eta_{j} > \eta_{s}$ for $1 \le j \le J$.

The only other equilibrium points of the $J$-dimensional uncoupled ODE model, 
which are relevant for transmission switching, are points with 
at least one $0$-value coordinate and at least one $\eta_{s}$-value coordinate. 
We refer to these equilibrium points as $\eta_{s}-0$ equilibrium points. 
Additionally, we refer to coordinates for which the equilibrium value is 
$\eta_{s}$ as $\eta_{s}$-value coordinates, and to coordinates for which 
the equilibrium value is $0$ as $0$-value coordinates. 
There are $2^{J}-2$ $\eta_{s}-0$ equilibrium points. For example, 
in the 3-dimensional uncoupled ODE model, the equilibrium points of this form are 
$(\eta_{s},0,0)$, $(0,\eta_{s},0)$, $(0,0,\eta_{s})$, $(\eta_{s},\eta_{s},0)$, 
$(\eta_{s},0,\eta_{s})$, and $(0,\eta_{s},\eta_{s})$.      
The $\eta_{s}-0$ equilibrium points exist provided that $\kappa > 4\eta^{2}/5$.  
They are stable for $\kappa > 8\eta^{2}/5$ and unstable for 
$4\eta^{2}/5 < \kappa \le 8\eta^{2}/5$. Thus, these points are stable for 
$\kappa$ values for which $(\eta,\eta,\dots,\eta)$ is unstable, 
and are unstable for $\kappa$ values for which $(\eta,\eta,\dots,\eta)$ is stable.    
As a result, in the leading-order approximation to the full LV model, 
these equilibrium points can serve as the final amplitude state for the 
$J$-sequence system in on-off transmission switching. Additionally, the 
$\eta_{s}-0$ equilibrium points play a role in the initial stage of off-on 
switching, as $\pmb{\eta}(z)$ tends to an equilibrium point of this 
form for $z<z_{s}$, i.e, before the switching. We also note that the 
trapping region for the $\eta_{s}-0$ equilibrium points for $\kappa > 8\eta^{2}/5$  
is $\eta_{j} > \eta$ for the $\eta_{s}$-value coordinates and $0 < \eta_{j} < \eta$ 
for the $0$-value coordinates. Therefore, in the leading-order approximation to the 
full LV model, the region in phase space, where on-off transmission switching can 
be realized is $\eta_{j} > \eta$ for the $\eta_{s}$-value coordinates and 
$0 < \eta_{j} < \eta$ for the $0$-value coordinates. As a simple example, in a three-sequence 
system, on-off switching of the third sequence brings the amplitudes state from an 
initial state close to $(\eta,\eta,\eta)$ for $z \lesssim z_{s}$, to a final state 
close to $(\eta_{s},\eta_{s},0)$. The leading-order approximation to the region in 
phase space, in which this switching can be implemented, is $\eta_{j} > \eta$ for 
$j=1,2$ and $0 < \eta_{j} < \eta$ for $j=3$.

The last example also illustrates a very important property of the 
switching processes that are introduced in the current paper. Namely, 
in each given switching process (off-on or on-off) only three equilibrium 
points out of the entire set of $3^{J}$ equilibrium points play an 
important role. The three equilibrium points are $(\eta,\eta,\dots,\eta)$, 
$(0,0,\dots,0)$, and one appropriate $\eta_{s}-0$ equilibrium point. 
This highly desirable property of the switching processes ensures their 
robustness and scalability. It is a consequence of the relatively simple 
form of the $J$-dimensional uncoupled ODE model (\ref{gl41}), and the smallness of the 
coupling constant $\sigma/T$ in the full LV model (\ref{gl13})-(\ref{gl15}).

\subsection{Approximate guiding principles for transmission switching setups}
\label{stability_4}

Based on the discussion in Sections \ref{stability_3} and \ref{stability_2}, 
we now formulate approximate guiding principles for transmission switching of a single 
soliton sequence in a $J$-sequence system. The generalization of these guiding principles 
to switching of two or more sequences is straightforward. We use the index $m$ 
as the index of the switched sequence, while the index $j$ runs from $1$ to $J$. 

(a) {\it Off-on transmission switching setups.} 
\begin{enumerate}  
\item The initial and final values of $\kappa$, $\kappa_{i}$ and $\kappa_{f}$, 
should satisfy $\kappa_{i} > \kappa_{c}$ and $\kappa_{th} < \kappa_{f} < \kappa_{c}$, 
where $\kappa_{c}$ is the exact bifurcation value at which the equilibrium 
point $(\eta,\eta,\dots,\eta)$ changes from unstable to stable in the full LV model, 
and $\kappa_{th}$ is the exact bifurcation value at which the equilibrium 
point $(0,0,\dots,0)$ changes from unstable to stable in the full LV model.   
$\kappa_{c}$ is determined by the solution of Eq. (\ref{gl24}) or Eq. (\ref{gl26}), 
and $\kappa_{th}$ is given by Eq. (\ref{gl21}). In the leading-order 
approximation to the full LV model, which is given by the uncoupled ODE model 
(\ref{gl41}), $\kappa_{c} \simeq 8\eta^{2}/5$ and $\kappa_{th} \simeq 4\eta^{2}/5$.

\item The initial amplitude values for the soliton sequences should satisfy  
\begin{equation}
\eta_{j}(0) > \eta \;\; \mbox{for} \;\; j \ne m \;\; \mbox{and} 
\;\; \eta_{sf} < \eta_{m}(0) < \eta ,
\label{gl43}
\end{equation}
where $\eta_{sf} = (5\kappa_{f}/4 - \eta^{2})^{1/2}$. Since 
$\kappa_{i} > \kappa_{c}$, $\pmb{\eta}(z)$ should tend to 
$(\eta_{si},\dots,\eta_{si},\eta_{m}=0,\eta_{si},\dots,\eta_{si})$ 
for $z \lesssim z_{s}$, where $\eta_{si} = (5\kappa_{i}/4 - \eta^{2})^{1/2}$. 
Note that we require $\eta_{m}(0) > \eta_{sf}$ to ensure consistency with 
condition (\ref{gl44}).

\item The amplitude values at the switching distance $z=z_{s}$ should satisfy 
\begin{equation}
\eta_{j}(z_{s}) > \eta_{sf} \;\; \mbox{for} \;\; 1 \le j \le J. 
\label{gl44}
\end{equation} 
As a result, by the leading-order approximation to the full LV model,  
$\pmb{\eta}(z)$ should tend to $(\eta,\eta,\dots,\eta)$ for $z>z_{s}$. 

\end{enumerate}

(b) {\it Basic on-off transmission switching setups.} 
\begin{enumerate}  
\item The initial and final values of $\kappa$ should satisfy 
$\kappa_{th} < \kappa_{i} < \kappa_{c}$ and $\kappa_{f} > \kappa_{c}$, 
where $\kappa_{c}$ is determined by the solution of Eq. (\ref{gl24}) 
or Eq. (\ref{gl26}), and $\kappa_{th}$ is given by Eq. (\ref{gl21}).

\item The initial amplitude values should satisfy 
\begin{equation}
\eta_{j}(0) > \eta \;\; \mbox{for} \;\; j \ne m \;\; \mbox{and} 
\;\; \eta_{si} < \eta_{m}(0) < \eta .
\label{gl45}
\end{equation}  
Since $\kappa_{th} < \kappa_{i} < \kappa_{c}$, $\pmb{\eta}(z)$ should tend 
to $(\eta,\eta,\dots,\eta)$ for $z \lesssim z_{s}$. Note that we require 
$\eta_{j}(0) > \eta$ for $j \ne m$ and $\eta_{m}(0) < \eta$ to ensure 
consistency with condition (\ref{gl46}).

\item The amplitude values at $z=z_{s}$ should satisfy 
\begin{equation}
\eta_{j}(z_{s}) > \eta \;\; \mbox{for} \;\; j \ne m \;\; \mbox{and} 
\;\; 0 < \eta_{m}(z_{s}) < \eta .
\label{gl46}
\end{equation} 
Therefore, by the leading-order approximation to the full LV model, $\pmb{\eta}(z)$ 
should tend to $(\eta_{sf},\dots,\eta_{sf},\eta_{m}=0,\eta_{sf},\dots,\eta_{sf})$ 
for $z>z_{s}$. 

\end{enumerate}

We emphasize again that Eqs. (\ref{gl43})-(\ref{gl46}) are only approximate 
guiding conditions for the design of waveguide setups for transmission switching. 
The {\it actual} ({\it exact}) theoretical conditions for transmission switching 
are determined by the numerical solution of the full LV model (\ref{gl13})-(\ref{gl15}). 
Nevertheless, due to the smallness of the coupling parameter $\sigma/T$, 
the conditions (\ref{gl43})-(\ref{gl46}) serve as an excellent staring point 
in the search for the exact regions in phase space, where transmission switching 
can be realized.

Another complication in the realization of on-off transmission 
switching and its resolution are discussed in the following paragraphs.

(c) {\it Improved on-off transmission switching setups.} 

Numerical simulations with the coupled-NLS model (\ref{gl1})-(\ref{gl3}) 
show that it is sometimes difficult to realize on-off transmission switching 
with the basic setups, described in item (b). 
The main reason for this is that the numerically obtained amplitude 
values for $z \lesssim z_{s}$ are close to $(\eta,\eta,\dots,\eta)$ 
and are sometimes oscillating. Due to these oscillations, the amplitude 
values at $z=z_{s}$, which are obtained by numerical solution of 
Eqs. (\ref{gl1})-(\ref{gl3}), do not satisfy the approximate switching condition 
(\ref{gl46}) and its exact counterpart, which is based on the numerical 
solution of the full LV model (\ref{gl13})-(\ref{gl15}). As a result, 
in this case, the desired on-off switching is not realized in the 
coupled-NLS simulation.

The shortcoming of the basic on-off transmission switching setups can be 
overcome by the introduction of a short intermediate waveguide span 
$(z_{i},z_{s}]$, in which the soliton sequences propagate in the presence 
of weak linear gain or weak linear loss. More specifically, in this  
interval, the sequences that should remain in an on state 
propagate in the presence of weak linear gain, while the sequences that 
should be turned off propagate in the presence of weak linear loss. 
Thus, the propagation in the interval $(z_{i},z_{s}]$ 
is described by: 
\begin{eqnarray} &&
i\partial_z\psi_{j}+\partial_{t}^2\psi_{j}+2|\psi_{j}|^2\psi_{j}=
s_{j}\epsilon_{1j}\psi_{j}/2 , 
\label{gl47}
\end{eqnarray}        
where $1 \le j \le J$, $0 < \epsilon_{1j} \ll 1$ is the linear gain 
or linear loss coefficient for the $j$th sequence in the intermediate interval, 
$s_j=1$ if the $j$th sequence should remain in an on state, and $s_j=-1$ 
if the $j$th sequence should be turned off. By the adiabatic perturbation theory 
for the cubic NLS soliton \cite{Hasegawa95,Iannone98,PC2020,Kaup91}, the dynamics 
of the $\eta_j$ in the intermediate interval is described by: 
\begin{eqnarray}&&  
\eta_{j}(z) = \eta_{j}(z_{i})\exp\left[s_{j}\epsilon_{1j}(z-z_{i})\right] .
\label{gl48}
\end{eqnarray}     
As will be shown in Section \ref{simu}, this simple modification of the 
basic on-off switching setups ensures that on-off transmission switching 
can be realized in the coupled-NLS simulations, even in the presence of 
substantial oscillations in the numerically obtained amplitude values.  
Furthermore, it is found that the improved method is not very sensitive 
to the choice of values for $z_{i}$ and $\epsilon_{1j}$.

In summary, in the improved on-off transmission switching setups, the propagation 
is divided into three intervals $0 \le z \le z_{i}$, $z_{i} < z \le z_{s}$, 
and $z > z_{s}$. Similar to the basic on-off switching setups, the propagation 
in the first and third intervals is described by Eqs. (\ref{gl1})-(\ref{gl3}) with   
$\kappa_{th} < \kappa_{i} < \kappa_{c}$ and $\kappa_{f} > \kappa_{c}$, respectively. 
Additionally, the propagation in the second interval is described by Eq. (\ref{gl47}), 
as detailed in the preceding paragraph.

\subsection{Extension of the calculations in Section \ref{stability_4} by 
application of the Lyapunov function method}
\label{stability_5}

In this subsection, we demonstrate that the Lyapunov function method can be 
used to obtain improved estimates for the trapping regions of equilibrium points 
of the full LV model, which are involved in transmission stabilization and 
switching. These estimates provide more accurate conditions on the regions 
in phase space, where transmission stabilization and switching can be achieved, 
compared with the conditions that were obtained in Section \ref{stability_4}, 
using the uncoupled ODE model.

We first provide a general description of the Lyapunov function method, 
as applied to the full $J$-dimensional LV model (\ref{gl13})-(\ref{gl15}). 
Following Lyapunov stability theorem \cite{Lyapunov92,Smale74,Meiss2007},  
we look for a Lyapunov function in the form $V_{L}(\pmb{\eta})=
\sum_{j=1}^{J}(\eta_{j}-\eta_{j}^{(eq)})^2$, where $\eta_{j}^{(eq)}$ 
with $j=1,\, \dots,\, J$ are the coordinates of one of the stable equilibrium 
points of the $J$-dimensional LV model, whose trapping region we want to find. 
$V_{L}(\pmb{\eta})$ obviously satisfies two of the three required properties 
of a Lyapunov function, $V_{L}(\pmb{\eta}^{(eq)})=0$ and 
$V_{L}(\pmb{\eta} \ne \pmb{\eta}^{(eq)}) \ne 0$. In addition, 
$dV_{L}/dz=2\sum_{j=1}^{J}(\eta_{j}-\eta_{j}^{(eq)})d\eta_{j}/dz$, 
where $d\eta_{j}/dz$ are given by Eqs. (\ref{gl13})-(\ref{gl15}).                
Thus, using Eqs. (\ref{gl13})-(\ref{gl15}), we can write  
$dV_{L}/dz=G_{L}(\pmb{\eta})$. We then find numerically 
the connected region around $\pmb{\eta}^{(eq)}$, in which 
$G_{L}(\pmb{\eta})<0$. This region is the numerically obtained 
estimate for the trapping region of the stable equilibrium point 
$\pmb{\eta}^{(eq)}$.

We now demonstrate the Lyapunov function method by employing it to evaluate 
the trapping region of the equilibrium point $(\eta,\eta,\eta)$, which plays a 
major role in transmission stabilization and switching with $J=3$ soliton 
sequences. We emphasize that in the same manner, the method can be used to 
estimate the trapping regions for the other stable equilibrium points of the 
$3$-dimensional and the $J$-dimensional LV models. We first note 
that the derivative along trajectories of the Lyapunov function for 
$(\eta,\eta,\dots,\eta)$ in the $J$-dimensional LV model can be written as  
\begin{equation}
dV_{L}/dz=G_{L}(\pmb{\eta})=G_{L1}(\pmb{\eta})+G_{L2}(\pmb{\eta}),  
\label{gl49}
\end{equation}   
where $G_{L1}(\pmb{\eta})$ is the term proportional to $\epsilon_{5}$, which 
is associated with single-sequence dynamics, and $G_{L2}(\pmb{\eta})$ is the 
term proportional to $\epsilon_{5}\sigma/T$, which is associated with dynamics 
due to intersequence interaction. Additionally, $G_{L1}(\pmb{\eta})$ can be written as 
\begin{equation}
G_{L1}(\pmb{\eta})=\frac{8}{3}\epsilon_{5}
\sum_{j=1}^{J}\eta_{j}(\eta_{j}-\eta)^2(\eta_{j}+\eta) 
\left[\kappa-\frac{4}{5}\left(\eta_{j}^2 + \eta^{2}\right)
\right].    
\label{gl50}        
\end{equation}
It has exactly the same functional form as $dV_{L}/dz$ for $(\eta,\eta,\dots,\eta)$ 
in the uncoupled ODE model (\ref{gl41}).

\begin{figure}[ptb]
\begin{center}
\epsfxsize=10cm  \epsffile{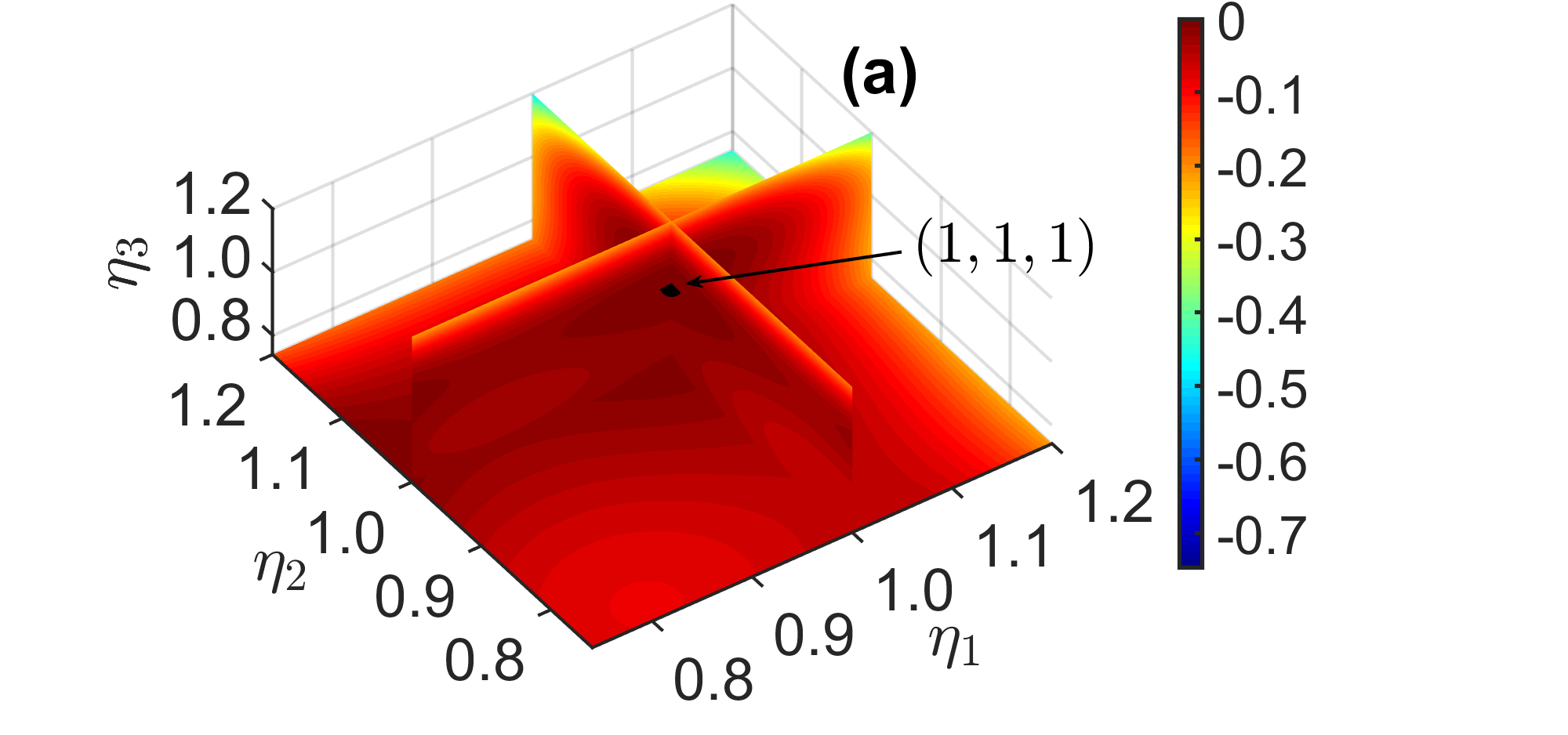}\\
\epsfxsize=10cm  \epsffile{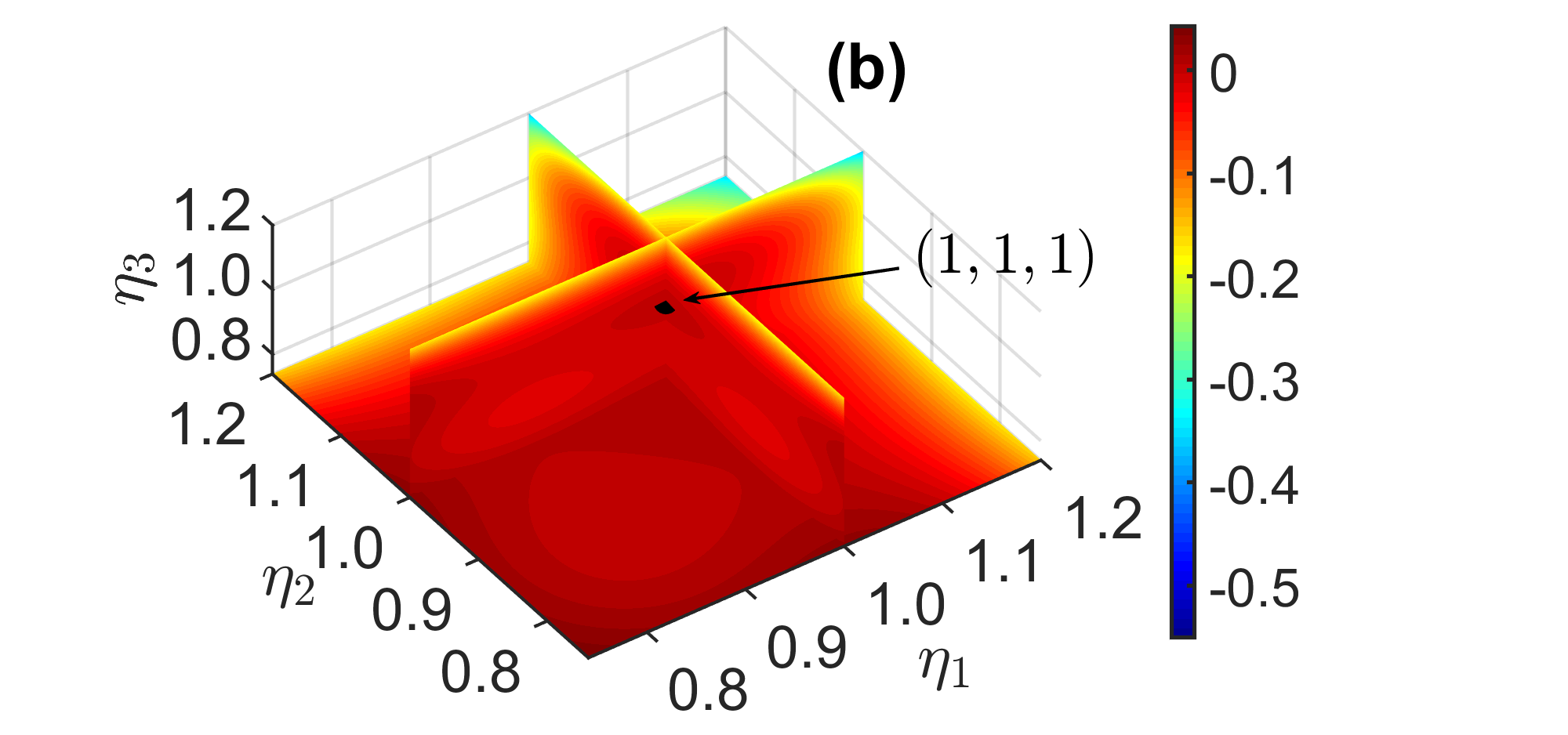}\\
\epsfxsize=10cm  \epsffile{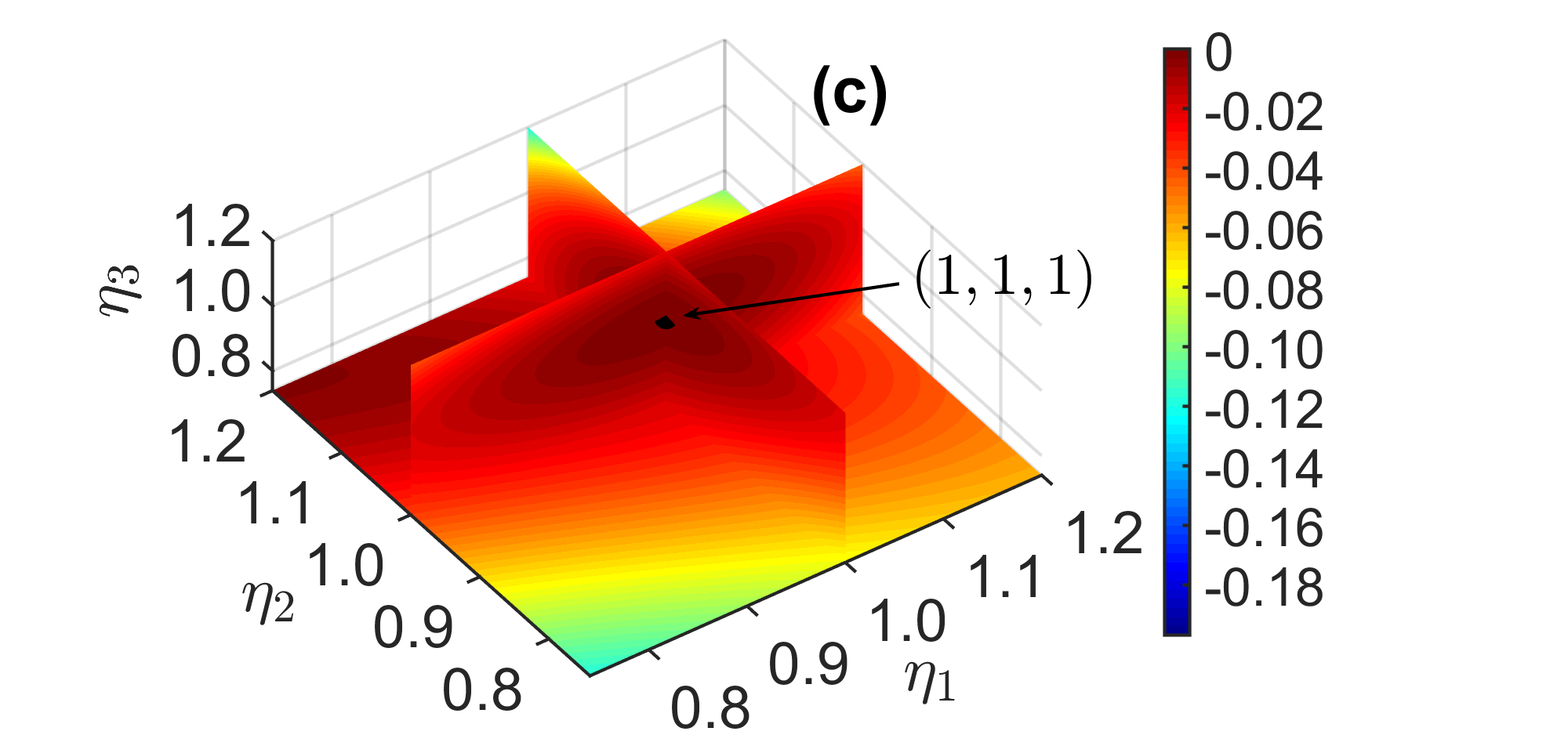}
\end{center}
\caption{(Color online) 
Contour plots of the functions $G_{L}(\pmb{\eta})$ (a), $G_{L1}(\pmb{\eta})$ (b), 
and $G_{L2}(\pmb{\eta})$ (c), defined in Eqs. (\ref{gl49})-(\ref{gl50}), 
in the box $0.74\le \eta_{j} \le 1.2$ with $j=1,\, 2,\, 3$ 
in the $3$-dimensional phase space. For clarity, the contour plots are 
shown on the three planes $\eta_1=1, \eta_2=1,$ and $\eta_3=0.74$. The parameter values 
are $\eta=1$, $\kappa=1.3$, $\sigma=0.1$, $T=15$, and $\epsilon_{5}=0.1$.}
\label{fig_add1}
\end{figure}

Figure \ref{fig_add1} shows the contour plots of $G_{L}(\pmb{\eta})$, $G_{L1}(\pmb{\eta})$, 
and $G_{L2}(\pmb{\eta})$ with $J=3$ near $(\eta,\eta,\eta)$ for the parameter values $\eta=1$, 
$\kappa=1.3$, $\sigma=0.1$, $T=15$, and $\epsilon_{5}=0.1$, which are also the 
values used in the coupled-NLS simulations for transmission stabilization and 
off-on switching. The trapping region of $(\eta,\eta,\eta)$ in the LV model is 
the region where $G_{L}(\pmb{\eta})<0$, and the trapping region in the 
uncoupled ODE model is the region where $G_{L1}(\pmb{\eta})<0$.    
We observe that the trapping region of $(\eta,\eta,\eta)$ in the 
LV model is noticeably larger than the corresponding trapping region in the uncoupled 
ODE model. More specifically, we find that the trapping region in the LV model contains the infinite 
box $\eta_{j}>0.739$ for $j=1,\, 2,\, 3$, while the trapping region in the uncoupled 
ODE model (with $\kappa=1.3$) is $\eta_{j}>0.791$ for $j=1,\, 2,\, 3$. 
Additionally, $G_{L2}(\pmb{\eta})<0$ everywhere in the box $\eta_{j}>0.739$ for $j=1,\, 2,\, 3$ 
except for at $(1,1,1)$, where it is equal to zero. As a result, $G_{L}(\pmb{\eta})<G_{L1}(\pmb{\eta})$ 
everywhere in the same box, except for at $(1,1,1)$, where both functions are equal to zero.      
The observed increase in the trapping region of $(\eta,\eta,\eta)$, which is 
interesting from both the dynamical and the application points of view, can be 
intuitively explained in the following manner. In the uncoupled ODE model, 
the combination of linear loss, cubic gain, and quintic loss in each ODE 
for $0 < \kappa < \kappa_{c}$ and $\eta_{j} > \eta_{s}$ is a stabilizing 
dynamical mechanism in the sense that $\eta_{j}(z)$ tends to the equilibrium 
value $\eta$ with increasing $z$. As a result, $(\eta,\eta,\eta)$ is a stable 
equilibrium point of the uncoupled ODE model. Additionally, the nonlinear 
intersequence interaction terms due to cubic gain and quintic loss in the 
full LV model (the terms proportional to $\epsilon_{5}\sigma/T$) have the same 
signs as the nonlinear single-sequence terms due to cubic gain and quintic loss 
(the terms proportional to $\epsilon_{5}$). Therefore, the inclusion of the 
nonlinear intersequence interaction terms in the LV model adds a second 
stabilizing mechanism to the dynamical model, and this causes the observed 
increase in the trapping region of $(\eta,\eta,\eta)$.

A similar estimate for the trapping region of $(\eta,\eta,\eta)$ can be obtained 
by a heuristic topological argument regarding the locations of the equilibrium 
points of the full LV model (\ref{gl13})-(\ref{gl15}), which lie away from the $\eta_{j}$ axes. 
The argument is motivated by the Hartman-Grobman theorem \cite{Smale74,Meiss2007}.  
It relies on the assumption that the phase portrait of the full LV model is a weakly 
deformed version of the phase portrait of the uncoupled ODE model (\ref{gl41}). 
This assumption is justified by the fact that the intersequence interaction terms 
in the full LV model are weak regular perturbation terms for the uncoupled ODE model. 
In the 3-dimensional models, there are seven equilibrium points other than 
$(\eta,\eta,\eta)$, which lie away from the axes. These equilibrium points 
are all unstable when $(\eta,\eta,\eta)$ is stable. 
For the LV model, using the parameter values in Fig. \ref{fig_add1}, 
we find that the seven equilibrium points are located at $M'_{3}=(0.682,0.591,0.682)$, 
$M'_{4}=(0.993, 1.033, 0.732)$, $M'_{5}=(1.043,0.668,1.043)$, $M'_{6}=(0.732, 1.033, 0.993)$, 
$M'_{7}=(1.047, 0.626, 0.683)$, $M'_{8}=(0.738,1.059,0.738)$, and $M'_{9}=(0.683, 0.626, 1.047)$.  
We see that these equilibrium points are slightly shifted relative to the following 
seven equilibrium points of the uncoupled ODE model: $M_{3}=(\eta_{s},\eta_{s},\eta_{s})$, 
$M_{4}=(\eta,\eta,\eta_{s})$, $M_{5}=(\eta,\eta_{s},\eta)$, $M_{6}=(\eta_{s},\eta,\eta)$, 
$M_{7}=(\eta,\eta_{s},\eta_{s})$, $M_{8}=(\eta_{s},\eta,\eta_{s})$, and 
$M_{9}=(\eta_{s},\eta_{s},\eta)$ with $\eta=1$ and $\eta_{s}=0.791$. 
We recall that the trapping region of $(\eta,\eta,\eta)$ in the 
uncoupled ODE model is $\eta_{j}>\eta_{s}$ for $j=1,\, 2,\, 3$. 
Using the weak deformation relation between the phase portraits of the two dynamical models, 
we can estimate the trapping region of $(\eta,\eta,\eta)$ in the full LV model as the 
infinite box $\eta_{j}>\eta'_{s}$ for $j=1,\, 2,\, 3$, where $\eta'_{s}$ is the maximal 
value of the $\eta_{s}$-shifted coordinates among all the seven equilibrium points 
$M'_{3}$-$M'_{9}$. For the parameter values used in Fig. \ref{fig_add1}, we find 
$\eta'_{s}=0.738$, in very good agreement with the value $\eta'_{s}=0.739$ that was 
obtained in the preceding paragraph by the Lyapunov function method.

In summary, in the current subsection, we demonstrated that the accuracy of the 
conditions for transmission stabilization and switching, obtained in Section 
\ref{stability_4}, can be improved by employing the Lyapunov function method for 
the stable equilibrium points of the full LV model. More specifically, we used Lyapunov 
function analysis to find more accurate estimates for the trapping regions of 
equilibrium points involved in transmission stabilization and switching. The 
improved estimates yield the regions in phase space, where transmission 
stabilization and switching in the full LV model can be realized. 
We also demonstrated that the trapping regions can be estimated by a simple 
topological argument about the locations of the equilibrium points of the full 
LV model, which is motivated by the Hartman-Grobman theorem.

\section{Numerical simulations with the perturbed coupled-NLS model}
\label{simu} 

\subsection{Introduction}
\label{simu_1} 

The LV model (\ref{gl13})-(\ref{gl15}) is based on a number of simplifying 
assumptions, whose validity might break down at intermediate and large 
propagation distances. Most importantly, Eqs. (\ref{gl13})-(\ref{gl15}) 
neglect the effects of radiation emission and pulse distortion, which are 
included in the full weakly perturbed coupled-NLS model (\ref{gl1})-(\ref{gl3}). 
These effects can lead to destabilization of the soliton sequences and to the 
breakdown of the LV model description \cite{PNC2010,PC2012,CPN2016,PNT2016,PNH2017A}. 
Therefore, it is important to check the predictions of the LV model (\ref{gl13})-(\ref{gl15}) 
for transmission stabilization and switching by numerical simulations with the 
full coupled-NLS model (\ref{gl1})-(\ref{gl3}). In the current section, we take 
on this important task.

We numerically solve the coupled-NLS system (\ref{gl1})-(\ref{gl3}) by the 
split-step method with periodic boundary conditions \cite{Agrawal2019,Yang2010}. 
Since we use periodic boundary conditions, the simulations describe 
propagation of the soliton sequences in a closed doughnut-shaped 
waveguide-array loop. The initial condition for the simulations is 
in the form of $J$ periodic sequences of $2K$ fundamental NLS solitons 
with amplitudes $\eta_{j}(0)$, frequencies $\beta_{j}(0)$, and zero phases, 
where the cases $J=3$, $J=4$, and $J=5$ are considered. 
Thus, the initial condition has the form 
\begin{eqnarray} &&
\psi_{j}(t,0)\!=\!\sum_{k=-K}^{K-1}
\frac{\eta_{j}(0)\exp\{i\beta_{j}(0)[t-kT-y_{j0}]\}}
{\cosh\{\eta_{j}(0)[t-kT-y_{j0}]\}},  
\label{gl51}
\end{eqnarray} 
where $1 \le j \le J$, $\Delta\beta=\beta_{j+1}(0)-\beta_{j}(0) \gg 1$, 
and $0 \le |y_{j0}| < T$. As an example, we present the simulations results 
for $K=1$, $T=15$, and $\Delta\beta=15$. We emphasize, however, that 
similar results are obtained with other physical parameter values that 
satisfy the validity conditions of the LV model.

In addition to $K=1$, $T=15$, and $\Delta\beta=15$, the following 
parameter values are used in the simulations discussed in the current section.    
\begin{enumerate}
\item $\eta=1$ and $\sigma=0.1$ are used in all the simulations.   
Further, all the simulations are run up to the final distance $z_{f}=1000$. 

\item In transmission stabilization simulations, we use the values 
$\epsilon_{5}=0.1$ and $\kappa=1.3$. 

\item In simulations of off-on switching, we use $\epsilon_{5i}=0.02$ and 
$\kappa_{i}=2.1$ in the initial (off) interval, and $\epsilon_{5f}=0.1$ 
and $\kappa_{f}=1.3$ in the final (on) interval. The switching distance 
is $z_{s}=25$.  

\item In simulations of on-off switching, we implement the improved 
setups discussed in part (c) of Section \ref{stability_4}. In these 
simulations, we use the values $\epsilon_{5i}=0.02$ and $\kappa_{i}=1.3$ 
in the initial (on) interval, the values $\epsilon_{1j}=0.01$ for 
$1 \le j \le J$ in the intermediate interval, and the values 
$\epsilon_{5f}=0.1$ and $\kappa_{f}=2.1$ in the final (off) interval. 
Additionally, $z_{i}=500$ and $z_{s}=502$. 
     
\end{enumerate}     
Note that since we use $\epsilon_{5}=0.1$ in transmission stabilization 
and $\epsilon_{5f}=0.1$ in both types of transmission switching, the 
stabilization and switching are realized over relatively short intervals 
($\Delta z \sim 10$) compared with the total propagation distance ($z_{f}=1000$).

\subsection{Simulations results for transmission stabilization and switching}
\label{simu_2}

\subsubsection{Three soliton sequences ($J=3$)}
\label{simu_21}

Let us describe the numerical simulations results for transmission  
stabilization and switching with three soliton sequences. 
The values of $\beta_{j}(0)$ and $y_{j0}$ in these simulations are 
$\beta_{1}(0)=-\Delta\beta$, $\beta_{2}(0)=0$, $\beta_{3}(0)=\Delta\beta$, 
$y_{10}=-T/2$, $y_{20}=0$, and $y_{30}=T/2$, where $\Delta\beta=15$ and $T=15$. 
For these setups, the values of the parameters $\kappa_{th}$ and $\kappa_{c}$, 
defined in Section \ref{stability}, are $\kappa_{th}=0.9630$ and $\kappa_{c}=1.6163$.

\begin{figure}[ptb]
\begin{center}
\epsfxsize=10cm  \epsffile{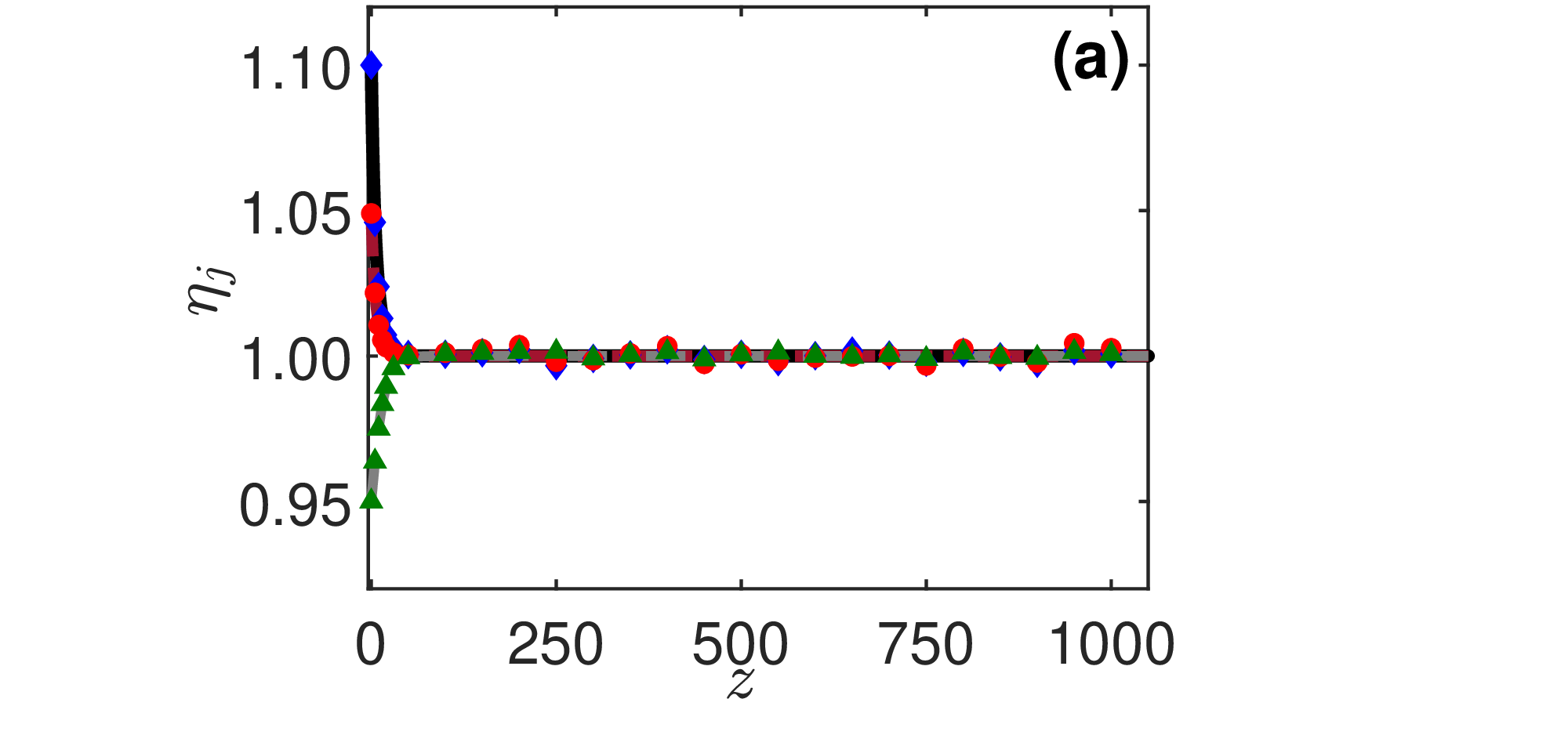}\\
\epsfxsize=10cm  \epsffile{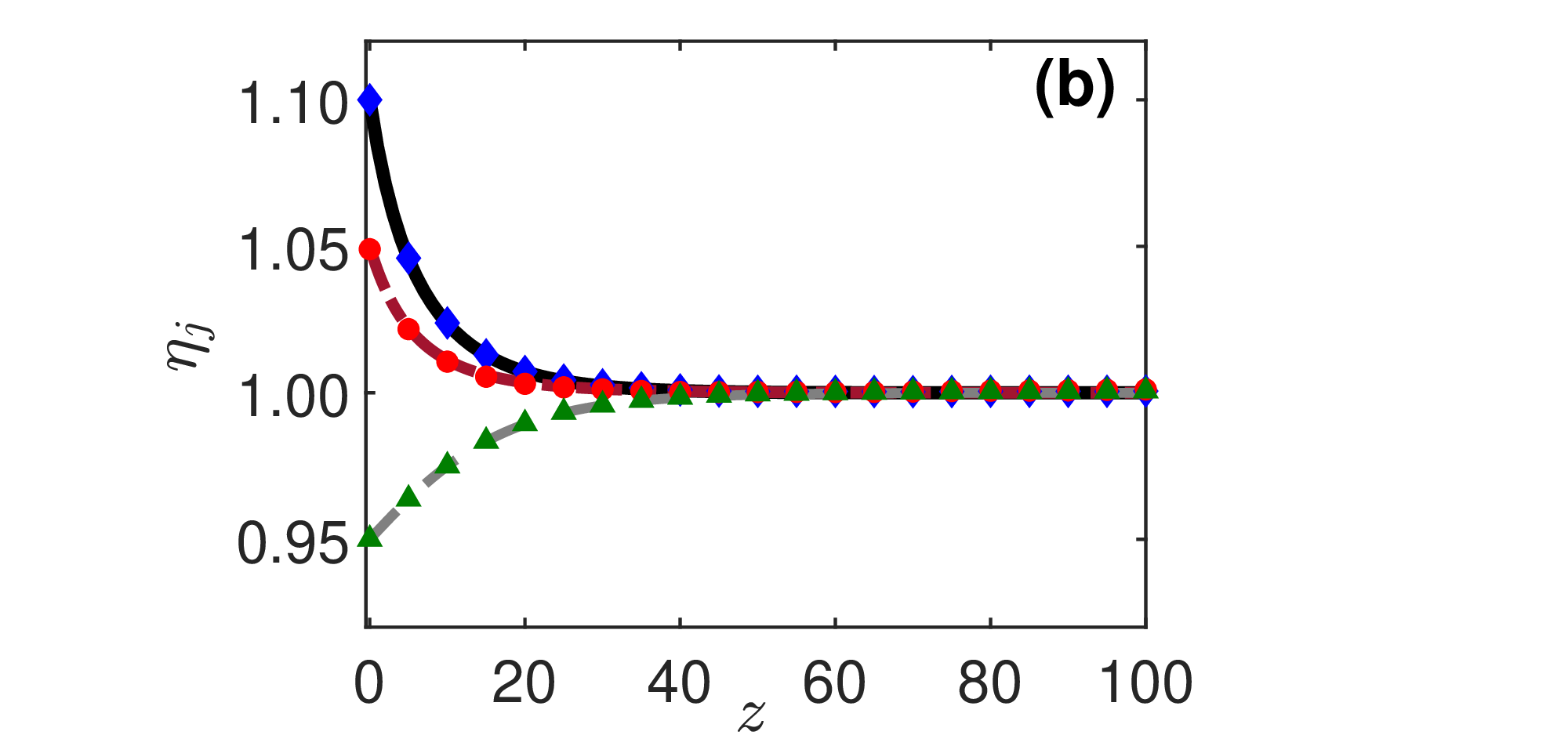}
\end{center}
\caption{(Color online) 
$\eta_{j}$ vs $z$ in transmission stabilization of three soliton 
sequences in a nonlinear waveguide array with a weak GL gain-loss 
profile and NN interaction (a). The main parameter values are 
$\eta=1$, $\sigma=0.1$, $\epsilon_{5}=0.1$, $\Delta\beta=15$, and $T=15$.   
Graph (b) is a magnified version of graph (a) for short distances. 
The blue diamonds, red circles, and green triangles represent  
$\eta_{1}(z)$, $\eta_{2}(z)$, and $\eta_{3}(z)$ obtained by the  
numerical simulation with Eqs. (\ref{gl1})-(\ref{gl3}). The solid black, 
dashed-dotted brown, and dashed gray curves correspond to   
$\eta_{1}(z)$, $\eta_{2}(z)$, and $\eta_{3}(z)$ obtained by 
the LV model (\ref{gl13})-(\ref{gl15}).}
\label{fig3}
\end{figure}

\begin{figure}[ptb]
\begin{center}
\epsfxsize=10cm  \epsffile{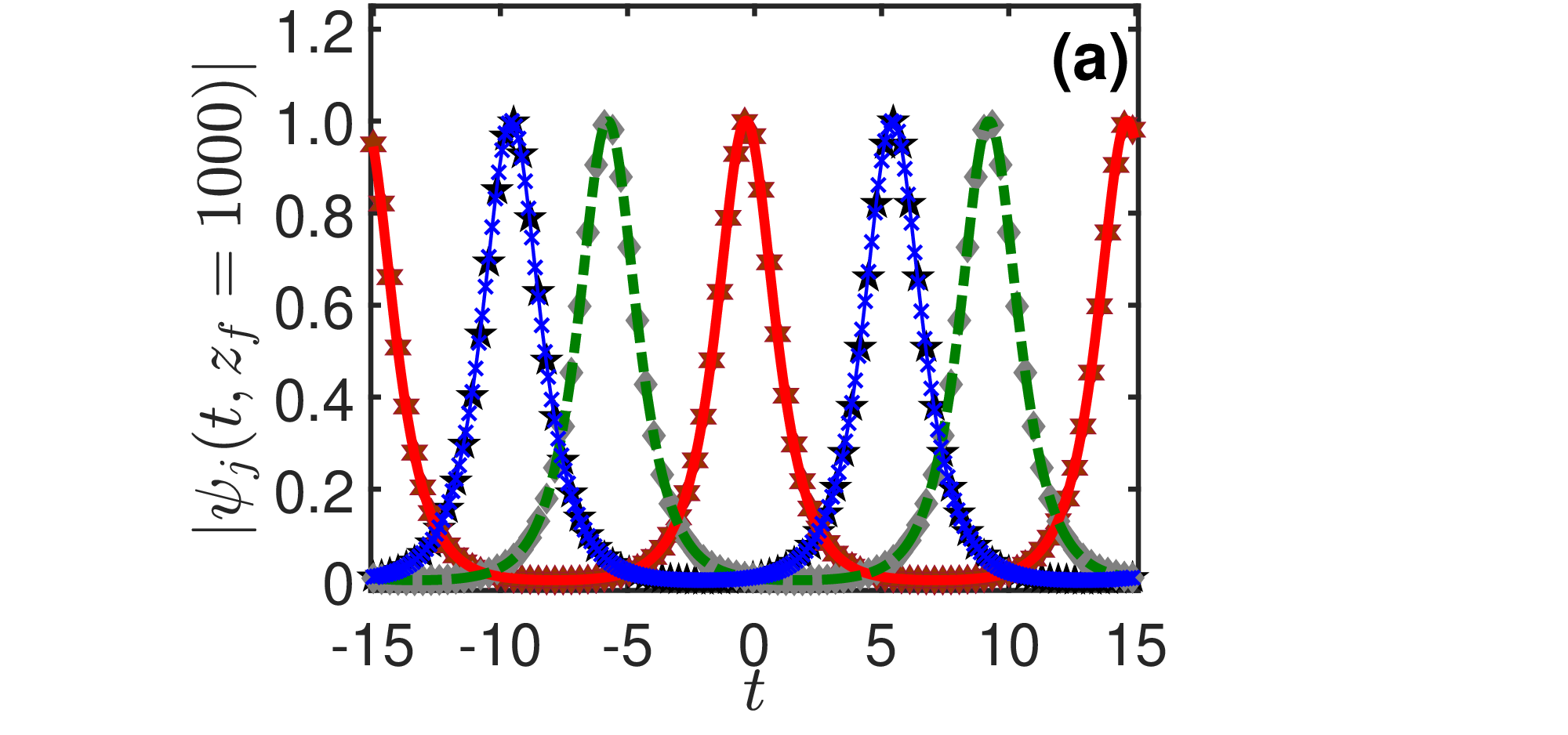}\\
\epsfxsize=10cm  \epsffile{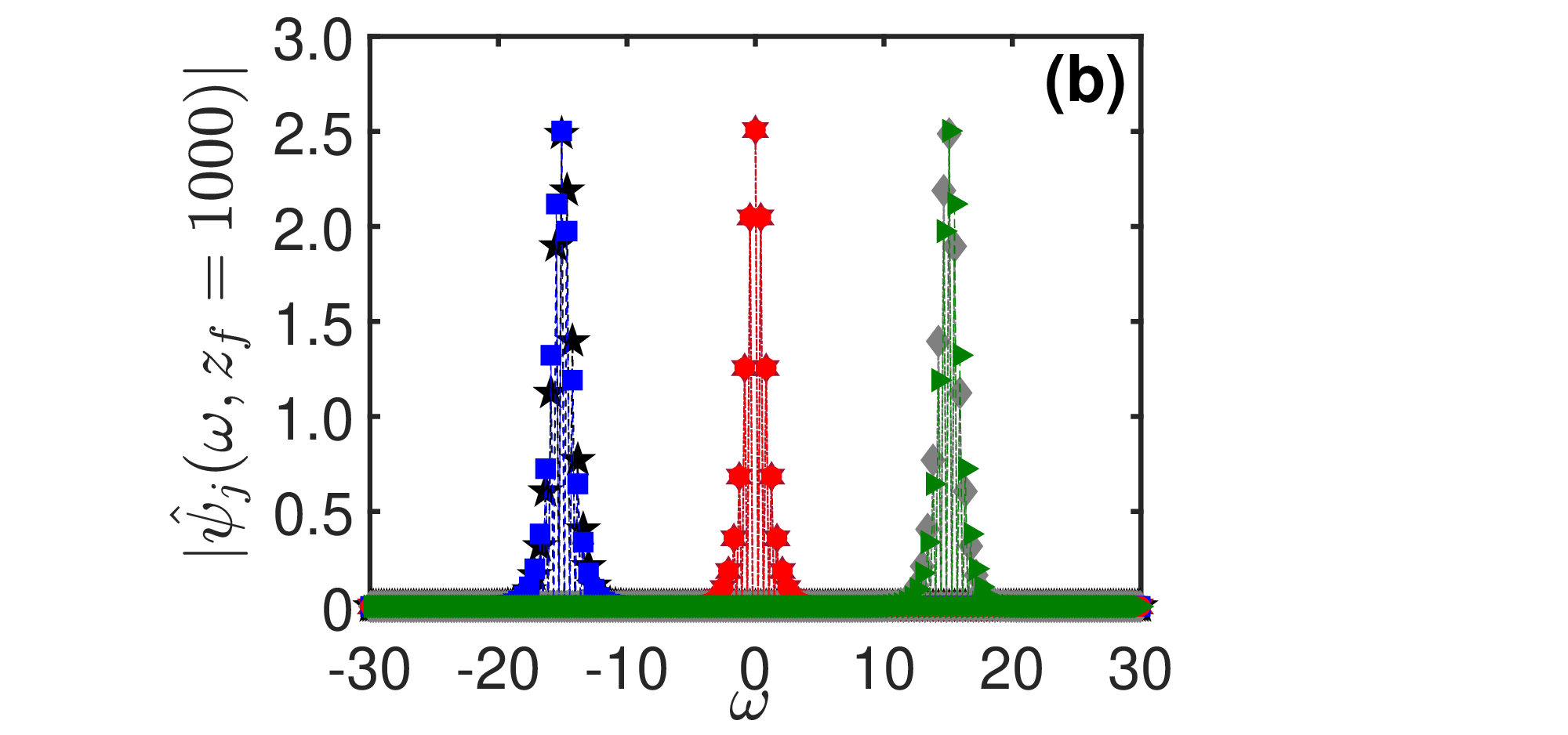}
\end{center}
\caption{(Color online) 
The final pulse patterns $|\psi_{j}(t,z_{f})|$ (a) and the final 
Fourier spectra $|\hat{\psi}_{j}(\omega,z_{f})|$ (b) of the three 
soliton sequences during transmission stabilization in a nonlinear 
waveguide array with a weak GL gain-loss profile. $z_{f}=1000$ 
and the other parameter values are the same as in Fig. \ref{fig3}. 
The solid-crossed blue curve, solid red curve, and dashed-dotted green curve 
in (a) represent $|\psi_j(t,z_f)|$ with $j=1, 2, 3$, obtained in the 
simulation with Eqs. (\ref{gl1})-(\ref{gl3}). The blue squares, red circles, 
and green right-pointing triangles in (b) represent 
$|\hat\psi_j(\omega,z_f)|$ with $j=1, 2, 3$, obtained in the simulation.
The black stars, brown six-pointed stars, and gray diamonds represent 
the theoretical prediction for $|\psi_{j}(t,z_f)|$ in (a) or for 
$|\hat\psi_{j}(\omega,z_f)|$ in (b) with $j=1, 2, 3$.}
\label{fig4}
\end{figure}

We start by discussing transmission stabilization for $J=3$. Since 
we use $\kappa=1.3$ in the numerical simulations, the condition 
$\kappa_{th} < \kappa < \kappa_{c}$ that is required for 
transmission stabilization is satisfied. Figure \ref{fig3} shows 
the $z$ dependence of the soliton amplitudes as obtained in the 
simulation with Eqs. (\ref{gl1})-(\ref{gl3}) with initial amplitudes $\eta_{1}(0)=1.1$, 
$\eta_{2}(0)=1.05$, and $\eta_{3}(0)=0.95$. Also shown is the prediction 
of the LV model (\ref{gl13})-(\ref{gl15}). We find that the 
amplitude values obtained with Eqs. (\ref{gl1})-(\ref{gl3}) tend to the equilibrium value 
$\eta=1$ with increasing distance, in very good agreement with the prediction 
of the LV model and with the linear stability analysis of Section \ref{stability_2}.         
We also find that amplitude stabilization takes place along a relatively short 
interval (of order $10^{1}$) compared with the total propagation distance, 
in accordance with the value of $\epsilon_{5}$ that is used ($\epsilon_{5}=0.1$).   
Additionally, the numerically obtained amplitude values exhibit weak oscillations 
around the equilibrium value $\eta=1$. Similar oscillatory behavior of soliton 
parameters was observed in earlier studies of propagation of NLS solitons in the 
presence of perturbations \cite{CPJ2013,Kuznetsov95,Pelinovsky98,NP2010B}. It is 
associated with the emission of radiation and with the interaction between the 
solitons and the emitted radiation  \cite{Kuznetsov95,Pelinovsky98,NP2010B}.       
Further insight into the dynamics is gained from the $t$ and $\omega$ 
dependences of the pulse patterns $|\psi_{j}(t,z)|$ and the Fourier 
spectra $|\hat{\psi}_{j}(\omega,z)|$. Figure \ref{fig4} shows   
the final pulse patterns $|\psi_{j}(t,z_{f})|$ and the corresponding 
Fourier spectra $|\hat{\psi}_{j}(\omega,z_{f})|$ that were obtained 
in the simulation together with the theoretical predictions.  
We observe that the solitons retain their shapes during the propagation, 
and that no resonant or nonresonant destabilizing features appear 
in the Fourier spectra of the soliton sequences at $z=z_{f}$. 
These observations are strongly supported by measurements of the 
pulse-pattern quality integrals $I_{j}(z)$, which are defined in 
Eq. (\ref{appendA3}) in Appendix \ref{appendA}. Indeed, the numerically 
measured values of the $I_{j}(z)$ are all smaller than $0.02$ 
throughout the propagation. Similar results to the ones shown 
in Figs. \ref{fig3} and \ref{fig4} are obtained with other 
initial conditions and with other sets of physical parameter 
values. Based on these findings we conclude that robust transmission stabilization 
with three soliton sequences is indeed possible in nonlinear optical waveguide 
arrays with a weak GL gain-loss profile and NN interaction. Furthermore, the numerical simulations 
confirm that it is indeed possible to use stability analysis for the equilibrium 
points of the LV model for designing these robust stabilizing waveguide-array setups.

\begin{figure}[ptb]
\begin{center}
\epsfxsize=10cm  \epsffile{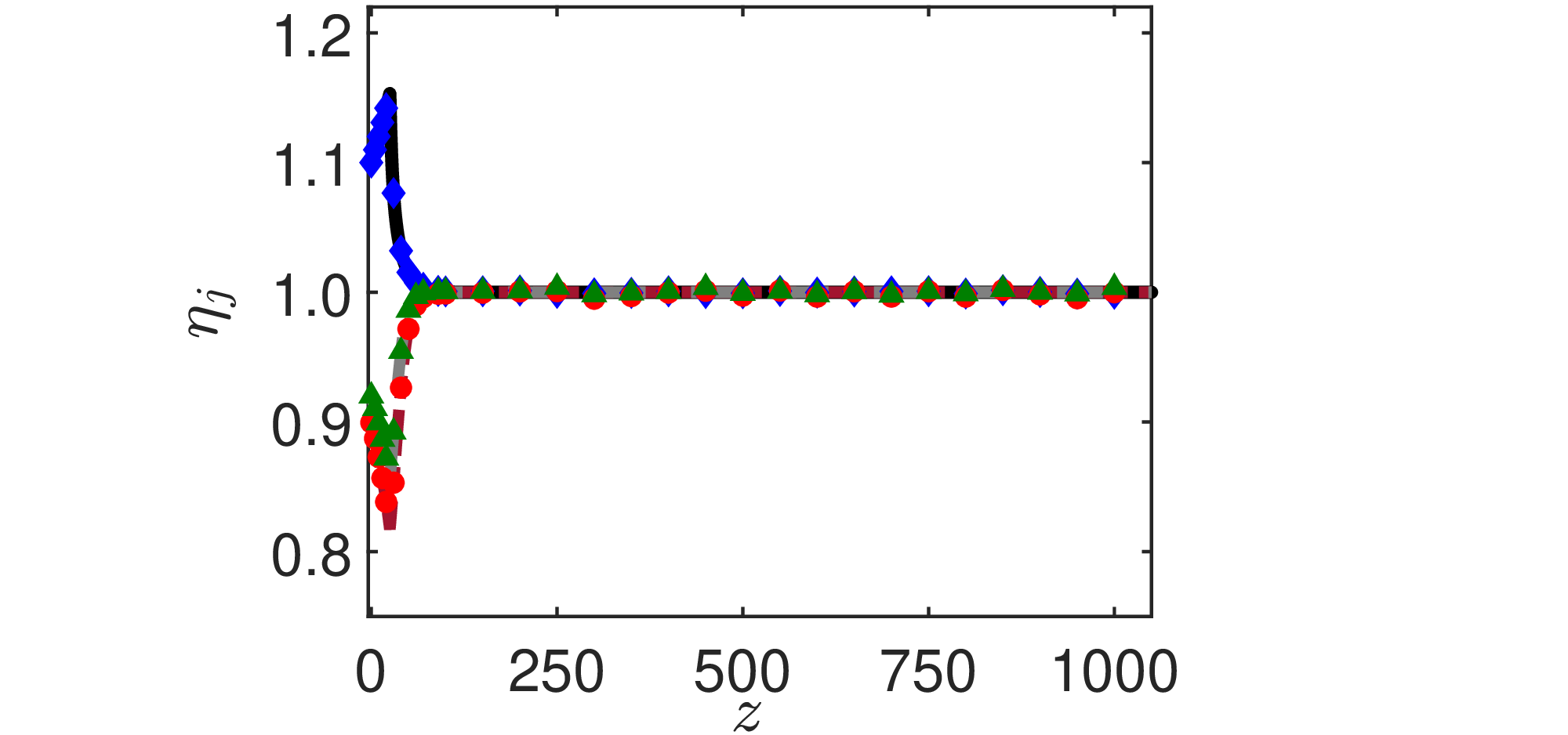}\\
\end{center}
\caption{(Color online) 
$\eta_{j}$ vs $z$ in off-on switching of sequences $j=2$ and $j=3$ 
in three-sequence transmission in a nonlinear waveguide array 
with a weak GL gain-loss profile. The switching distance is $z_{s}=25$.  
The blue diamonds, red circles, and green triangles represent  
$\eta_{1}(z)$, $\eta_{2}(z)$, and $\eta_{3}(z)$ obtained by   
numerical solution of Eqs. (\ref{gl1})-(\ref{gl3}). The solid black, 
dashed-dotted brown, and dashed gray curves correspond to   
$\eta_{1}(z)$, $\eta_{2}(z)$, and $\eta_{3}(z)$ obtained by 
the LV model (\ref{gl13})-(\ref{gl15}).}
\label{fig5}
\end{figure}

We now turn to describe the results of the simulations for transmission 
switching with three soliton sequences. As an example, we consider switching on 
and switching off of two out of the three sequences, and present the results 
for the simultaneous switching of sequences $j=2$ and $j=3$. We begin with 
the case of off-on switching. The values of $\kappa_{i}$ and $\kappa_{f}$ in the 
simulation are $\kappa_{i}=2.1$ and $\kappa_{f}=1.3$, and therefore, the conditions 
$\kappa_{i} > \kappa_{c}$ and $\kappa_{th} < \kappa_{f} < \kappa_{c}$ 
for stable off-on transmission switching are satisfied.  
Figure \ref{fig5} shows the $z$ dependence of the $\eta_{j}$ obtained in  
the simulation with Eqs. (\ref{gl1})-(\ref{gl3}) with initial amplitudes $\eta_{1}(0)=1.1$, 
$\eta_{2}(0)=0.9$, and $\eta_{3}(0)=0.92$, which satisfy the condition (\ref{gl43}). 
The prediction of the LV model (\ref{gl13})-(\ref{gl15}) is also shown.   
We observe very good agreement between the coupled-NLS simulation and 
the LV model's prediction. More specifically, for $z<z_{s}$ (before the switching), 
the value of $\eta_{1}$ increases with increasing $z$ while the values 
of $\eta_{2}$ and $\eta_{3}$ decrease with increasing $z$, such that  
sequences $j=2$ and $j=3$ are in an off state. For $z>z_{s}$ (after the switching),  
the values of all three amplitudes tend to 1 and the transmission of sequences 
$j=2$ and $j=3$ is turned on in full accordance with the prediction of the LV model.

\begin{figure}[ptb]
\begin{center}
\epsfxsize=10cm  \epsffile{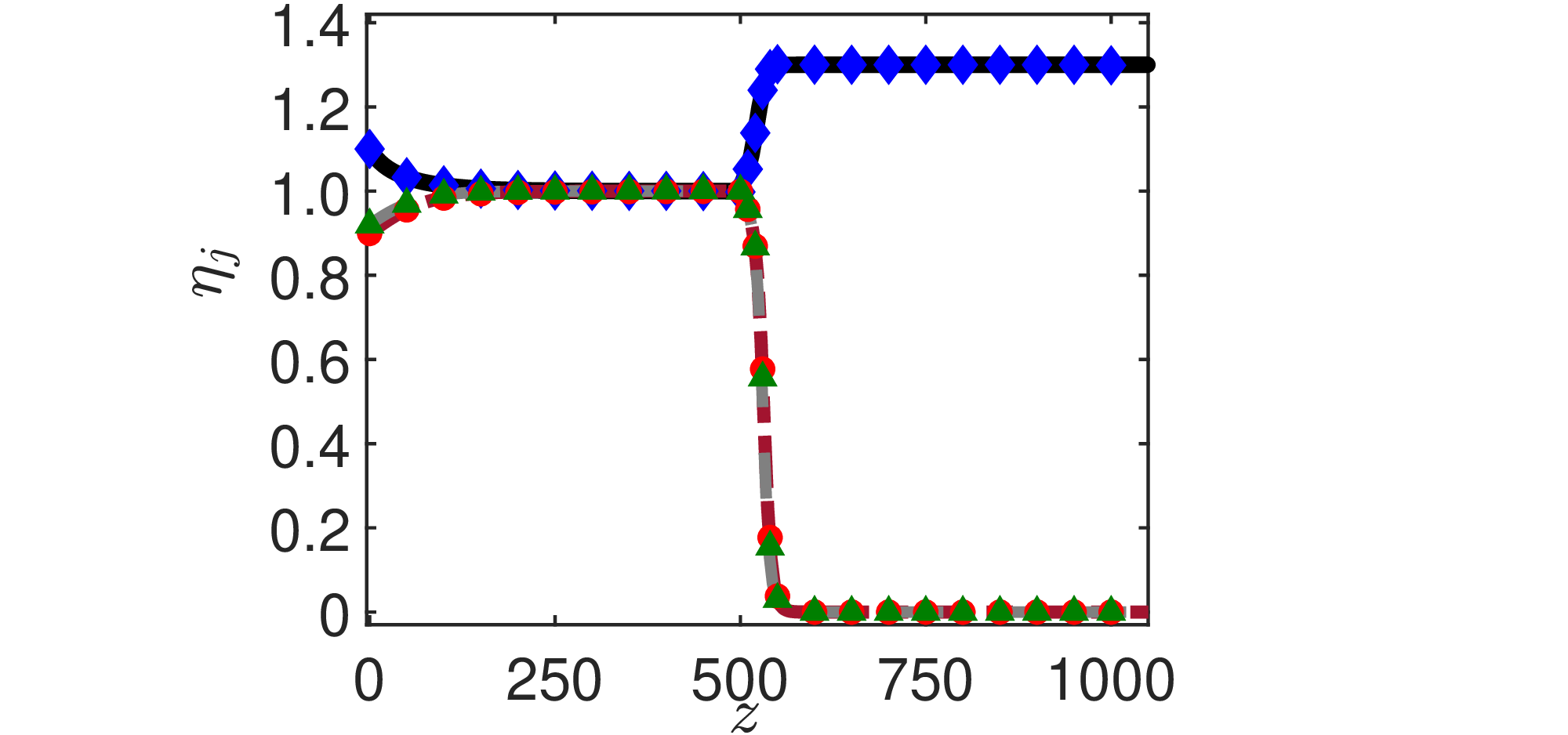}\\
\end{center}
\caption{(Color online) 
$\eta_{j}$ vs $z$ in on-off switching of sequences $j=2$ and $j=3$ 
in three-sequence transmission in a nonlinear waveguide array 
with a weak GL gain-loss profile. The intermediate and switching 
distances are $z_{i}=500$ and $z_{s}=502$, respectively.  
The blue diamonds, red circles, and green triangles represent  
$\eta_{1}(z)$, $\eta_{2}(z)$, and $\eta_{3}(z)$ obtained by   
numerical solution of Eqs. (\ref{gl1})-(\ref{gl3}). The solid black, 
dashed-dotted brown, and dashed gray curves correspond to   
$\eta_{1}(z)$, $\eta_{2}(z)$, and $\eta_{3}(z)$ obtained by 
the LV model (\ref{gl13})-(\ref{gl15}).}
\label{fig6}
\end{figure}

Next, we describe the numerical simulations results for on-off switching. 
Since we use $\kappa_{i}=1.3$ and $\kappa_{f}=2.1$, the conditions 
$\kappa_{th} < \kappa_{i} < \kappa_{c}$ and $\kappa_{f} > \kappa_{c}$ 
for stable on-off transmission switching are fulfilled.   
Figure \ref{fig6} shows the $z$ dependence of the soliton amplitudes obtained in  
the simulation with Eqs. (\ref{gl1})-(\ref{gl3}) with initial amplitudes $\eta_{1}(0)=1.1$, 
$\eta_{2}(0)=0.9$, and $\eta_{3}(0)=0.92$, which satisfy condition (\ref{gl45}). 
A comparison with the prediction of the LV model (\ref{gl13})-(\ref{gl15}) is also shown. 
The agreement between the coupled-NLS simulation and the LV model's prediction is very good. 
In particular, for $0<z<z_{i}$ (before the switching), the numerically obtained amplitude 
values approach 1 with increasing $z$, and all three sequences are in an on state. 
For $z>z_{s}$ (after the switching), the value of $\eta_{1}$ tends to 
$\eta^{(num)}_{1}=1.3042$, while the values of $\eta_{2}$ and $\eta_{3}$ 
tend to zero. Thus, after the switching, the transmission of sequences 
$j=2$ and $j=3$ is turned off in full accordance with the LV model's 
predictions and with the stability analysis in Sections \ref{stability_2} 
and \ref{stability_3}. We also note that the numerically obtained equilibrium 
value of $\eta_{1}$, $\eta^{(num)}_{1}=1.3042$, is in very good agreement 
with the equilibrium value predicted by the LV model ($\eta^{(th)}_{1}=1.3001$) 
and is also quite close to the prediction of the uncoupled ODE 
model ($\eta^{(un)}_{1}=1.2748$). The results shown in Figs.  
\ref{fig5} and \ref{fig6} together with results obtained with other sets 
of the physical parameter values clearly demonstrate that it is possible 
to realize robust off-on and on-off transmission switching with three 
soliton sequences in nonlinear waveguide arrays with a weak GL gain-loss profile 
and NN interaction. Moreover, the results show that the design of waveguide 
setups for robust transmission switching can indeed be based on stability 
and bifurcation analysis for the equilibrium points of 
the LV model (\ref{gl13})-(\ref{gl15}).

\subsubsection{Four soliton sequences ($J=4$)}
\label{simu_22}

In the numerical simulations for transmission stabilization and switching 
with four soliton sequences, we use $\beta_{1}(0)=-3\Delta\beta/2$, 
$\beta_{2}(0)=-\Delta\beta/2$, $\beta_{3}(0)=\Delta\beta/2$, and 
$\beta_{4}(0)=3\Delta\beta/2$ with $\Delta\beta=15$. In addition, 
$y_{10}=-T/2$, $y_{20}=0$, $y_{30}=0$, and $y_{40}=T/2$, where $T=15$. 
Thus, the values of $\kappa_{th}$ and $\kappa_{c}$ are $\kappa_{th}=0.9630$ 
and $\kappa_{c}=1.6183$.

\begin{figure}[ptb]
\begin{center}
\epsfxsize=10cm  \epsffile{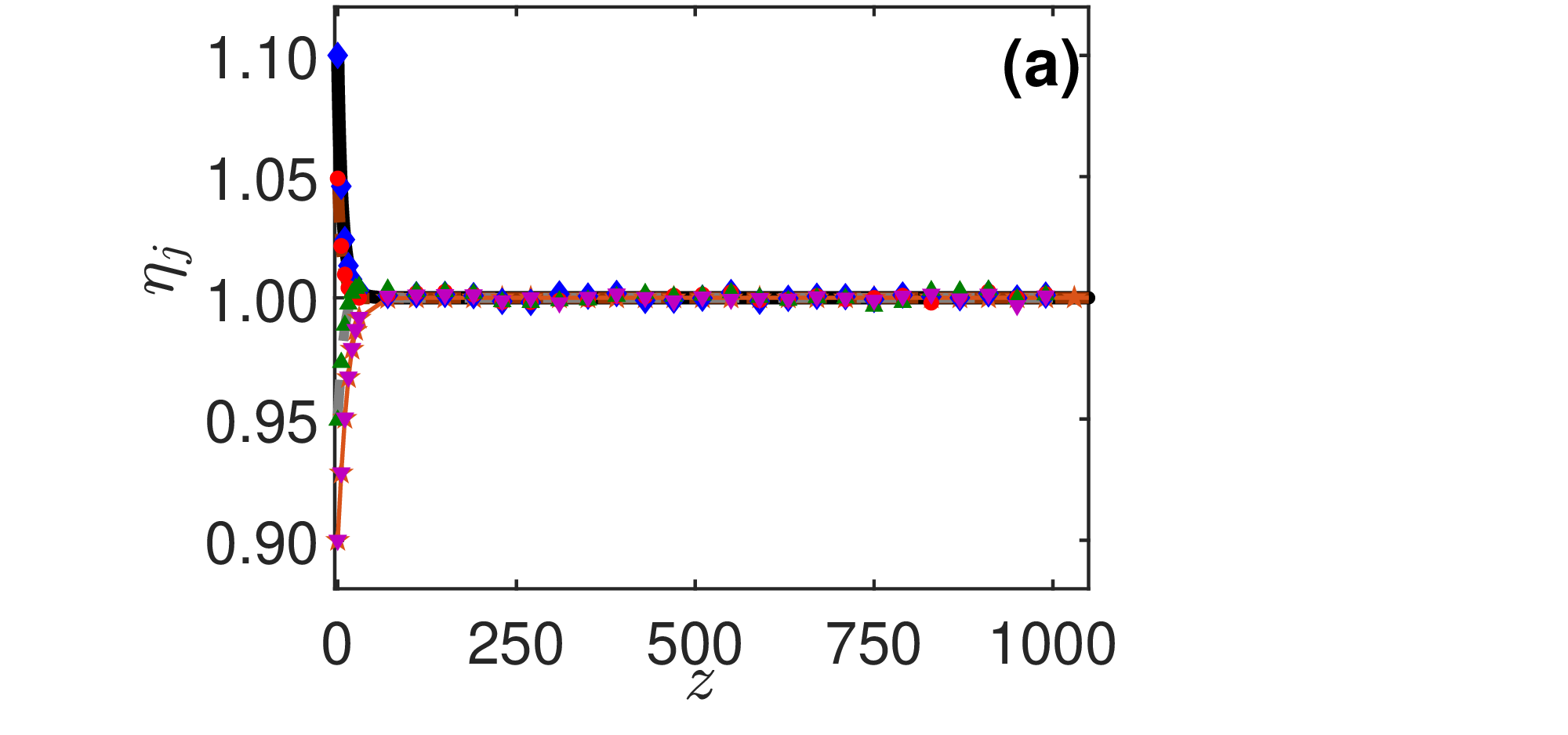}\\
\epsfxsize=10cm  \epsffile{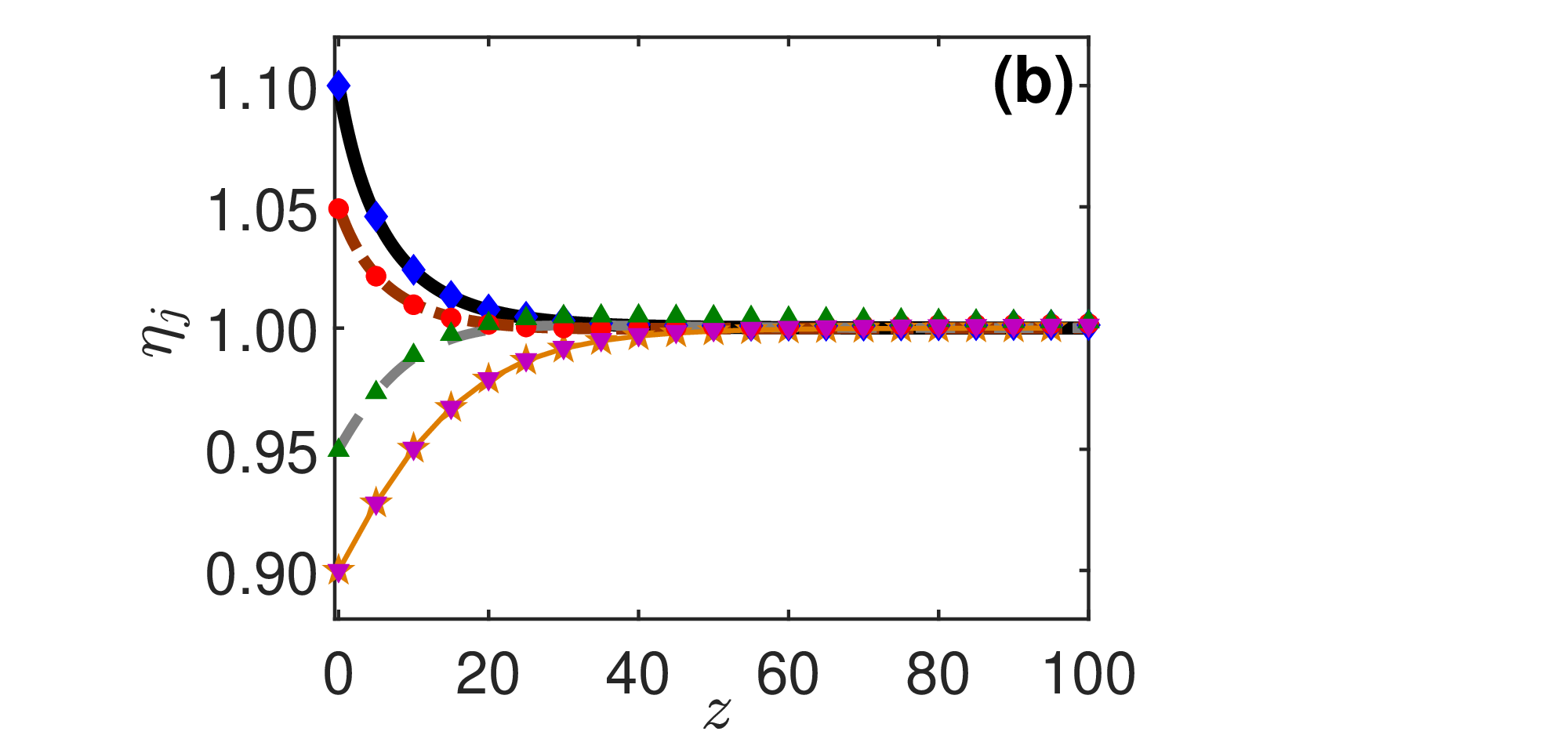}
\end{center}
\caption{(Color online) 
$\eta_{j}$ vs $z$ in transmission stabilization of four soliton 
sequences in a nonlinear waveguide array with a weak GL gain-loss 
profile and NN interaction (a). The main parameter values are 
$\eta=1$, $\sigma=0.1$, $\epsilon_{5}=0.1$, $\Delta\beta=15$, and $T=15$.   
Graph (b) is a magnified version of graph (a) for short distances. 
The blue diamonds, red circles, green up-pointing triangles, and 
magenta down-pointing triangles represent $\eta_{1}(z)$, $\eta_{2}(z)$, 
$\eta_{3}(z)$ and $\eta_{4}(z)$ obtained by numerical solution of Eqs. (\ref{gl1})-(\ref{gl3}). 
The solid black, dashed-dotted brown, dashed gray, and solid-starred orange 
curves correspond to $\eta_{1}(z)$, $\eta_{2}(z)$, $\eta_{3}(z)$, and $\eta_{4}(z)$ 
obtained by the LV model (\ref{gl13})-(\ref{gl15}).}
\label{fig7}
\end{figure}

\begin{figure}[ptb]
\begin{center}
\epsfxsize=10cm  \epsffile{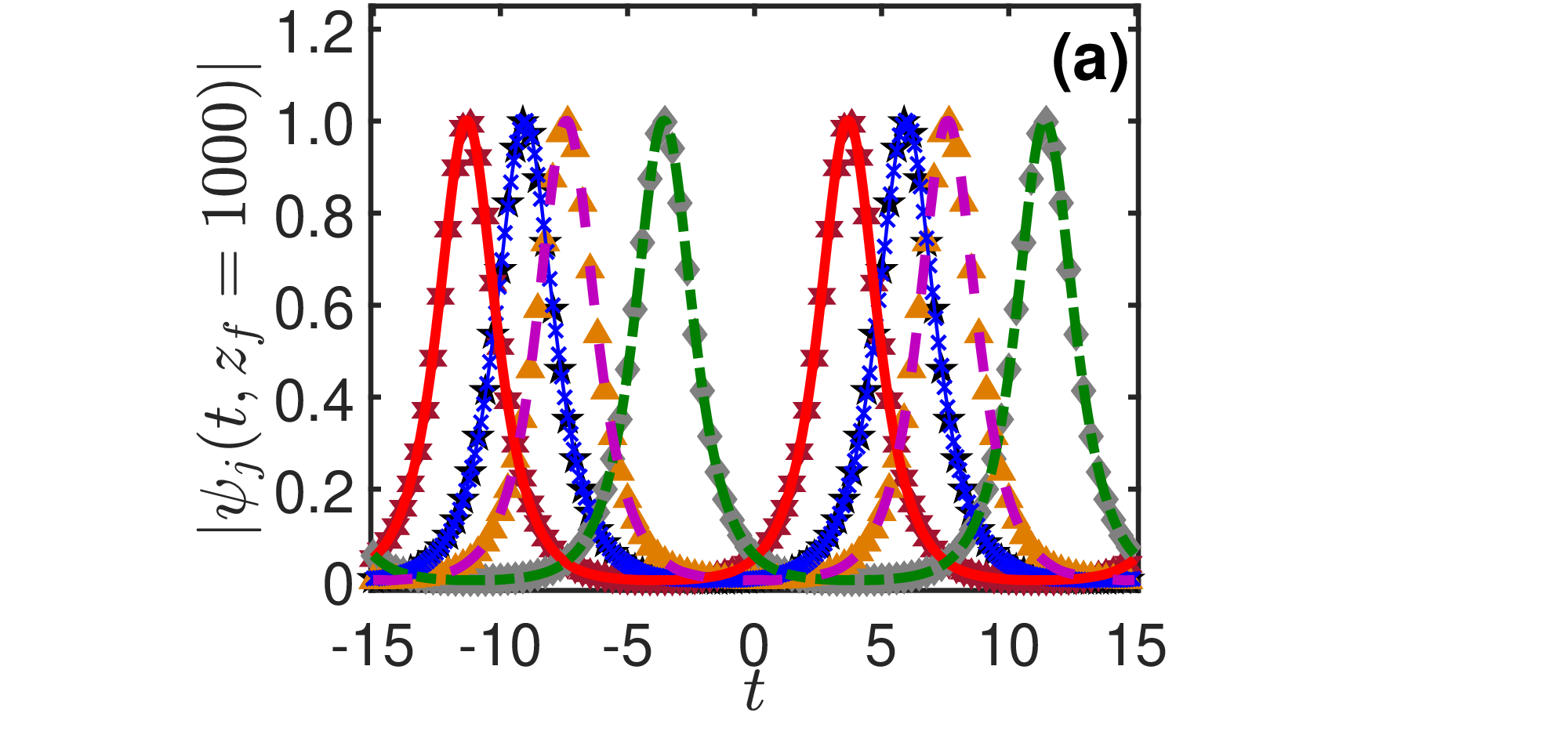}\\
\epsfxsize=10cm  \epsffile{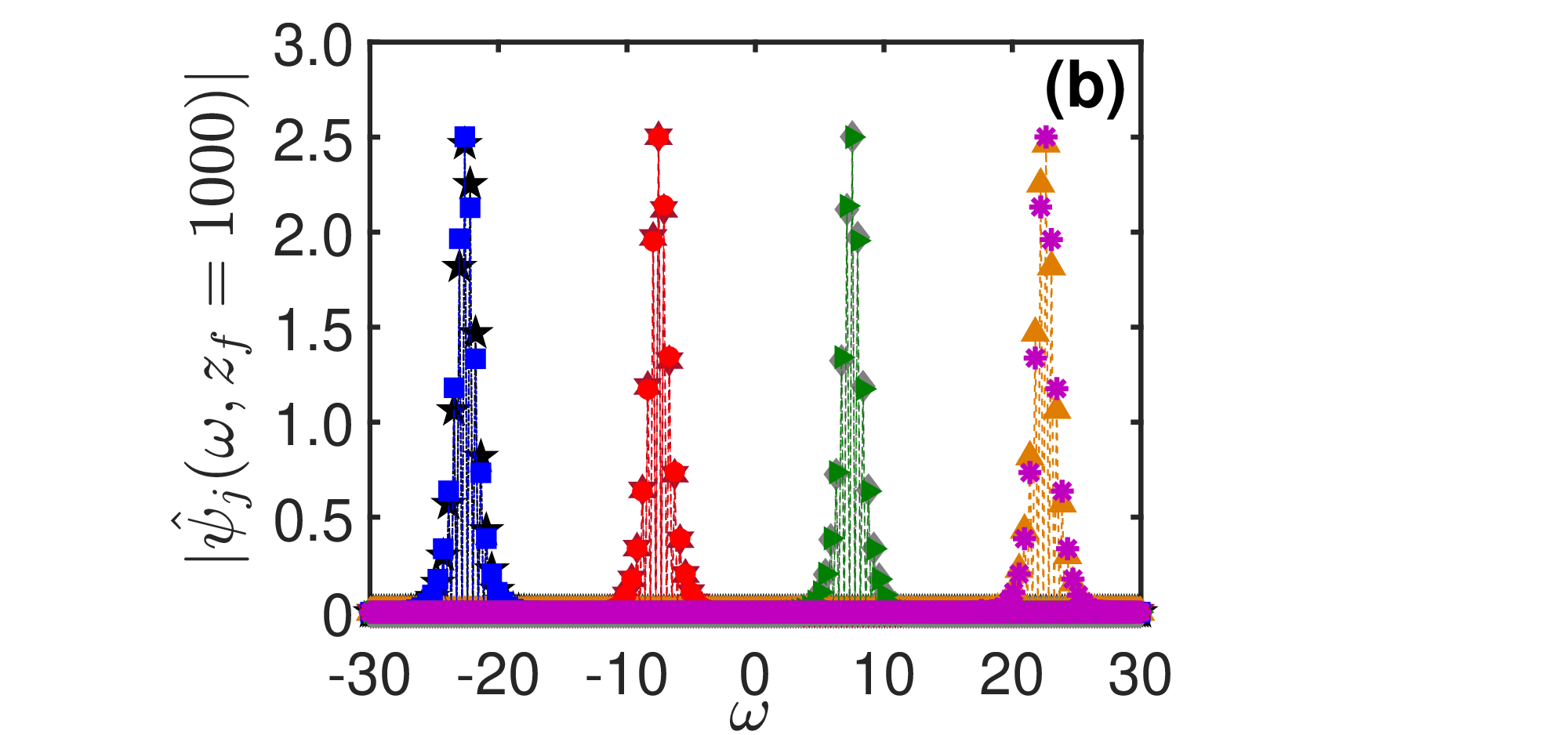}
\end{center}
\caption{(Color online) 
The final pulse patterns $|\psi_{j}(t,z_{f})|$ (a) and the corresponding  
Fourier spectra $|\hat{\psi}_{j}(\omega,z_{f})|$ (b) of the four  
soliton sequences during transmission stabilization in a nonlinear 
waveguide array with a weak GL gain-loss profile. $z_{f}=1000$ 
and the other parameter values are the same as in Fig. \ref{fig7}. 
The solid-crossed blue curve, solid red curve, dashed-dotted green curve, 
and dashed magenta curve in (a) represent $|\psi_j(t,z_f)|$ with $j=1, 2, 3, 4$, 
obtained in the simulation with Eqs. (\ref{gl1})-(\ref{gl3}). The blue squares, red circles, 
green right-pointing triangles, and magenta asterisks in (b) represent 
$|\hat\psi_j(\omega,z_f)|$ with $j=1, 2, 3, 4$, obtained in the simulation.
The black stars, brown six-pointed stars, gray diamonds, and orange up-pointing triangles
represent the theoretical prediction for $|\psi_{j}(t,z_f)|$ in (a) or for 
$|\hat\psi_{j}(\omega,z_f)|$ in (b) with $j=1, 2, 3, 4$.}
\label{fig8}
\end{figure}

In the simulations for transmission stabilization we use the value 
$\kappa=1.3$, and as a result, the required condition $\kappa_{th} < \kappa < \kappa_{c}$    
is met. The $z$ dependence of the soliton amplitudes obtained in the simulation 
with Eqs. (\ref{gl1})-(\ref{gl3}) with initial amplitudes $\eta_{1}(0)=1.1$, $\eta_{2}(0)=1.05$, 
$\eta_{3}(0)=0.95$, and $\eta_{4}(0)=0.9$ is shown in Fig. \ref{fig7} together 
with the prediction of the LV model (\ref{gl13})-(\ref{gl15}). We observe 
that the numerically obtained amplitude values tend to the equilibrium value 
of $1$ with increasing distance, in very good agreement with the LV model's prediction. 
Additionally, transmission stabilization is realized within a relatively short 
interval, $\Delta z \sim 10$, in accordance with the value of $\epsilon_{5}$ 
that is used, $\epsilon_{5}=0.1$. Stabilization of the four soliton sequences 
is also evident in Fig. \ref{fig8}, which shows the final pulse patterns 
$|\psi_{j}(t,z_{f})|$ and the final Fourier spectra $|\hat{\psi}_{j}(\omega,z_{f})|$. 
We see that the solitons preserve their shapes throughout the propagation.  
Furthermore, no destabilizing radiative features are present in the 
Fourier spectra at $z=z_{f}$. These observations are also backed up by the 
numerically measured values of the $I_{j}(z)$ integrals, which are all smaller 
than 0.02 for $0 \le z \le z_{f}$. Similar results are obtained with other 
sets of physical parameter values.

\begin{figure}[ptb]
\begin{center}
\epsfxsize=10cm  \epsffile{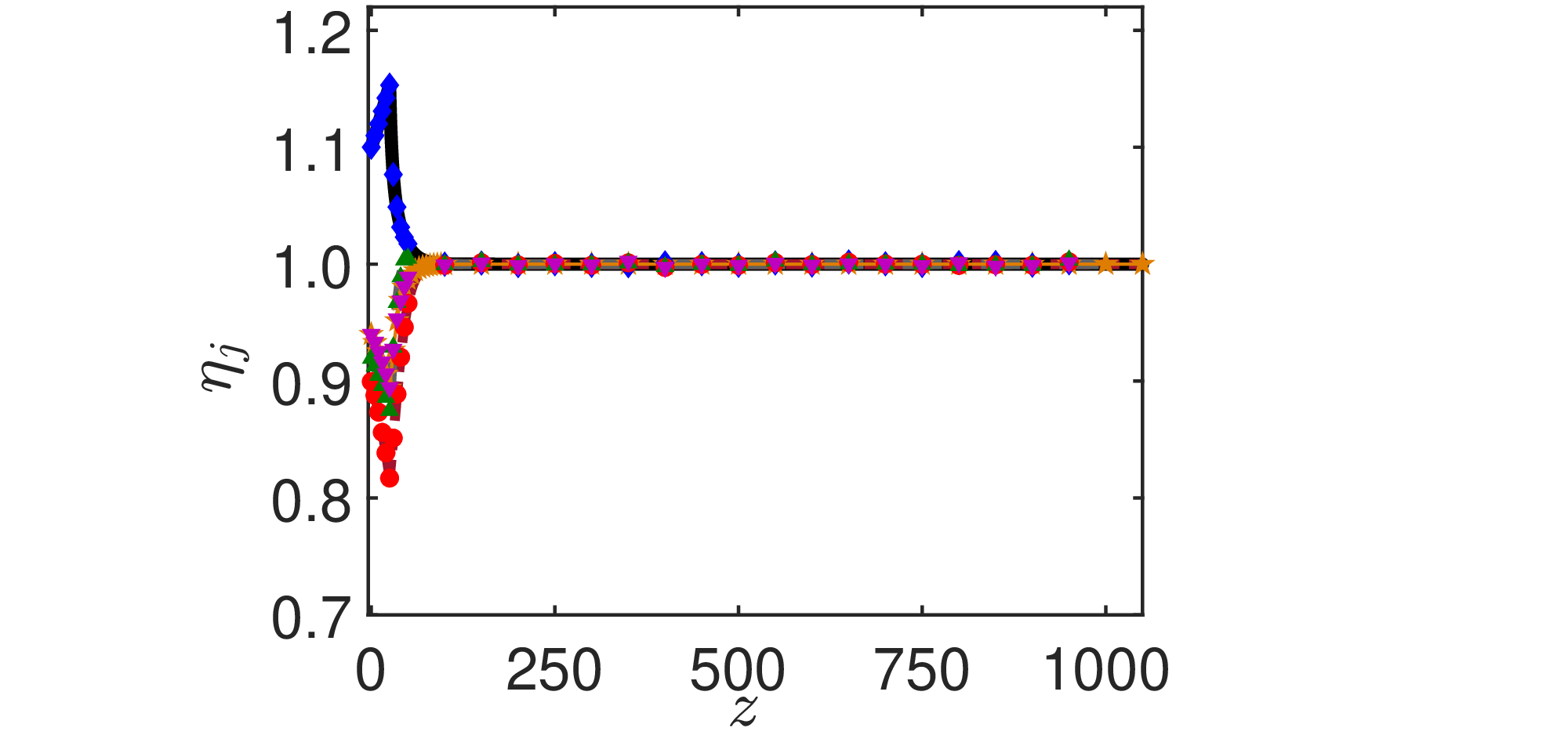}\\
\end{center}
\caption{(Color online) 
$\eta_{j}$ vs $z$ in off-on switching of sequences $j=2$, $j=3$, 
and $j=4$ in four-sequence transmission in a nonlinear waveguide array 
with a weak GL gain-loss profile. The switching distance is $z_{s}=25$.  
The blue diamonds, red circles, green up-pointing triangles, and magenta 
down-pointing triangles represent $\eta_{1}(z)$, $\eta_{2}(z)$, $\eta_{3}(z)$, 
and $\eta_{4}(z)$ obtained by numerical solution of Eqs. (\ref{gl1})-(\ref{gl3}). 
The solid black, dashed-dotted brown, dashed gray, and solid-starred orange 
curves correspond to $\eta_{1}(z)$, $\eta_{2}(z)$, $\eta_{3}(z)$, and $\eta_{4}(z)$ 
obtained by the LV model (\ref{gl13})-(\ref{gl15}).}
\label{fig9}
\end{figure}

We now describe the results of the simulations for transmission switching, 
considering as an example, the switching of three out of the four soliton sequences. 
We present the simulations results for switching of sequences $j=2$, $j=3$, and $j=4$, 
and start with the case of off-on switching. Since $\kappa_{i}=2.1$ and $\kappa_{f}=1.3$ 
are used in the simulation, the conditions $\kappa_{i} > \kappa_{c}$ and 
$\kappa_{th} < \kappa_{f} < \kappa_{c}$ for stable off-on transmission switching 
are satisfied. The $z$ dependence of the soliton amplitudes obtained by   
numerical solution of Eqs. (\ref{gl1})-(\ref{gl3}) with initial amplitudes $\eta_{1}(0)=1.1$, 
$\eta_{2}(0)=0.9$, $\eta_{3}(0)=0.92$, and $\eta_{4}(0)=0.94$, 
which satisfy condition (\ref{gl43}), is shown in Fig. \ref{fig9}.   
A comparison with the prediction of the LV model (\ref{gl13})-(\ref{gl15}) 
is also shown. The agreement between the coupled-NLS simulation and the 
LV model's prediction is very good. In particular, before the switching (for $z<z_{s}$), 
the value of $\eta_{1}$ increases with increasing $z$ while the values of 
$\eta_{2}$, $\eta_{3}$, and $\eta_{4}$ decrease with increasing $z$, and as a result,   
sequences $j=2$, $j=3$, and $j=4$ are in an off state. After the switching (for $z>z_{s}$),  
the values of all four amplitudes tend to 1 and therefore, the transmission of sequences 
$j=2$, $j=3$, and $j=4$ is turned on, in full agreement with the LV model's prediction.

\begin{figure}[ptb]
\begin{center}
\epsfxsize=10cm  \epsffile{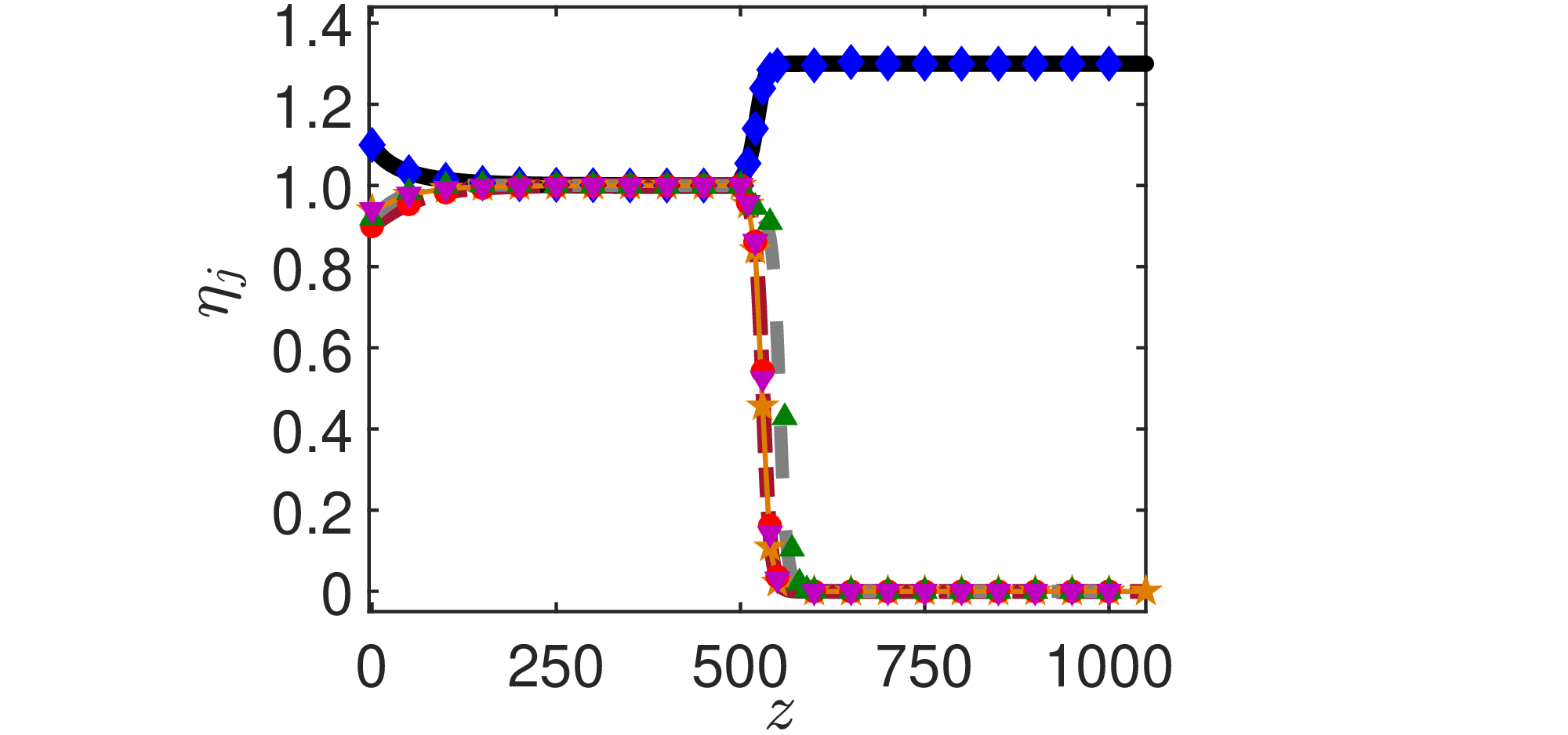}\\
\end{center}
\caption{(Color online) 
$\eta_{j}$ vs $z$ in on-off switching of sequences $j=2$, $j=3$, 
and $j=4$ in four-sequence transmission in a nonlinear waveguide array 
with a weak GL gain-loss profile. The intermediate and switching 
distances are $z_{i}=500$ and $z_{s}=502$, respectively.  
The blue diamonds, red circles, green up-pointing triangles, and magenta 
down-pointing triangles represent $\eta_{1}(z)$, $\eta_{2}(z)$, $\eta_{3}(z)$, 
and $\eta_{4}(z)$ obtained by numerical solution of Eqs. (\ref{gl1})-(\ref{gl3}). 
The solid black, dashed-dotted brown, dashed gray, and solid-starred orange 
curves correspond to $\eta_{1}(z)$, $\eta_{2}(z)$, $\eta_{3}(z)$, and $\eta_{4}(z)$ 
obtained by the LV model (\ref{gl13})-(\ref{gl15}).}
\label{fig10}
\end{figure}

In the numerical simulation for on-off switching of sequences $j=2$, $j=3$, 
and $j=4$, we use the parameter values $\kappa_{i}=1.3$ and $\kappa_{f}=2.1$. 
As a result, the conditions $\kappa_{th} < \kappa_{i} < \kappa_{c}$ and 
$\kappa_{f} > \kappa_{c}$ for stable on-off transmission switching are met. 
The $z$ dependence of the $\eta_{j}$ obtained in the simulation with 
Eqs. (\ref{gl1})-(\ref{gl3}) with initial amplitudes $\eta_{1}(0)=1.1$, $\eta_{2}(0)=0.9$, 
$\eta_{3}(0)=0.92$, and $\eta_{4}(0)=0.94$, which satisfy condition (\ref{gl45}), 
is shown in Fig. \ref{fig10}. Also shown is the prediction of the LV model 
(\ref{gl13})-(\ref{gl15}). We find very good agreement between the coupled-NLS 
simulation and the LV model's prediction. Indeed, before the switching (for $0<z<z_{i}$), 
the numerically obtained amplitude values approach 1 with increasing $z$, 
and all four soliton sequences are in an on state. Additionally, after the switching 
(for $z>z_{s}$), the value of $\eta_{1}$ tends to $\eta^{(num)}_{1}=1.3001$, 
while the values of $\eta_{2}$, $\eta_{3}$, and $\eta_{4}$ tend to zero, 
in full alignment with the LV model's prediction. Thus, after the switching, 
the transmission of sequences $j=2$, $j=3$, and $j=4$ is turned off.   
We also point out that the numerically obtained equilibrium 
value of $\eta_{1}$, $\eta^{(num)}_{1}=1.3001$, is in excellent agreement 
with the equilibrium value predicted by the LV model, $\eta^{(th)}_{1}=1.3001$.  
Similar results to the ones shown in Figs. \ref{fig7}-\ref{fig10} are obtained 
with other sets of initial conditions and with other physical parameter values. 
Thus, based on all these results, we conclude that the design of robust setups for transmission 
stabilization and switching with four soliton sequences can indeed be based on 
stability and bifurcation analysis for the equilibrium points of the LV model 
(\ref{gl13})-(\ref{gl15}).

\subsubsection{Five soliton sequences ($J=5$)}
\label{simu_23}

We now turn to describe the results of the simulations for transmission stabilization 
and switching with five soliton sequences. We remark that this is the first instance, 
where simulations of long-distance multisequence propagation of NLS solitons with more 
than four sequences are performed and analyzed. The values of $\beta_{j}(0)$ and $y_{j0}$ 
used in the simulations are $\beta_{1}(0)=-2\Delta\beta$, $\beta_{2}(0)=-\Delta\beta$, 
$\beta_{3}(0)=0$, $\beta_{4}(0)=\Delta\beta$, $\beta_{5}(0)=2\Delta\beta$, $y_{10}=-T/2$, 
$y_{20}=0$, $y_{30}=0$, $y_{40}=0$, and $y_{50}=T/2$, where $\Delta\beta=15$ 
and $T=15$. As a result, the values of $\kappa_{th}$ and $\kappa_{c}$ are 
$\kappa_{th}=0.9630$ and $\kappa_{c}=1.6195$.

\begin{figure}[ptb]
\begin{center}
\epsfxsize=10cm  \epsffile{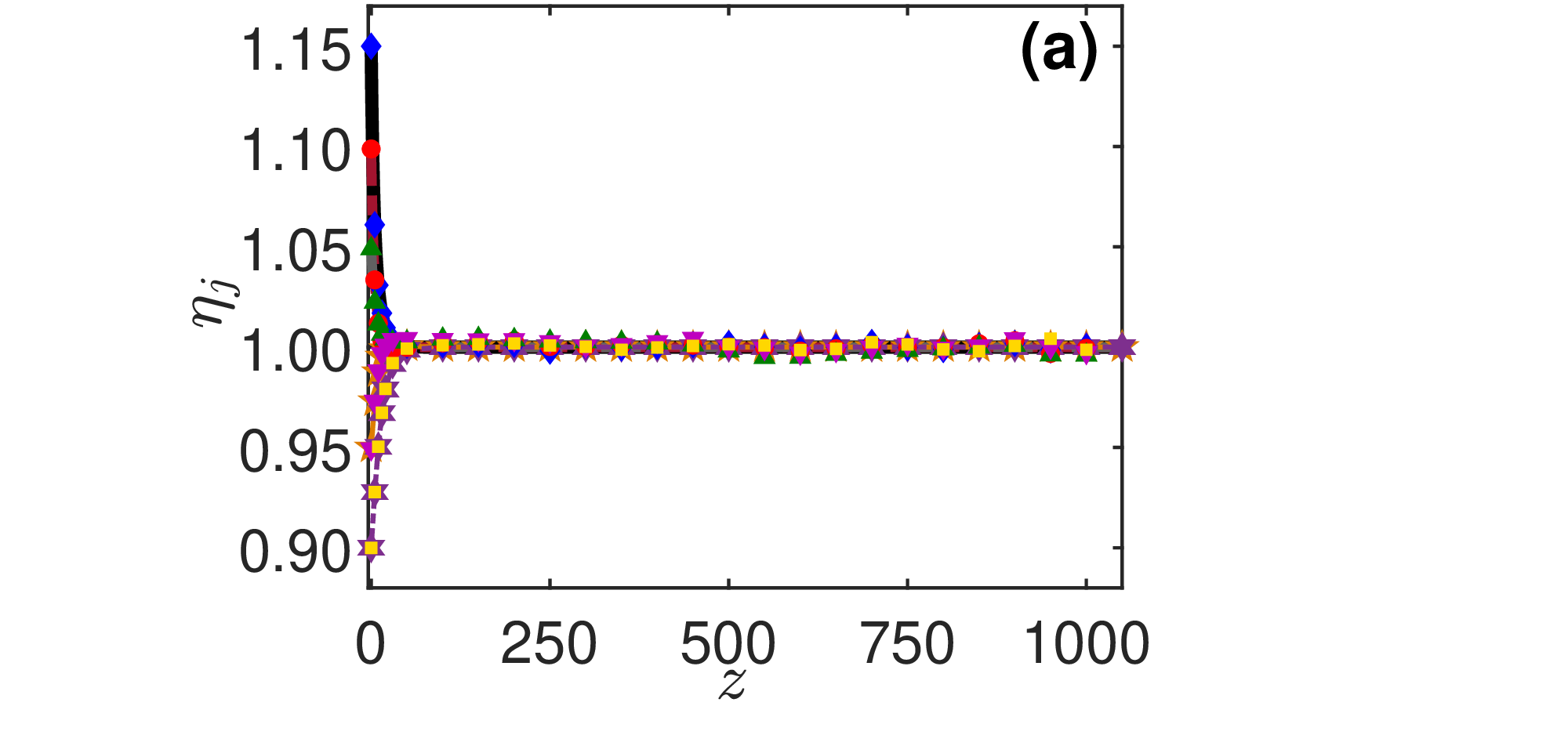}\\
\epsfxsize=10cm  \epsffile{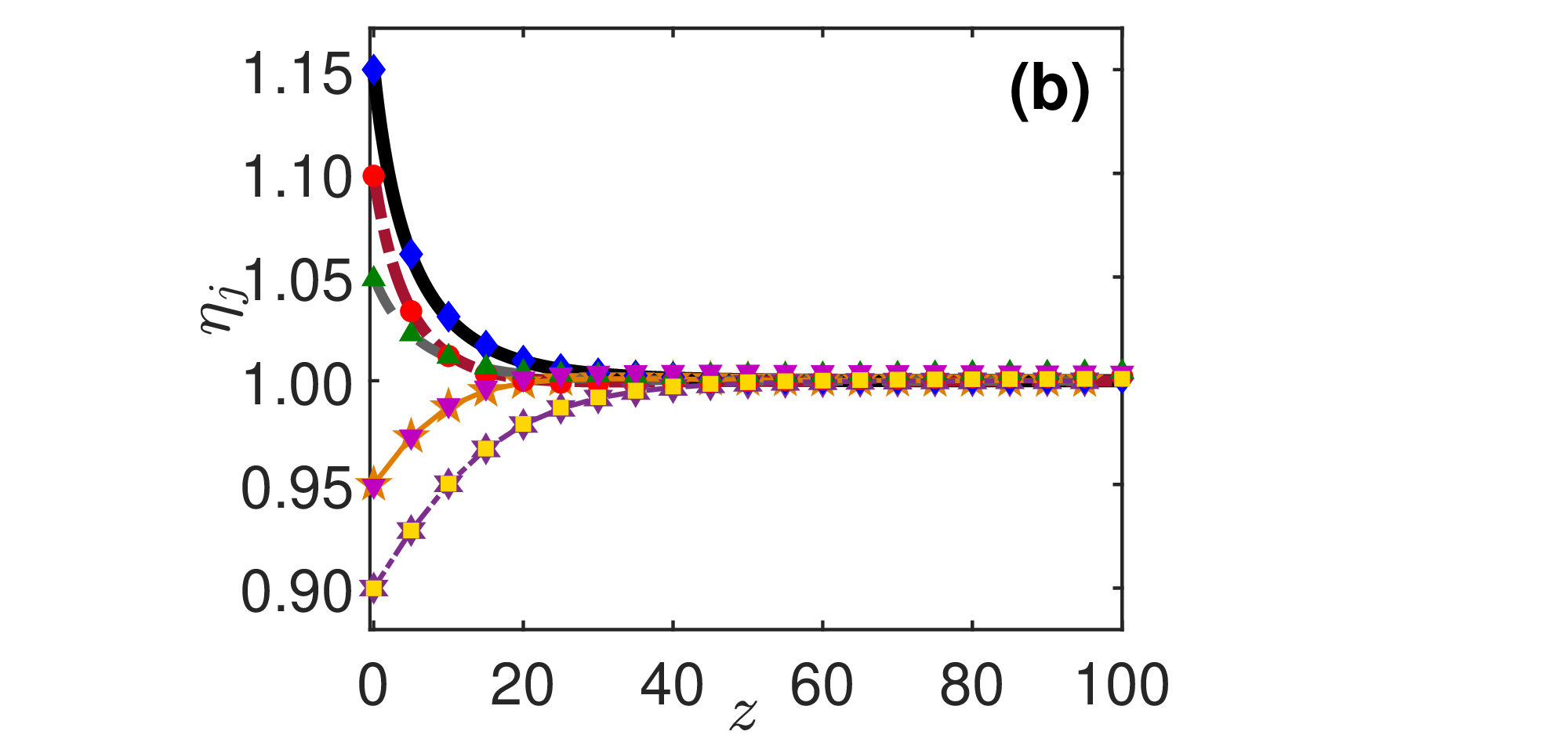}
\end{center}
\caption{(Color online) 
$\eta_{j}$ vs $z$ in transmission stabilization of five soliton 
sequences in a nonlinear waveguide array with a weak GL gain-loss 
profile and NN interaction (a). The main parameter values are 
$\eta=1$, $\sigma=0.1$, $\epsilon_{5}=0.1$, $\Delta\beta=15$, and $T=15$.   
Graph (b) is a magnified version of graph (a) for short distances. 
The blue diamonds, red circles, green up-pointing triangles, 
magenta down-pointing triangles, and yellow squares represent $\eta_{j}(z)$ 
with $j=1, 2, 3, 4, 5$, obtained by the simulation with Eqs. (\ref{gl1})-(\ref{gl3}). 
The solid black, dashed-dotted brown, dashed gray, solid-starred orange, 
and dashed-dotted six-pointed starred magenta curves correspond to $\eta_{j}(z)$ 
with $j=1, 2, 3, 4, 5$, obtained by the LV model (\ref{gl13})-(\ref{gl15}).}
\label{fig11}
\end{figure}

\begin{figure}[ptb]
\begin{center}
\epsfxsize=10cm  \epsffile{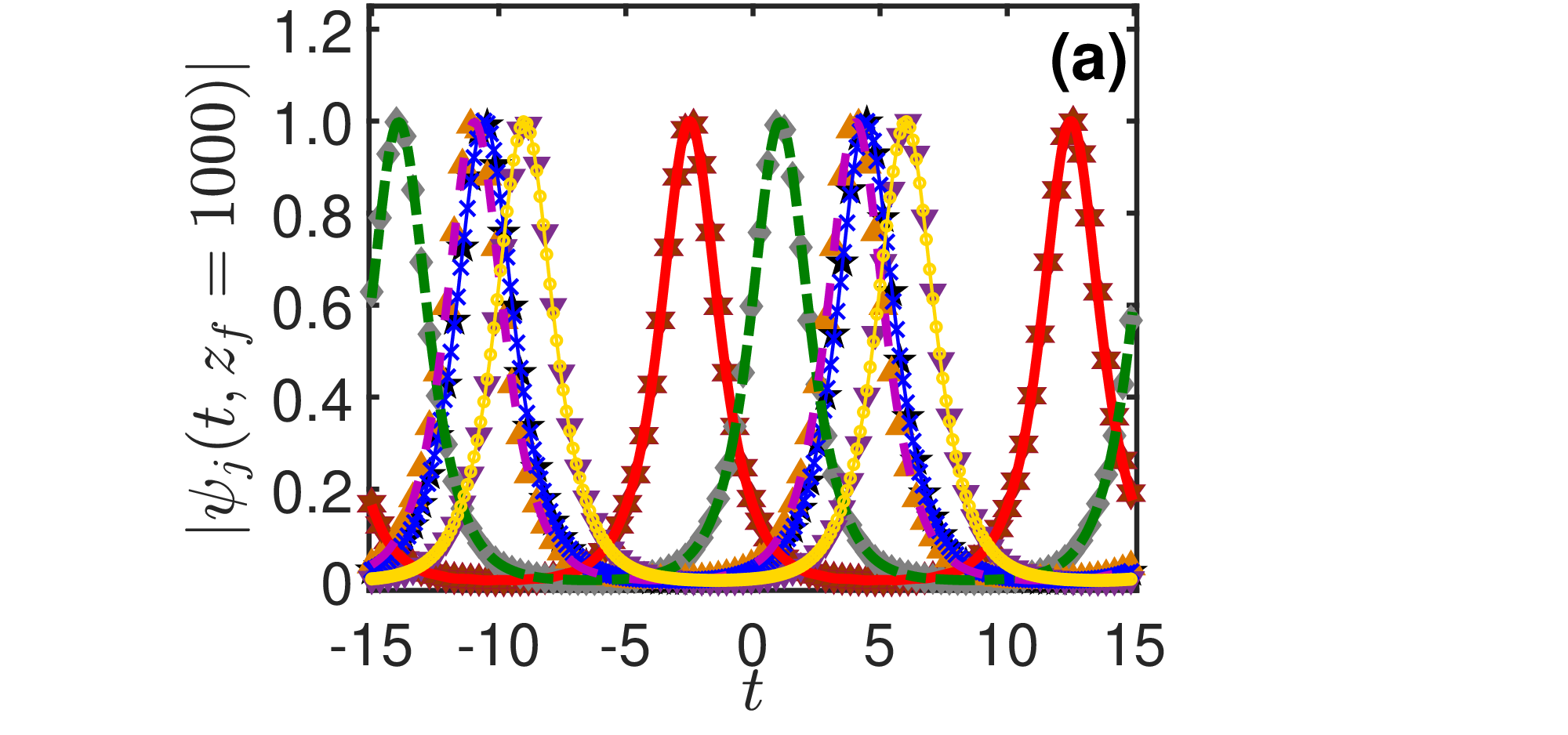}\\
\epsfxsize=10cm  \epsffile{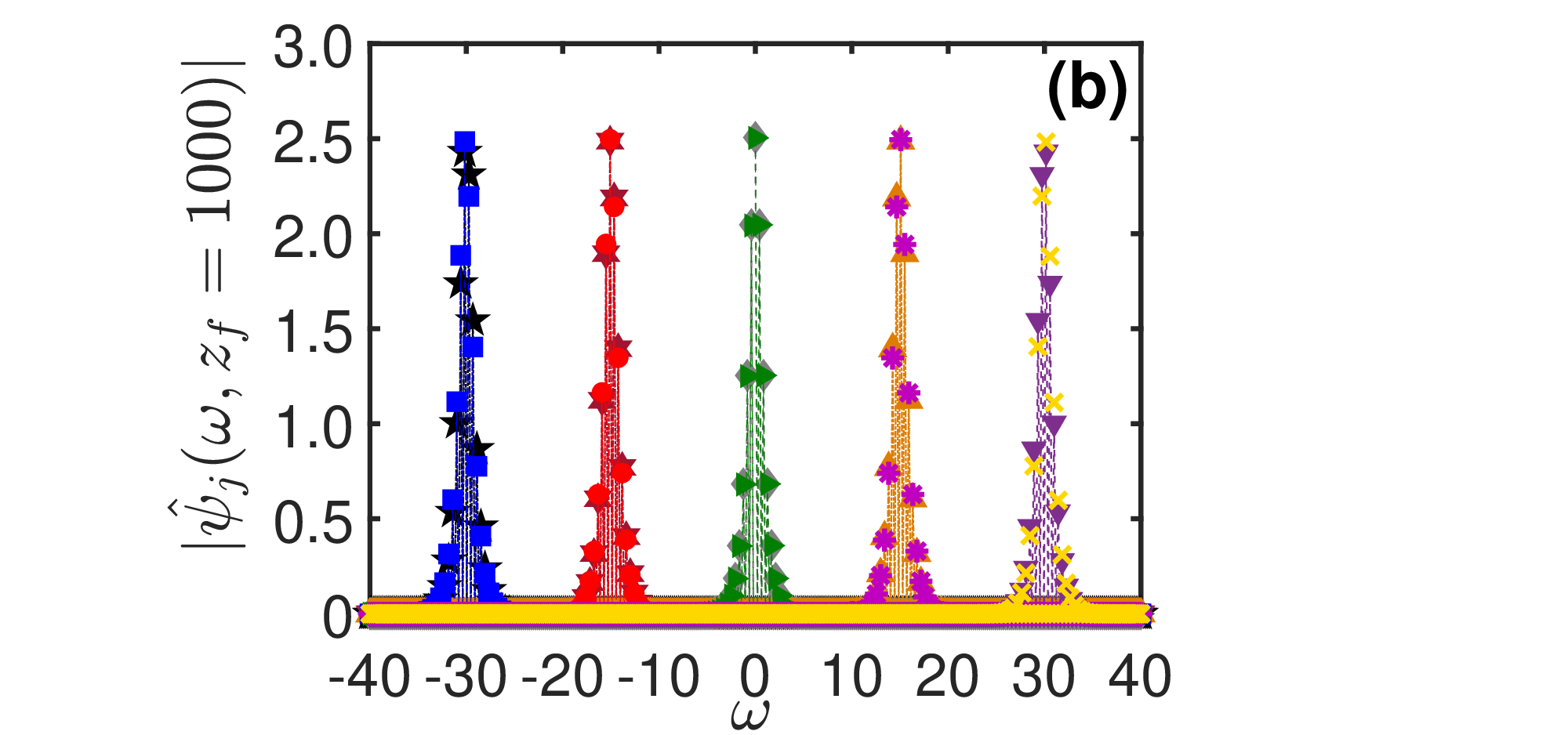}
\end{center}
\caption{(Color online) 
The final pulse patterns $|\psi_{j}(t,z_{f})|$ (a) and the corresponding  
Fourier spectra $|\hat{\psi}_{j}(\omega,z_{f})|$ (b) of the five   
soliton sequences during transmission stabilization in a nonlinear 
waveguide array with a weak GL gain-loss profile. $z_{f}=1000$ 
and the other parameter values are the same as in Fig. \ref{fig11}. 
The solid-crossed blue curve, solid red curve, dashed-dotted green curve, 
dashed magenta curve, and solid-circled yellow curve in (a) 
represent $|\psi_j(t,z_f)|$ with $j=1, 2, 3, 4, 5$, obtained in 
the simulation with Eqs. (\ref{gl1})-(\ref{gl3}). The blue squares, red circles, 
green right-pointing triangles, magenta asterisks, and yellow crosses in (b) 
represent $|\hat\psi_j(\omega,z_f)|$ with $j=1, 2, 3, 4, 5$, obtained in the simulation.
The black stars, brown six-pointed stars, gray diamonds, orange up-pointing triangles, 
and dark magenta down-pointing triangles represent the theoretical prediction for 
$|\psi_{j}(t,z_f)|$ in (a) or for $|\hat\psi_{j}(\omega,z_f)|$ in (b) 
with $j=1, 2, 3, 4, 5$.}
\label{fig12}
\end{figure}

We consider first transmission stabilization with five pulse sequences. 
The parameter value $\kappa=1.3$ is used in the simulation, and therefore, 
the required condition $\kappa_{th} < \kappa < \kappa_{c}$ is satisfied. 
The $\eta_{j}(z)$ curves obtained in the simulation with Eqs. (\ref{gl1})-(\ref{gl3}) 
with initial amplitudes $\eta_{1}(0)=1.15$, $\eta_{2}(0)=1.1$, $\eta_{3}(0)=1.05$, 
$\eta_{4}(0)=0.95$, and $\eta_{5}(0)=0.9$ are shown in Fig. \ref{fig11} 
together with the prediction of the LV model (\ref{gl13})-(\ref{gl15}).    
We find that the amplitude values obtained by numerical solution of 
Eqs. (\ref{gl1})-(\ref{gl3}) approach the equilibrium value of $1$ with increasing distance, 
in excellent agreement with the LV model's prediction. 
Furthermore, stabilization is achieved within a relatively short interval, 
$\Delta z \sim 10$, compared with the final propagation distance, $z_{f}=1000$. 
Additional insight into stabilization dynamics is gained from Fig. \ref{fig12}, 
which shows the final pulse patterns $|\psi_{j}(t,z_{f})|$ and the corresponding 
Fourier spectra $|\hat{\psi}_{j}(\omega,z_{f})|$. We see that the solitons 
preserve their shapes during the propagation and that no destabilizing features 
appear in the Fourier spectra at $z=z_{f}$. These findings are strongly supported 
by the values of the $I_{j}(z)$ integrals measured in the simulation, which are 
all smaller than 0.02 for $0 \le z \le z_{f}$. The results obtained with other 
initial conditions and with other sets of physical parameter values 
are similar to the results shown in Figs. \ref{fig11} and \ref{fig12}.

\begin{figure}[ptb]
\begin{center}
\epsfxsize=10cm  \epsffile{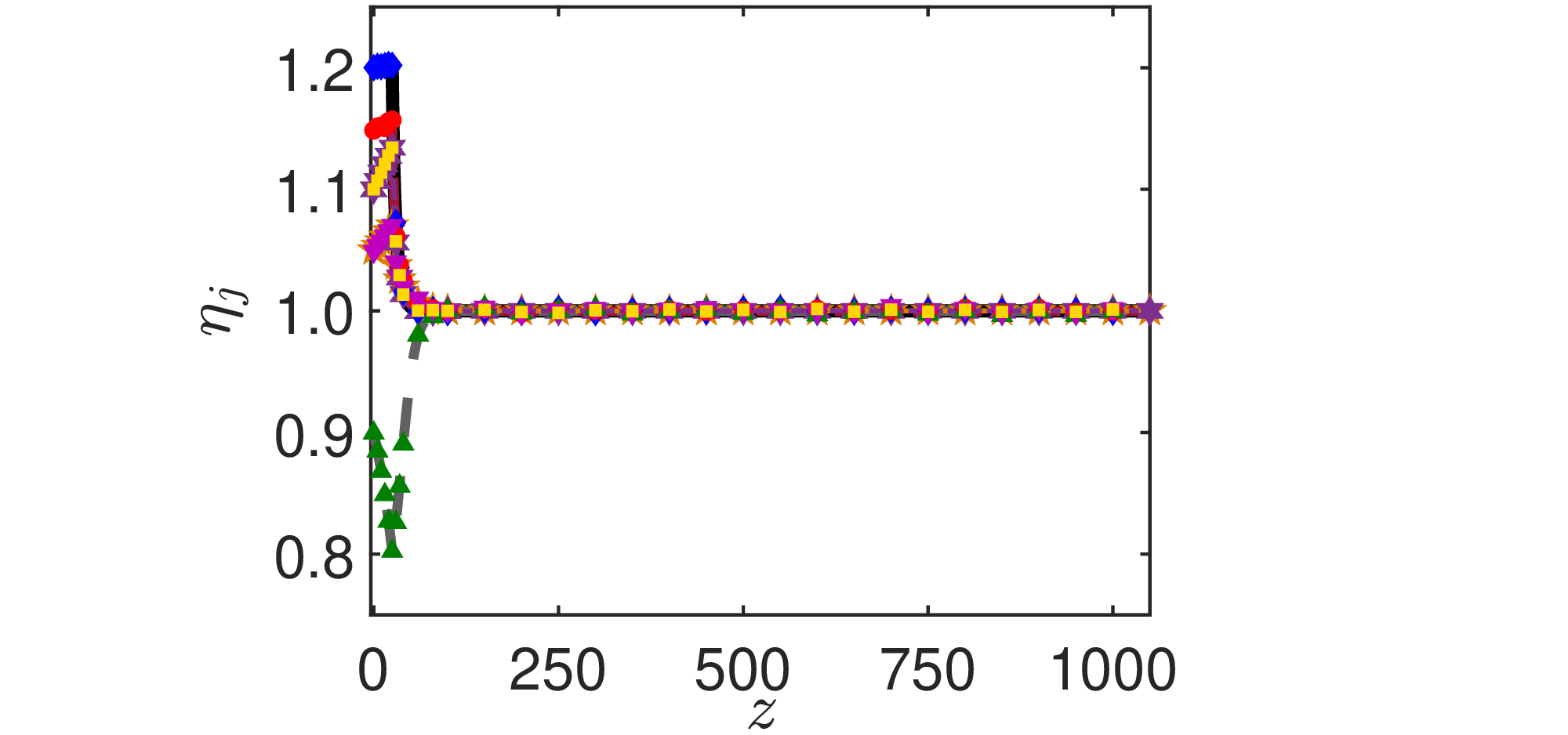}\\
\end{center}
\caption{(Color online) 
$\eta_{j}$ vs $z$ in off-on switching of the sequence $j=3$ in 
five-sequence transmission in a nonlinear waveguide array with 
a weak GL gain-loss profile. The switching distance is $z_{s}=25$.  
The blue diamonds, red circles, green up-pointing triangles, 
magenta down-pointing triangles, and yellow squares represent $\eta_{j}(z)$ 
with $j=1, 2, 3, 4, 5$, obtained by numerical solution of Eqs. (\ref{gl1})-(\ref{gl3}). 
The solid black, dashed-dotted brown, dashed gray, solid-starred orange, 
and dashed-dotted six-pointed starred magenta curves correspond to $\eta_{j}(z)$ 
with $j=1, 2, 3, 4, 5$, obtained by the LV model (\ref{gl13})-(\ref{gl15}).}
\label{fig13}
\end{figure}

We now move to describe the simulations results for transmission switching 
with five soliton sequences. We consider as an example the switching of one 
out of the five sequences, and present the simulations results for switching 
of the sequence $j=3$. We start with the case of off-on switching. 
Figure \ref{fig13} shows the $z$ dependence of the soliton amplitudes 
obtained in the simulation with Eqs. (\ref{gl1})-(\ref{gl3}) with initial amplitudes 
$\eta_{1}(0)=1.2$, $\eta_{2}(0)=1.15$, $\eta_{3}(0)=0.9$, $\eta_{4}(0)=1.05$, 
and $\eta_{5}(0)=1.1$, which satisfy condition (\ref{gl43}). The prediction of 
the LV model (\ref{gl13})-(\ref{gl15}) is also shown. The agreement between the 
coupled-NLS simulation and the LV model's prediction is very good. 
More precisely, before the switching (for $z<z_{s}$), the values of $\eta_{1}$, 
$\eta_{2}$, $\eta_{4}$, and $\eta_{5}$ increase with increasing $z$ while 
the value of $\eta_{3}$ decreases with increasing $z$, and as a result,  
sequence $j=3$ is in an off state. After the switching (for $z>z_{s}$),  
the values of all five amplitudes tend to 1 and therefore, the transmission 
of sequence $j=3$ is turned on, in full alignment with the LV model's prediction 
and with the linear stability analysis of Section \ref{stability_2}.

\begin{figure}[ptb]
\begin{center}
\epsfxsize=10cm  \epsffile{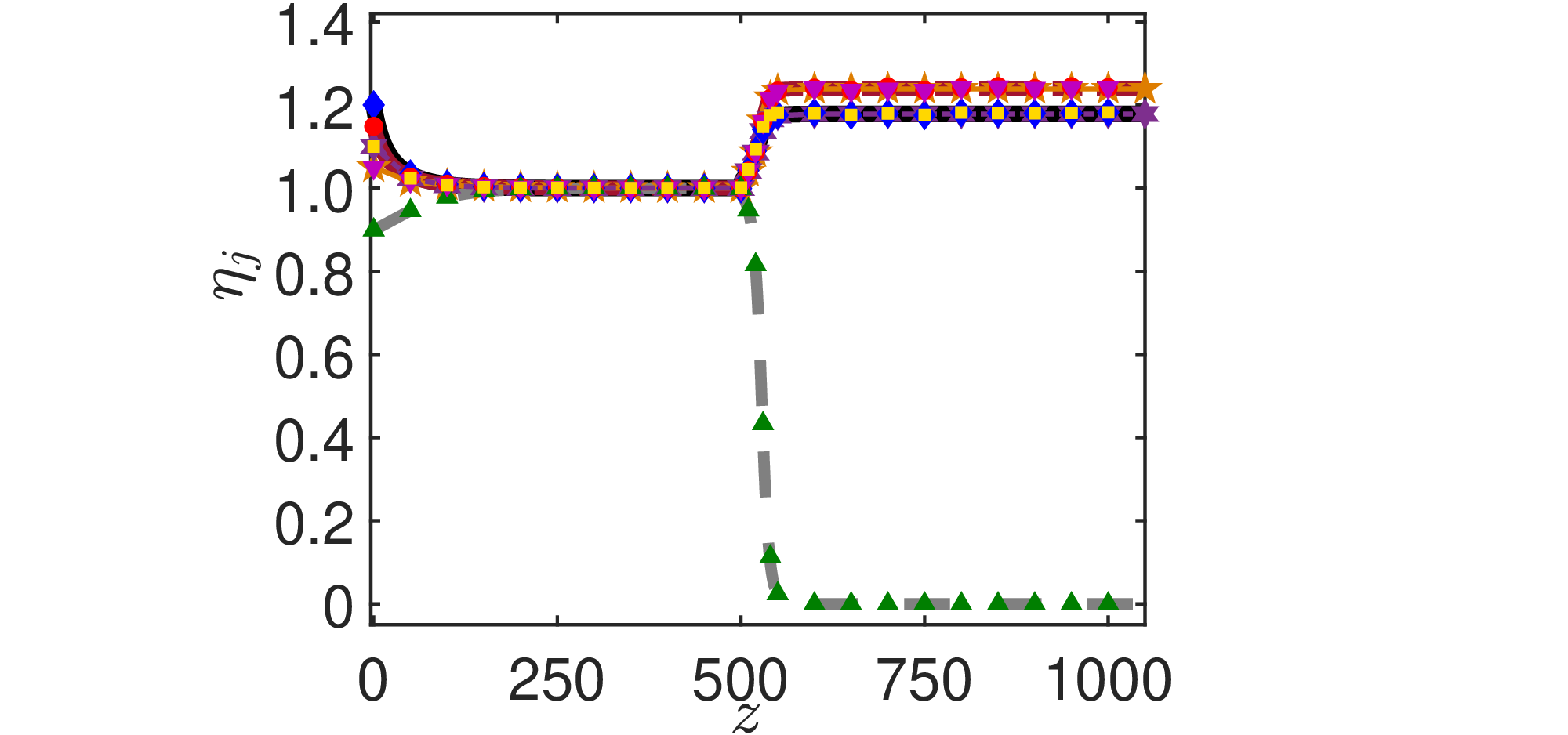}\\
\end{center}
\caption{(Color online) 
$\eta_{j}$ vs $z$ in on-off switching of the sequence $j=3$ in 
five-sequence transmission in a nonlinear waveguide array with 
a weak GL gain-loss profile. The intermediate and switching 
distances are $z_{i}=500$ and $z_{s}=502$, respectively.  
The blue diamonds, red circles, green up-pointing triangles, 
magenta down-pointing triangles, and yellow squares represent $\eta_{j}(z)$ 
with $j=1, 2, 3, 4, 5$, obtained by numerical solution of Eqs. (\ref{gl1})-(\ref{gl3}). 
The solid black, dashed-dotted brown, dashed gray, solid-starred orange, 
and dashed-dotted six-pointed starred magenta curves correspond to $\eta_{j}(z)$ 
with $j=1, 2, 3, 4, 5$, obtained by the LV model (\ref{gl13})-(\ref{gl15}).}
\label{fig14}
\end{figure}

Finally, we describe the results of the numerical simulations for on-off 
switching of the sequence $j=3$. The $z$ dependence of the soliton amplitudes  
obtained in the simulation with Eqs. (\ref{gl1})-(\ref{gl3}) with initial amplitudes 
$\eta_{1}(0)=1.2$, $\eta_{2}(0)=1.15$, $\eta_{3}(0)=0.9$, $\eta_{4}(0)=1.05$, 
and $\eta_{5}(0)=1.1$, which satisfy condition (\ref{gl45}), is shown in 
Fig. \ref{fig14}. A comparison with the prediction of the LV model 
(\ref{gl13})-(\ref{gl15}) is also shown. We observe very good agreement 
between the result of the coupled-NLS simulation and the LV model's prediction.  
More specifically, before the switching (for $0<z<z_{i}$), the numerically obtained 
values of the $\eta_{j}$ approach 1 with increasing $z$, such that all five sequences 
are in an on state. After the switching (for $z>z_{s}$), the values of $\eta_{1}$, 
$\eta_{2}$, $\eta_{4}$, and $\eta_{5}$ tend to new nonzero equilibrium values, 
while the value of $\eta_{3}$ tends to zero. Thus, after the switching, the transmission of 
sequence $j=3$ is turned off, in full agreement with the LV model's prediction. 
The results shown in Figs. \ref{fig11}-\ref{fig14} are very representative, in 
the sense that similar behavior is observed with other sets of the physical 
parameter values and with other initial conditions. It follows that one can 
indeed use stability and bifurcation analysis for the LV model 
(\ref{gl13})-(\ref{gl15}) for designing robust setups for transmission 
stabilization and switching with five soliton sequences in 
nonlinear waveguide arrays. Moreover, the results of our numerical simulations 
with 3, 4, and 5 pulse sequences show that soliton stability and the agreement 
between the simulations results and the LV model's predictions do not decrease 
with an increasing number of sequences. Therefore, these results strongly 
indicate that stable transmission control of the soliton sequences can 
be realized with an arbitrary number of pulse sequences.

\section{Conclusions}
\label{conclusions}

We studied propagation of $J$ colliding soliton sequences in a nonlinear optical 
waveguide array with generic weak GL gain-loss and NN interaction. The propagation was 
described by a system of $J$ weakly perturbed coupled-NLS equations. The GL gain-loss 
with cubic gain, quintic loss, and linear loss with appropriately chosen coefficients 
enables stabilization of the propagation against collision-induced changes in the soliton 
amplitudes and against emission of radiation \cite{PC2012,CPJ2013,NPT2015,PNH2017A}. 
However, in the presence of quintic loss, three-pulse interaction effects become 
important, and the complex nature of these effects limits the stabilization to 
two-sequence systems \cite{PC2012,CPJ2013,NPT2015}, or to systems with a nongeneric 
GL gain-loss \cite{PNH2017A}. The NN interaction property of the optical waveguides 
and the corresponding coupled-NLS models in the current paper leads to the complete 
elimination of collisional three-pulse interaction effects. Therefore, this property 
opens the way for the first investigation of robust control of multiple colliding 
sequences of NLS solitons with generic GL gain-loss and with an arbitrary number 
of sequences, which was carried out in the current paper.

In order to develop waveguide setups for robust transmission stabilization 
and switching, we first derived a reduced model for the dynamics of the soliton 
amplitudes. More specifically, using the results of single-collision analysis in 
Refs. \cite{PNC2010,PC2012} together with collision-rate calculations, we showed that amplitude 
dynamics in a $J$-sequence transmission system can be described by a generalized $J$-dimensional 
LV model with NN interaction, whose form is given by Eqs. (\ref{gl13})-(\ref{gl15}).   
We then carried out linear stability analysis and bifurcation analysis for the 
equilibrium points $(0,0,\dots,0)$ and $(\eta,\eta,\dots,\eta)$ of the LV model, 
which play the key role in transmission stabilization and switching. We found 
that the condition for linear stability of $(0,0,\dots,0)$, inequality (\ref{gl21}), 
is independent of the number of soliton sequences $J$. Furthermore, we obtained 
a simplified form for the characteristic equation of the linearization of the LV model 
about $(\eta,\eta,\dots,\eta)$, which is valid for a general $J$ value. We then used 
the latter equation to obtain the conditions for linear stability of 
$(\eta,\eta,\dots,\eta)$ for $J=3$, $4$, and $5$ soliton sequences. 
Additionally, we used the properties of the equilibrium points of 
the uncoupled nonlinear ODE model (\ref{gl41}) to obtain 
approximate conditions for the regions in phase space, where 
transmission switching can be implemented. Moreover, we showed that the 
conditions for transmission switching can be made more accurate 
by employing the Lyapunov function method for the relevant equilibrium 
points of the full LV model (\ref{gl13})-(\ref{gl15}).                
A similar improvement in the transmission switching conditions was obtained 
by a simple topological argument regarding the locations of the equilibrium 
points of the LV model, which was motivated by the Hartman-Grobman theorem.     
The Lyapunov function analysis also demonstrated that stability of the 
equilibrium points of the LV model is stronger than linear.

The LV model (\ref{gl13})-(\ref{gl15}) is based on several major approximations, 
whose validity might break down at intermediate and large propagation 
distances. For this reason, it is important to check the predictions 
of the LV model by numerical simulations with the weakly perturbed coupled-NLS 
model. We carried out extensive numerical simulations with the coupled-NLS 
model for transmission stabilization and for transmission switching with 
3, 4, and 5 soliton sequences. In all cases, we found very good agreement 
between the simulations results and the predictions of the LV model. 
Furthermore, the quality of the agreement between the LV model's predictions 
and the coupled-NLS simulations was independent of $J$, which is a remarkable 
improvement compared with all previous works on multisequence soliton propagation. 
Based on our results we concluded that robust transmission stabilization and transmission 
switching with an arbitrary number of soliton sequences can indeed be realized in 
nonlinear waveguide arrays with generic weak GL gain-loss and NN interaction.       
Moreover, the results clearly demonstrated that the design of the 
waveguide arrays can be based on stability and bifurcation analysis 
for the equilibrium points of the LV model.

It is worth emphasizing the broader impact of our results, beyond waveguide 
arrays with generic weak GL gain-loss and NN interaction. First, the same methods 
that were developed and used in the current work can be employed for other types of 
waveguide arrays with NN interaction. In particular, they can be used for waveguides, 
in which the collision-induced amplitude shifts are due to delayed Raman 
response \cite{Agrawal2019,Agrawal2020,NP2010,PNT2016}. Second, our results 
open the way for investigating the dynamics of periodic trains of interacting 
coherent patterns in other systems with NN interaction. A major example 
is provided by the dynamics of density pulses in traffic flow through 
multilane highways, where the assumption of NN interaction between pulses 
moving in different lanes is fairly reasonable \cite{Whitham99}. 
Third, our results are also important in the context of research on the many 
systems that are described by the complex GL equation \cite{Hohenberg92,Kramer2002}. 
Indeed, in our previous work in Ref. \cite{PNH2017A}, we provided the 
first example for stable long-distance propagation of multiple periodic 
soliton sequences with more than two sequences in a complex GL system. 
However, the results of Ref. \cite{PNH2017A} were limited, since the GL 
gain-loss profile considered in this work was narrowband, and therefore 
nongeneric, and since the cubic gain and quintic loss did not affect 
the collisional changes in soliton amplitudes at all. In the current 
work, we enhanced the results of Ref. \cite{PNH2017A} significantly 
by providing the first demonstration of stable long-distance propagation 
of an arbitrary number of soliton sequences in systems described by the 
complex GL equation with a {\it generic} (broadband) gain-loss profile. 
In this case, the cubic gain and quintic loss affected both the amplitude 
changes due to single-soliton propagation and the amplitude changes 
induced by intersequence soliton collisions.


\appendix
\section{The pulse-pattern quality integrals}
\label{appendA}   

In this Appendix, we present the theoretical predictions for the pulse patterns 
and their Fourier spectra, and the definition of the $z$-dependent pulse-pattern 
quality integrals $I_{j}(z)$. These quantities were used in Section \ref{simu}, 
in stability analysis for the soliton sequences.

The theoretical predictions for the pulse patterns and for the corresponding 
Fourier spectra are based on the adiabatic perturbation theory for the soliton 
of the cubic NLS equation \cite{PC2020,Kaup91,Chertkov2003,Kaup76}. 
According to this perturbation theory, one expresses the solution $\psi_{j}(t,z)$ 
to the perturbed NLS equation as the sum $\psi_{j}(t,z)=\psi_{js}(t,z)+\nu_{jr}(t,z)$, 
where $\psi_{js}(t,z)$ is the soliton part, and $\nu_{jr}(t,z)$ is the radiation 
part \cite{PC2020,Kaup91,Chertkov2003}. In the current work, the soliton part 
$\psi_{js}$ is just the sum of $2K$ fundamental soliton solutions of the unperturbed 
cubic NLS equation with slowly varying parameters, whose peaks are separated by 
a constant integer multiple of $T$ \cite{PC2020,Kaup91,Chertkov2003}. 
We assume that $|\psi_{js}(t,z)| \gg |\nu_{jr}(t,z)|$ for any $t$ and $z$. 
We therefore take $\psi_{js}(t,z)$ as the theoretical prediction for 
$\psi_{j}(t,z)$, i.e., $\psi_{j}^{(th)}(t,z) \equiv \psi_{js}(t,z)$.  
It follows that $\psi_{j}^{(th)}(t,z)$ is given by \cite{PNT2016}:   
\begin{eqnarray}&&
\!\!\!\!\!\!\!\!\!\!\!\!\!\!\!\!\!\!\!\!\!\!
\psi_{j}^{(th)}(t,z) 
= \eta_{j}(z)e^{i\theta_{j}(z)}
\sum_{k = -K}^{K-1}\frac{\exp\{ i\beta_{j}(z) 
\left[ t-y_{j}(z)-kT \right] \}}
{\mathrm{cosh}\{\eta_{j}(z)\left[t-y_{j}(z)-kT\right]\}},  
\label{appendA1}
\end{eqnarray}  
where $\eta_{j}(z)$ is the common amplitude of the $j$th sequence solitons, 
$\beta_{j}(z)$ is the common frequency, $\theta_{j}(z)$ is the common overall phase, 
$y_{j}(z)=\Delta y_{j}(z) + y_{j0}$, and $\Delta y_{j}(z)$ is the common overall 
position shift. The theoretical prediction for $\hat{\psi}_{j}(\omega,z)$ is taken 
as the Fourier transform of $\psi_{js}(t,z)$ \cite{PNT2016}: 
\begin{eqnarray} &&
\!\!\!\!\!\!\!\!\!\!\!\!\!\!\!\!\!\!\!\!\!\!
\hat{\psi}^{(th)}_{j}(\omega,z)=\left(\frac{\pi}{2}\right)^{1/2}
\mbox{sech}\left\{\frac{\pi\left[\omega - \beta_{j}(z)\right]}{2\eta_{j}(z)}\right\}
e^{i\theta_{j}(z) - i\omega y_{j}(z)}{\sum_{k=-K}^{K-1}e^{-ikT\omega}}.
\;\;\;
\label{appendA2}
\end{eqnarray}       
The theoretical pulse pattern of the $j$th sequence, $|\psi_{j}^{(th)}(t,z)|$, 
is then calculated by using Eq. (\ref{appendA1}), while the theoretical Fourier spectrum 
of the $j$th sequence, $|\hat{\psi}_{j}^{(th)}(\omega,z)|$, is obtained with 
Eq. (\ref{appendA2}). In these calculations, $\eta_{j}(z)$ is obtained by the 
LV model (\ref{gl13})-(\ref{gl15}), $\beta_{j}(z)=\beta_{j}(0)$, and $y_{j}(z)$ 
is measured from the numerical simulation with Eqs. (\ref{gl1})-(\ref{gl3}).

The pulse-pattern quality integral for the $j$th sequence $I_{j}(z)$ measures 
the deviation of the numerically obtained pulse pattern $|\psi_{j}^{(num)}(t,z)|$ 
from the theoretical prediction $|\psi_{j}^{(th)}(t,z)|$. More precisely, we 
define $I_{j}(z)$ by \cite{PNT2016}:    
\begin{eqnarray}&&
\!\!\!\!\!
I_{j}(z)=
\left[ \int\limits_{-KT}^{KT} 
\left| \psi _j^{(th)}\left( {t,z} \right) \right|^2 dt \right]^{-1/2}
\nonumber \\&&
\times
\left\{\int\limits_{-KT}^{KT} 
\left[\;\left| \psi _j^{(th)}\left( {t,z} \right) \right| - 
\left| \psi _j^{(num)}\left( {t,z} \right) \right| \; \right]^2 dt 
\right\}^{1/2},   
\label{appendA3}
\end{eqnarray}   
where $1 \le j \le J$. Therefore, the $I_{j}(z)$ integrals measure 
both distortions in the shape of the pulses, and deviations of the numerically 
obtained values of the soliton parameters from the values predicted by the 
adiabatic perturbation theory and by the LV model (\ref{gl13})-(\ref{gl15}).




\begin{thebibliography}{}
\bibitem{Multisequence} Multisequence transmission is also known as multichannel transmission 
and as wavelength-division-multiplexed transmission.
\end{thebibliography}
\end{document}